\documentclass[12pt]{article}
\pdfoutput=1
\usepackage[letterpaper, margin=1in]{geometry}
\usepackage{amsmath, amssymb}
\usepackage{cite}
\usepackage{graphicx}
\usepackage[font=small,labelfont=bf]{caption}
\usepackage[normalem]{ulem}

\usepackage{xcolor}
\usepackage{slashed}
\usepackage{cancel}
\usepackage{bm}
\usepackage{verbatim}
\usepackage{tabularx}

\newcommand{\myfrac}[3][0pt]{\genfrac{}{}{}{}{\raisebox{#1}{$#2$}}{\raisebox{-#1}{$#3$}}}
\newcommand{\lab}[1]{{\mathrm{#1}}}

\definecolor{DarkRed}{rgb}{0.6,0,0}
\definecolor{DarkGreen}{rgb}{0,0.6,0}
\definecolor{DarkBlue}{rgb}{0,0,0.6}

\usepackage{setspace}
\usepackage{authblk}
\usepackage{physics}
\usepackage{caption}
\usepackage{subcaption}
\usepackage{multirow}

\usepackage{bbm}

\usepackage{dsfont}

\usepackage{etoolbox}

\let\OLDthebibliography\thebibliography
\renewcommand\thebibliography[1]{
	\OLDthebibliography{#1}
	\setlength{\parskip}{3.0pt plus 2.5pt minus 1.0pt}
	\setlength{\itemsep}{3.0pt plus 2.5pt minus 1.0pt}
}

\usepackage[colorlinks=true,linktocpage=true,linkcolor=DarkBlue,citecolor=DarkGreen,urlcolor=DarkGreen]{hyperref}

\definecolor{goodgreen}{rgb}{0,.6,0.4}

\usepackage[greek,english]{babel}

\newcommand{\mug}{(g-2)_\mu}
\newcommand{\bknn}{ B^+\rightarrow K^+ \bar{\nu}\nu }
\newcommand{\bknnbr}{\mathrm{BR} \left( B^+\rightarrow K^+ \bar{\nu}\nu \right)}
\def\hc{\textrm{h.c.}}

\newcommand{\sbar}{{\scalebox{1.0}[1.05]{$\mid$}}}
\newcommand{\scharges}[2]{\sbar \raisebox{0.1em}{\scriptsize$[#1] + [#2]$}\sbar}
\newcommand{\schargesm}[2]{\sbar \raisebox{0.1em}{\scriptsize$[#1] - [#2]$}\sbar}

\begin{document}

\title{\bf\Large Wrinkles in the Froggatt--Nielsen Mechanism and Flavorful New Physics}

\author{Pouya Asadi$^{1}$,~
Arindam Bhattacharya$^{2}$,~
Katherine Fraser$^{2}$,\\\vskip -0.625em
Samuel Homiller$^{2}$,~ and 
Aditya Parikh$^{3}$\\\vskip 0.5em
{\small \color{gray} 
\texttt{pasadi (@uoregon.edu), arindamb, kfraser, shomiller (@g.harvard.edu)},\\\vskip -0.25em 
\texttt{aditya.parikh (@stonybrook.edu)}} \\\vskip0.5em
\small \textit{${}^1$Institute for Fundamental Science and Department of Physics,}\\
\textit{University of Oregon, Eugene, OR 97403, USA} \\\vskip 0.5em
\small \textit{${}^2$Department of Physics, Harvard University, Cambridge, MA 02138, USA} \\\vskip 0.5em
\small \textit{${}^3$C.N. Yang Institute for Theoretical Physics,}\\
\textit{Stony Brook University, Stony Brook, NY 11794, USA}
}

\date{\today}

\maketitle

\begin{abstract}
\begin{spacing}{1.05}\noindent
When the Froggatt--Nielsen mechanism is used to explain the Standard Model flavor hierarchy, new physics couplings are also determined by the horizontal symmetry. 
However, additional symmetries or dynamics in the UV can sometimes lead to a departure from this na\"ive scaling for the new physics couplings. 
We show that an effective way to keep track of these changes is by using the new spurions of the $\mathrm{U}(3)^5$ global flavor symmetry, where we parameterize extra suppression or enhancement factors, referred to as {\it wrinkles}, using the same power counting parameter as in the original Froggatt--Nielsen model. As a concrete realization, we consider two flavor spurions of the $S_1$ leptoquark, and demonstrate that wrinkles can be used to make an enhanced value of $\textrm{BR}(B^+ \to K^+\nu\bar{\nu})$ consistent with other flavor observables. 
We also present example UV models that realize wrinkles, and comment on choosing consistent charges in ordinary Froggatt--Nielsen models without the typical monotonicity condition. 
\end{spacing}
\end{abstract}

\begin{spacing}{1.15}

\newpage
\tableofcontents

\vskip 1cm

\section{Aperitif}
\label{sec:intro}

Flavor physics has been a harbinger of physics beyond the Standard Model (BSM) at various points in time, from predicting the existence of the charm quark~\cite{Bjorken:1964gz, Glashow:1970gm} to estimating the mass of the top quark \cite{Campbell:1981rg, Shifman:1987nd, Ellis:1987mm, Martin:1989ux} long before its discovery at the Tevatron~\cite{CDF:1995wbb, D0:1995jca}. 
Precision experiments, in particular, help establish or find violations of the Standard Model (SM) symmetry structures, and prove to be noteworthy indirect probes of new physics whose mass scale lies beyond the reach of direct collider searches; see Refs.~\cite{EuropeanStrategyforParticlePhysicsPreparatoryGroup:2019qin,Artuso:2022ouk} for reviews of many such experiments.

A primary goal of flavor physics is to understand the appearance of large hierarchies in the masses and mixing angles of the SM fermions. 
The two most popular solutions to this puzzle are (i) the Froggatt--Nielsen (FN) mechanism and its variations~\cite{Froggatt:1978nt, Leurer:1992wg, Leurer:1993gy,Pouliot:1993zm},
and (ii) extra dimensional models where an $\mathcal{O}(1)$ difference in the bulk masses of fermions gives rise to an exponential hierarchy between the observed masses in the IR~\cite{Arkani-Hamed:1999ylh, Gherghetta:2000qt, Huber:2000ie, Kaplan:2001ga, Ahmed:2019zxm}.  
Other notable possibilities include generating the mass hierarchy via running to the IR in extensions of the SM with scale invariant sectors in the UV~\cite{Nelson:2000sn}, 
or radiatively generating the Yukawas with the hierarchy governed by powers of the loop expansion parameter~\cite{Weinberg:1972ws, Georgi:1972hy, Barr:1976bk,Balakrishna:1988ks}. 
A review of these and other dynamical solutions to the flavor puzzle can be found in Refs.~\cite{Babu:2009fd, Feruglio:2015jfa, Feruglio:2019ybq}. 
In what follows, we focus our attention on the FN mechanism.

In the FN mechanism, the hierarchies in the SM fermion sector arise as different powers of a small expansion parameter. This expansion parameter is given by the ratio of the vacuum expectation value (vev) of a scalar field, known as the flavon, over a heavy mass scale. 
The SM Yukawa couplings are generated by non-renormalizable operators involving the chiral SM fermions, the Higgs, and the flavon. The dimensionality of these operators---and the resulting power of the expansion parameter that appears---is dictated by the charges of the SM fermions under a new Abelian horizontal symmetry, $\lab{U}(1)_H$, which is broken by the flavon.
As we will discuss, there is additional freedom in the assignment of these charges that was overlooked in Ref.~\cite{Froggatt:1978nt}.
In the original FN paper, it was supposed that these irrelevant operators are generated by ``chains'' including heavy vector-like matter, also charged under $\lab{U}(1)_H$. 
A number of variations to this model have been proposed, including ``inverted'' models \cite{Smolkovic:2019jow}, where the flavon vev is larger than the heavy mass scale.

One of the drawbacks of invoking the FN mechanism is that the new dynamics responsible for the SM hierarchies can exist at scales far above the weak scale, beyond the reach of direct experimental probes. 
Nevertheless, given the other shortcomings of the SM---the electroweak hierarchy problem in particular---there is ample reason to expect new physics at or near the TeV scale.
If the new physics is {\em flavorful} (i.e., it involves non-universal couplings to SM matter fields), its flavor structure may also be dictated by the FN dynamics.
This argument can also be run in reverse: given the stringent constraints from precision measurements of the SM, for new physics to exist at the TeV scale it must either be flavor-blind or incorporate some symmetry arguments to suppress flavor-violation~\cite{DAmbrosio:2002vsn, Egana-Ugrinovic:2018znw}.
This reasoning is familiar in the supersymmetric context, where it is understood that squarks must either be degenerate or flavor-aligned~\cite{Nir:1993mx}.

In this light, it is clearly worthwhile to study the application of the FN mechanism to the couplings of new BSM fields.
This is particularly true when flavorful new physics is invoked to explain potential discrepancies between experimental results and the SM expectations: should one of these discrepancies become an unambiguous signal of new physics, we might glean information about the dynamics associated with flavor in the UV.
This approach was advocated in Refs.~\cite{Feldmann:2006jk, Bordone:2019uzc, Bordone:2020lnb}, and we will review it extensively in this work.
An immediate consequence of this framework is that many different experimental observables become correlated. 
These correlations challenge some of the simplest solutions to various flavor anomalies, as the couplings and masses required to explain the discrepancy violate bounds set by other observables such as lepton flavor violating (LFV) processes or flavor changing neutral currents.

The goal of this work is to explore how these considerations can change if the FN setup is amended with additional symmetries or structure in the UV.
We do this by working in an effective field theory (EFT) framework, including the SM and new BSM fields, with their couplings to fermions treated as spurions under the $\lab{U}(3)^5$ flavor symmetry of the SM.
In this framework, we can introduce controlled deviations from the size of these spurions dictated by the horizontal charges. 
We refer to these deviations as {\em wrinkles}, since they appear in the UV as changes in the length of the chain diagrams responsible for the Yukawas in the IR. 
Wrinkles can exist in SM or BSM spurions, and allow us to relax the correlations between different observables, permitting sizable new physics contributions to some observables while satisfying other experimental bounds.

Importantly, while wrinkles allow for much greater flexibility in the couplings of BSM fields to SM fermions, this flexibility is not without bound.
If the effective theory is to be faithfully embedded in the FN mechanism, radiative corrections must not spoil the relationship between the couplings in the IR and the non-renormalizable operators in the UV.
This requirement has been previously formulated as a consistency condition in 
the context of minimal flavor violation EFTs \cite{Feldmann:2006jk} (see also Ref.~\cite{Bordone:2019uzc}). 
While these conditions are trivially satisfied in ordinary FN models, we show that they put 
meaningful bounds on wrinkled FN setups.

Since this wrinkled FN setup can be applied to any new physics, we will illustrate its application in an example, where the SM is extended by a single leptoquark, denoted $S_1$ in the nomenclature of Ref.~\cite{Dorsner:2016wpm}. See Refs.~\cite{Baver:1994hh,Bordone:2020lnb} for previous discussions of the $S_1$ leptoquark model with horizontal symmetries.
We will use this leptoquark to enhance the branching ratio of $\bknn$, which currently shows a small discrepancy with SM predictions \cite{Belle-II:2021rof} and will be precisely measured at the Belle~II experiment in the coming years.
Without wrinkles, the charges and masses required to generate a large $\bknn$ signal also imply the existence of large signals in other correlated observables, such as LFV decays or leptonic meson decays.
We will show a simple example where a wrinkled FN setup evades these bounds while satisfying the consistency conditions alluded to above. 
As we will see, the bound on the wrinkles implies other correlated signals are generated near detection thresholds in this example, and could potentially be seen in the near future.

In the coming years, troves of new data from colliders and small-scale experiments searching for signs of flavorful new physics will begin stress-testing the delicate flavor structure of the SM. 
Given the substantial motivation for BSM physics, this structure could break and potentially start showing signs of deviations from the SM expectation.
In preparation for such deviations, it is timely to develop new model-building tools which enable embedding their solutions in UV complete frameworks. 
Wrinkles in an FN Ansatz are a flexible, bottom-up tool that allow for a broader exploration of the complementarity of different flavor probes, while reliably parameterizing more sophisticated UV models of flavor. 
As such, they present a natural setup to search for a consistent IR picture of new physics with flavor, should any deviations from the SM come to light.

This paper is organised as follows: in \S\ref{sec:fn_intro}, we review the FN mechanism, its solution to the flavor hierarchy problem in SM and how it furnishes suitable Ans\"atze for couplings arising from new BSM physics. 
Next, in \S\ref{sec:wrinkles}, we introduce the concept of wrinkles for the FN mechanism, discuss constraints on them, and provide examples for how they can arise from UV complete models. 
In \S\ref{sec:bknunu}, we provide a concrete example of applying wrinkles to the $S_1$ scalar leptoquark embedded in a FN model. 
We demonstrate that wrinkles allow one to simultaneously explain bounds on BSM physics from current precision flavor observables, while also retaining predictive power for potential future measurements. 
We conclude in \S\ref{sec:conclusion}. Appendix~\ref{app:consistency} provides details about bounds on wrinkles arising from consistency conditions. 
Appendix~\ref{app:obs} provides details on flavor observable computations in the $S_1$ leptoquark model.

\section{Amuse-bouche: Froggatt--Nielsen and BSM Physics}
\label{sec:fn_intro}

The lepton and quark Yukawas and mixing angles present a clear generational hierarchy, with the charged particle masses ranging over five orders of magnitude. 
This hierarchy implores an explanation in the UV. 
Searches for flavorful new physics are carried out in pursuit of such an explanation. 
Hence, if any anomaly emerges in these experiments, it is well-motivated to  embed its BSM solutions within UV models that explain the flavor hierarchy as well. 

The FN mechanism~\cite{Froggatt:1978nt} provides a four-dimensional, field-theoretic explanation for this hierarchy, replacing the small dimensionless parameters with a power counting in powers of an inverse mass scale, fixed by a symmetry.
In this section, we review how this mechanism can explain the parameters in the SM matter sector, with an emphasis on the EFT point of view. We will then discuss how this perspective can naturally be extended to BSM physics.

\subsection{Review of the Froggatt--Nielsen Mechanism}
\label{subsec:fn_review}

The basic idea of the FN mechanism is to introduce a horizontal symmetry, $\lab{U}(1)_H$, under which different generations of the SM fermions have different charges. The horizontal symmetry is assumed to be spontaneously broken by the vacuum expectation value of a SM singlet scalar field, $\varphi$---the {\em flavon}. 
Assuming our EFT is valid up to some cutoff scale $M$, we are led to a natural expansion parameter $\lambda = \langle \varphi \rangle / M$, which appears in non-renormalizable operators involving the SM fermions. Later on, we will associate this scale $M$ with the mass of new heavy fermions. Without loss of generality, we take the SM Higgs to be neutral under $\lab{U}(1)_H$ and take the flavon charge to be $-1$.

At scales just below the cutoff, the lowest dimension operators involving the SM fermions and the Higgs take the form
\begin{equation}
\label{eq:fn_eft}
\mathcal{L} \,\supset\, 
r^u_{ij} \frac{\varphi^{(\dagger) m_{ij}}}{M^{m_{ij}}} Q_i H \bar{u}_j 
+ r^d_{ij} \frac{\varphi^{(\dagger) n_{ij}}}{M^{n_{ij}}} Q_i H^c \bar{d}_j 
+ r^e_{ij} \frac{\varphi^{(\dagger) l_{ij}}}{M^{l_{ij}}} L_i H^c \bar{e}_j + \hc
\end{equation}
where $(Q_i,~\bar{u}_i,~\bar{d}_i,~L_i,~\bar{e}_i)$ are different SM fermions, subscripts on fermion fields refer to different generations, $r_{ij}$ are $\mathcal{O}(1)$ couplings,  
\begin{equation}
m_{ij} = \big| [Q_i] + [\bar{u}_j] \big|, \qquad
n_{ij} = \big| [Q_i] + [\bar{d}_j] \big|, \qquad
l_{ij} = \big| [L_i] + [\bar{e}_j] \big|, 
\label{eq:SMdiffs}
\end{equation}
and the square brackets indicate the $\lab{U}(1)_H$ charge. The hermitian conjugate on $\varphi$ appears if the sum of charges inside the absolute value is negative.
At energies below $\langle \varphi \rangle$, these operators appear as the Yukawa couplings of the SM Higgs, with the coupling matrices given by
\begin{equation}
\label{eq:fn_sm_Yukawas}
Y_{Q\bar{u}}^{ij} = r^u_{ij}\frac{\langle\varphi^{(\dagger)}\rangle^{m_{ij}}}{M^{m_{ij}}} \sim \lambda^{m_{ij}}, \qquad 
Y_{Q\bar{d}}^{ij} = r^d_{ij}
\frac{\langle\varphi^{(\dagger)}\rangle^{n_{ij}}}{M^{n_{ij}}} \sim \lambda^{n_{ij}}, \qquad
Y_{L\bar{e}}^{ij} = r^e_{ij}
\frac{\langle\varphi^{(\dagger)}\rangle^{l_{ij}}}{M^{l_{ij}}} \sim \lambda^{l_{ij}} .
\end{equation}
This scaling implies that even modest differences in horizontal charges give rise to exponential hierarchies in Yukawa couplings. 
To connect with the observed flavor structure of the SM, we identify $\lambda$ with the Cabbibo angle, $\sim 0.2$, so that the CKM matrix hierarchies follow naturally from the Wolfenstein parameterization~\cite{Wolfenstein:1983yz}. 
We refer to this setup as vanilla FN.

At the $\mathcal{O}(1)$ level, the masses and mixing angles are
\begin{equation}
\label{eq:qlmixing}
V_{ij} \sim \lambda^{\schargesm{Q_i}{Q_j}}, \quad
U_{ij} \sim \lambda^{\schargesm{L_i}{L_j}}, 
\end{equation}
\begin{equation}
\label{eq:qlmass}
m^u_{i} \sim \lambda^{\scharges{Q_i}{\bar{u}_i}}, \quad
m^d_{i} \sim\lambda^{\scharges{Q_i}{\bar{d}_i}}, \quad 
m^{l}_{i} \sim \lambda^{\scharges{L_i}{\bar{e}_i}}, 
\end{equation} 
where $V$ ($U$) is the CKM \cite{Cabibbo:1963yz,Kobayashi:1973fv} (PMNS \cite{Pontecorvo:1957qd,Maki:1962mu}) matrix. 

\begin{table}
\renewcommand\arraystretch{1.2}
\begin{center}
\begin{tabular}{|c|c|c|c|}
\hline 
\quad\quad\quad	& Gen. 1 & Gen. 2 & Gen. 3 \\ 
\hline 
$Q$ & $-q_0-3 X$ & $-q_0 -  2X$ & $-q_0$ \\ 
\hline 
$\bar{u}$ & $q_0+3 X \pm 7$ & $q_0 - X$ & $q_0$ \\ 
\hline 
$\bar{d}$ & $q_0+3 X \pm 6$ & $q_0 - 3X$ & $q_0 - 2X$ \\ 
\hline 
$L$ & $l_0 + Y$ & $l_0$ & $l_0$\\ 
\hline 
$\bar{e}$ & $-l_0 - Y \pm 8$ & $-l_0 + 5Y$ & $-l_0 + 3Y$ \\ 
\hline 
\end{tabular} 
\caption{The most general horizontal charge assignment that explains the SM masses and mixings in FN with $\lambda \sim 0.2$. $q_0$ and $l_0$ denote general shifts in quark and lepton charges, respectively, that leave the IR masses and mixings unchanged. $X,Y=\pm 1$ denote the correlations between different charges that are required by the CKM and PMNS matrices. For every value of $(q_0,l_0)$, we have $2^5$ choices for the charge assignments. In supersymmetric theories, holomorphy sets $X=-Y=-1$ and picks the positive sign for first generation RH fermions.}
\label{tab:gen_charges}
\end{center}
\end{table}

The most general horizontal charge-assignment that gives rise to the observed structure of the CKM and PMNS matrices and SM fermion masses is given in Table~\ref{tab:gen_charges}.\footnote{In general, shifts of $\pm 1$ in most of these charges can be tolerated when random $\mathcal{O}(1)$ Yukawa couplings in the UV model are taken into account and the fact that the expansion parameter $\lambda$ is not particularly small is considered. The anarchic structure of the PMNS matrix, in particular, leaves room for such small changes in the charges; see Refs.~\cite{Aloni:2021wzk,Cornella:2023zme} for further exploration of these shifts.} 
We have the overall freedom to shift the charges of all quarks (leptons) by the same amounts
$q_0$ ($l_0$), respectively. 
Once these shifts are chosen, the CKM and PMNS structure constrain the other LH quarks' and leptons' charges. 
As indicated in Eq.~\eqref{eq:qlmixing}, these mixing matrices only fix the absolute value of the difference between charges, hence the freedom in choosing $X,Y=\pm 1$ in the table. 
The appearance of $X,Y$ in multiple entries captures the correlation between those charges. 
To find the RH fermion charges we use the measured values of masses in the SM. As in the case of mixing, Eq.~\eqref{eq:qlmass} only fixes the absolute value of the charge difference between LH and RH fermions, leaving the sign undetermined. 
We choose the signs so that the eigenvector associated with the heaviest (lightest) mass eigenstate has the biggest overlap with the third (first) generation for each type of fermion. 
To check this, we generated 10000 mass matrices for each charge assignment, drawing new random numbers $r^{u,d,e}_{ij} \in (0.2,1)$ for each test. For every charge assignment, we confirmed that a substantial fraction of trials yield the correct mixing patterns and 
mass eigenvalues that are within a factor of two of the experimentally-measured values.  

In the original FN proposal, it was assumed that the charges of all five types of fermions ($Q$, $\bar{u}$, $\bar{d}$, $L$, and $\bar{e}$) are ordered monotonically between different generations. Table~\ref{tab:gen_charges} indicates that, while some correlations between LH and RH fermions of the second and third generation (captured by $X,Y$) are needed to generate the correct mass eigenstates, the monotonicity condition can be removed for first generation RH fermions without distorting the model's prediction for SM masses. 
This manifests itself as a binary choice in the charge of each first generation RH fermion ($\bar{u},~\bar{d},~\bar{e}$).

It is also popular to consider supersymmetric variations of FN models. In the supersymmetric case, holomorphy of the superpotential forbids terms with $\varphi^{\dagger}$ instead of $\varphi$~\cite{Leurer:1992wg,Leurer:1993gy}. This eliminates a great deal of the freedom in charge assignments tabulated in Table~\ref{tab:gen_charges}. Specifically, it fixes $X=-Y=-1$ and picks the positive sign for first generation RH fermions, leaving only the separate overall shifts in the quark and lepton charges, $q_0$ and $l_0$. It also enforces the monotonicity of the horizontal charges across different generations. However, since we do not explore the supersymmetric case in detail in the rest of this paper, we do not need to enforce these constraints.

The simplest UV completion of this effective theory (and the one imagined by Froggatt and Nielsen \cite{Froggatt:1978nt}) is to introduce a set of vector-like fermions $F$ with mass $M$ that live in an SM representation permitting Yukawa couplings between the Higgs and SM fermions. We assume the existence of heavy fermions with all horizontal charges necessary to complete the 
SM Yukawas with Yukawa couplings to the flavon $\sim \varphi F \bar{F}'$. The flavon Yukawa couplings are assumed to be $\mathcal{O}(1)$, leading to the effective theory in Eq.~\eqref{eq:fn_eft} with $\mathcal{O}(1)$ Wilson coefficients denoted by $r_{ij}$.

\begin{figure}[t]
\centering
\includegraphics[height=3.5cm]{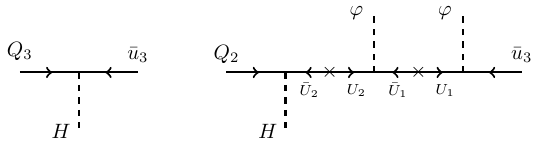}
\caption{Example diagrams leading to the effective operators for the up-type Yukawa couplings with vector-like heavy fermions $U$ and $\bar{U}$, where $\bar{U}$ has the same SM quantum numbers as $\bar{u}$. The subscripts on the $U$ fields refer to the horizontal charge of ${U}$, and we have taken charges from Table~\ref{tab:gen_charges} with $X = -1$ and $q_0 = 0$.}
\label{fig:fn_chain_example}
\end{figure}

As an example, the up-type Yukawa couplings can be generated by ``chain'' diagrams such as those shown in Figure~\ref{fig:fn_chain_example}. 
The top Yukawa arises at the renormalizable level, but the suppressed couplings arise by introducing the vector-like pair $U$ and $\bar{U}$, where $\bar{U}$ has the same quantum numbers under the SM gauge groups as $\bar{u}$. The subscripts indicate the $\lab{U}(1)_H$ charge of $U$. For instance, the chain shown on the right side of Figure~\ref{fig:fn_chain_example} gives rise to a $\lambda^2$ suppression in the coupling of $Q_2 \bar{u}_3$. 
If there exist heavy fermions with SM charges similar to $Q$ and the correct horizontal charges, chain diagrams with the Higgs and flavon insertions interchanged will contribute as well.
Similar chains give rise to the Yukawa couplings for other SM fermions.

Models of FN constructions with additional symmetries, multiple  expansion parameters, or expansion parameters that are allowed to freely vary have also been developed in the literature, e.g. see \cite{Leurer:1992wg,Leurer:1993gy,Nir:1996am,Aloni:2021wzk,Nakai:2021mha,Fedele:2020fvh}. For simplicity, however, in this work we focus on FN setups with only one expansion parameter, which we identify with the Cabbibo angle, and develop a systematic way for small deviations from them. We can straightforwardly generalize our discussions below to more baroque FN setups.

As a final note, in a UV complete model, quantum gravity considerations require that the horizontal symmetry be embedded in a gauge symmetry~\cite{Banks:2010zn}, which in turn demands the cancellation of all its anomalies.\footnote{The lack of evidence for the (pseudo-)Nambu--Goldstone boson associated with the spontaneous breaking of the horizontal symmetry is also often used as motivation for gauging it. However, models with a potentially viable Goldstone exist.
See Refs.~\cite{Wilczek:1982rv,Ema:2016ops,Calibbi:2016hwq,Bonnefoy:2019lsn} for examples where the Goldstone is identified with the QCD axion.} 
We have checked that the general charge assignment of Table~\ref{tab:gen_charges} can not cancel all gauge anomalies in the typical FN UV completion, see also Ref.~\cite{Bonnefoy:2019lsn} for a similar conclusion. This conclusion is also corroborated by Refs.~\cite{Allanach:2018vjg,Costa:2019zzy}, which deduce that the general charge assignments that can explain the SM Yukawa hierarchy can not be anomaly-free by studying general extensions of the SM with a new anomaly-free $\lab{U}(1)$ gauge group. 
As a result, in such a construction one should resort to either introducing new heavy chiral fermions (and subsequently extending the scalar sector so as to generate a mass for these fermions) or the Green-Schwarz mechanism to cancel anomalies \cite{Green:1984sg}. We will leave further investigations of anomaly cancellation for future work.

\subsection{Froggatt--Nielsen and Flavorful New Physics \label{sec:FNplusBSM}}

When introducing new physics, some assumptions must be made about the couplings of SM fields to new particles. These couplings are generically non-universal unless governed by additional structure such as new gauge symmetries.
Given the hierarchies that exist in the SM fermion couplings, it is a priori unclear what a ``natural'' size for such non-universal couplings should be. However, if one assumes a UV explanation of the flavor hierarchy such as the FN mechanism, there is a natural Ansatz for the new physics couplings as well.

The phenomenological significance of such an Ansatz lies in the fact that it correlates predictions of a BSM model for various flavorful observables in the IR. 
Thus, depending on the Ansatz, a model built for explaining a discrepancy in the data will give rise to correlated signals in other constraining observables. 
For instance, any solutions of the $\mug$ anomaly with non-minimal flavor Ansatz gives rise to unacceptably large contributions to various LFV decays, especially $\tau \rightarrow \mu \gamma$.

To better understand such Ans\"atze, it is useful to organize our thinking in terms of the global flavor symmetry of the SM:
\begin{equation}
G_{\textrm{flavor}} = \lab{SU}(3)_Q \times \lab{SU}(3)_u \times \lab{SU}(3)_d \times \lab{SU}(3)_L \times \lab{SU}(3)_e \times \lab{U}(1)^5 ,
\label{eq:Gflavor}
\end{equation}
where three of the $\lab{U}(1)$ factors can be identified with hypercharge, baryon number and lepton number.
This symmetry acts on the generation indices of the chiral matter in the SM, with the unbarred (barred) fields transforming as triplets (anti-triplets), respectively. The symmetry is broken explicitly by the Yukawa matrices, but formal invariance under $G_{\textrm{flavor}}$ can be restored if we promote the Yukawas to transform as spurions:
\begin{equation}
Y_{Q\bar{u}} \sim (\mathbf{\bar{3}}_Q, \mathbf{3}_u), \qquad
Y_{Q\bar{d}} \sim (\mathbf{\bar{3}}_Q, \mathbf{3}_d), \qquad
Y_{L\bar{e}} \sim (\mathbf{\bar{3}}_L, \mathbf{3}_e) .
\label{eq:GflavorSM}
\end{equation}
This formalism can be extended in a straightforward way to new physics with any new spurions of $G_{\textrm{flavor}}$ \cite{Feldmann:2006jk, Bordone:2019uzc,Bordone:2020lnb}. 
New fields are taken to be singlets of the $\lab{SU}(3)^5$ part of the SM flavor group, and their couplings to SM fermions then have definite transformation properties under $G_{\textrm{flavor}}$.

As an example, consider the scalar leptoquark $S_1$, a color anti-fundamental with hypercharge $Y = 1/3$. This allows for the renormalizable couplings to SM fields,\footnote{The SM gauge symmetries also permit the couplings $S_1 \bar{u} \bar{d}$ and $S_1 Q^{\dagger}Q^{\dagger}$, which lead to proton decay. We can forbid these couplings by enforcing conservation of baryon number and endowing the leptoquark with a baryon number of $-1/3$, or by potentially gauging some discrete subgroup. Therefore, in the rest of this work, we ignore these couplings.
We note that the ``wrinkles'' introduced in \S\ref{sec:wrinkles} cannot entirely alleviate the proton decay constraint, necessitating a symmetry-based explanation.}

\begin{equation}
\label{eq:S1lagrangian}
\mathcal{L} \supset - \Delta_{QL}^{ij} \epsilon^{ab} S_1 Q_{bi} L_{aj} - \Delta_{\bar{u}\bar{e}}^{ij} S_1^{\dagger} \bar{u}_i \bar{e}_j + \hc ,
\end{equation}
where the spinor indices are implicit, $a, b$ are $\lab{SU}(2)_L$ fundamental indices, $\epsilon^{12} = +1$, and $(i,j)$ are flavor indices. 
The $\Delta_{QL}$ coupling also appears in R-parity violating supersymmetric models, where $S_1$ is identified with a down squark; these models have $\Delta_{\bar{u}\bar{e}} = 0$~\cite{Barbier:2004ez}.
The new Yukawa couplings $\Delta_{QL}$ and $\Delta_{\bar{u}\bar{e}}$ transform as
\begin{equation}
\Delta_{QL} \sim (\mathbf{\bar{3}}_Q, \mathbf{\bar{3}}_L), \qquad
\Delta_{\bar{u}\bar{e}} \sim (\mathbf{3}_u, \mathbf{3}_e).
\label{eq:GflavorLQ}
\end{equation}

In the absence of any flavor Ansatz, the matrices $\Delta_{QL}$ and $\Delta_{\bar{u}\bar{e}}$ are arbitrary $3\times 3$ complex matrices. 
However, when embedded in a vanilla FN setup, and assuming the $S_1$ leptoquark is neutral under $\lab{U}(1)_H$, we find an Ansatz for the hierarchies present in the spurions $\Delta_{QL}$ and $\Delta_{\bar{u}\bar{e}}$.  
In analogy with Eq.~\eqref{eq:fn_sm_Yukawas}, we find:
\begin{equation}
\Delta_{QL}^{ij} \sim 
\lambda^{\scharges{Q_i}{L_j}} , \qquad
\Delta_{\bar{u}\bar{e}}^{ij} \sim 
\lambda^{\scharges{\bar{e}_j}{\bar{u}_i}} .
\label{eq:FNscalingLQ}
\end{equation}

Put differently, the SM charges and flavor symmetries are enough to determine how the new $S_1$ field should be embedded in the effective theory below $M$.  
The power counting of the effective theory then dictates that the expected FN scaling above holds, up to the $\mathcal{O}(1)$ Wilson coefficients of the effective theory (analogous to the $r_{ij}$ in Eq.~\eqref{eq:fn_eft}). This Ansatz generalizes to arbitrary new spurions of $G_{\mathrm{flavor}}$ that can arise in other leptoquark models. A complete list of these spurions is given in Ref.~\cite{Bordone:2019uzc}.

Once the effective theory is known, we can make predictions for the contributions of new physics to various observables. Because the same spurion contributes to multiple observables, these predictions are correlated by a FN Ansatz. These correlations can lead to inconsistencies with experimental results. 
Consequently, it is useful to have a systematic way of deviating from this scaling while still maintaining the predictivity of FN models. We discuss a systematic way of doing this in the next section.
Specifically, we show how modifications of the UV spectrum of a FN construction can allow a controlled deviation from correlations between various observables in the IR, alleviating violations of experimental bounds.

\section{Plat Principal: Wrinkles in Froggatt--Nielsen}
\label{sec:wrinkles}

As described in the previous section, the FN mechanism provides a natural Ansatz for the hierarchies of new flavor spurions coupled to the SM quarks and leptons. 
However, given our lack of knowledge about the dynamics underlying the flavor structure of the SM, it is worth exploring how this Ansatz could change within the general framework of horizontal symmetry explanations for the SM flavor pattern.

In this spirit, we introduce the notion of ``wrinkles'', as a way of parametrically changing the FN Ansatz for the flavor spurions that is described above without introducing additional scales. In \S\ref{subsec:wrinkles}, we will define them precisely, and argue that they allow for more flexibility in correlations between different flavor observables. While this flexibility inherently makes our Ansatz less predictive, the freedom to introduce wrinkles is not absolute: there is a bound on the number of wrinkles imparted by radiative corrections, which we will discuss in \S\ref{subsec:radiative}. In \S\ref{subsec:uv_completions}, we give several explicit examples of realizations of wrinkles in UV models.

\subsection{Wrinkled Froggatt--Nielsen Chains}
\label{subsec:wrinkles}

In \S\ref{sec:FNplusBSM}, we described how the FN Ansatz leads to a natural power counting for new flavor spurions in powers of $\lambda \equiv \langle \varphi \rangle / M$, which we identify with the Cabbibo angle.  Here, we generalize this power counting by considering modifications to the power of $\lambda$ that appears in the spurion. 

\begin{figure}[t]
\centering
\includegraphics[height=3.5cm]{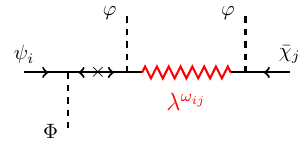}
\caption{A cartoon illustrating a ``wrinkle'' in the Yukawa coupling $\Phi \psi_i \bar{\chi}_j$, which leads to a change in the predicted scaling from the FN Ansatz.}
\label{fig:wrinkle_cartoon}
\end{figure}

Consider a flavor spurion $Y_{\psi\bar{\chi}}$, where $\psi$, $\bar{\chi}$ are given SM matter fields. We introduce what we call ``wrinkles'' to modify the scaling of a given element of $Y_{\psi\bar{\chi}}$:
\begin{equation}
\label{eq:wrinkledef}
Y_{\psi\bar{\chi}}^{ij} \sim W_{\psi\bar{\chi}}^{ij} \lambda^{|[\psi_i]+[\bar{\chi}_j]|} \equiv \lambda^{\omega_{ij} + |[\psi_i] + [\bar{\chi}_j]|} .
\end{equation}
Here we denote the power of $\lambda$ that appears in $W_{\psi\bar{\chi}}^{ij}$ by $\omega_{\psi\bar{\chi}}^{ij}$ which, for simplicity, is assumed to be an integer. This additional scaling is motivated by allowing for additional structure in the UV, such as symmetries inducing obstructions in the heavy fermion chains which generate the non-renormalizable operators, and is illustrated schematically in Figure~\ref{fig:wrinkle_cartoon}. In general, any modification of the UV theory that gives rise to deviations from predictions of the vanilla FN setup without changing the number of power counting parameters can be considered a wrinkle. Different UV completions can lead to different correlated patterns of matrix entries $\omega_{\psi\bar{\chi}}^{ij}$ as we will discuss in \S\ref{subsec:uv_completions}, but from the IR perspective, these correlations are not apparent. 

To be concrete, consider the example of the spurions $\Delta_{QL}$ and $\Delta_{\bar{u}\bar{e}}$ for the $S_1$ leptoquark, as in Eqs.~\eqref{eq:S1lagrangian} and \eqref{eq:GflavorLQ}. With additional wrinkles, the couplings in Eq.~\eqref{eq:FNscalingLQ} are modified to
\begin{equation}
\Delta_{QL}^{ij} \sim \lambda^{\omega_{QL}^{ij} + |[Q_i] + [L_j]|}, 
\qquad
\Delta_{\bar{u}\bar{e}}^{ij} \sim \lambda^{\omega_{\bar{u}\bar{e}}^{ij} + | [\bar{u}_i]| + [\bar{e}_j]}
\end{equation}
where $\omega_{QL}$ and $\omega_{\bar{u}\bar{e}}$ are   
matrices of integers, whose elements $\omega_{QL}^{ij}$ and $\omega_{\bar{u}\bar{e}}^{ij}$ can vary across generations independently for both fermions. 
The idea of wrinkles can also be extended to models with additional scales by allowing wrinkles for each power counting parameter. Here we will focus on the case with a single expansion parameter and not discuss the case of multiple parameters further.

Note that there are two distinct possibilities allowed by introducing wrinkles. 
The most straightforward one is that the number of factors of $\lambda$ in some couplings of a {\em new} flavor spurion are modified, 
suppressing or enhancing their contributions to some flavor observables. 
For instance, wrinkles could suppress BSM contributions to observables such as electric and magnetic dipole moments (EDMs and MDMs) or light meson decays, which are generally strongly constrained, and allow for spurions with smaller mass scales to contribute to other observables. We will discuss this possibility thoroughly, again in the case of the $S_1$ leptoquark, in \S\ref{sec:bknunu}. 

The second possibility is that wrinkles could exist in SM chains---i.e., $Y_{Q\bar{u}}$, $Y_{Q\bar{d}}$, or $Y_{L\bar{e}}$ could have fewer or additional factors of $\lambda$. In the IR, the SM Yukawa matrices must still match the measured masses and mixing angles of the quarks and leptons. Wrinkles in SM chains therefore necessitate different horizontal charges than the ones shown in Table~\ref{tab:gen_charges}.
This changes the expected scaling for BSM spurions, leading to different couplings than expected in a na\"{i}ve FN Ansatz between the SM fermions and new particles. We will not comment in detail on particular phenomenological applications of this scenario, but highlight that this is an interesting direction for further exploration.

\subsection{Bounds on Wrinkles from Radiative Corrections}
\label{subsec:radiative}

Allowing for wrinkles would appear to entirely eliminate the predictivity of the FN Ansatz. However, there is a natural bound on the size of the wrinkles that arises from demanding that the observed flavor structure in the IR arises predominantly from {\em tree-level} contributions to the effective operators below the scale $M$. Requiring that the tree-level contribution (including wrinkles) to the Yukawa coupling is larger than any subleading corrections from loops leads to a number of {\em consistency conditions} on the Yukawas, which in turn set a bound on the wrinkles. Provided these conditions are satisfied, the flavor structure in the IR is still determined by the FN mechanism in a predictive way, with departures from the minimal implementation parameterized by the wrinkles.

\begin{figure}[t]
\centering
\includegraphics[height=3.5cm]{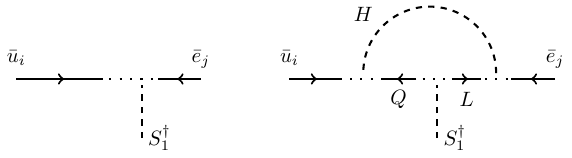}
\caption{Left: the tree level $S_1^{\dagger}\bar{u}_i \bar{e}_j$ coupling. Right: A loop contribution to the same spurion, leading to the spurion contribution described by Eq.~\eqref{eq:loop_contribution}. In both diagrams the dots indicate a chain of flavon and heavy fermion vertices, whose length is determined by the horizontal charges of the particles, which we suppress for clarity.}
\label{fig:loop_correction_S1_yuk}
\end{figure}

To illustrate these constraints, consider the Yukawa coupling matrix between the right-handed up-type quarks and the right-handed charged leptons for the $S_1$ leptoquark model in Eq.~\eqref{eq:S1lagrangian}, $\Delta_{\bar{u}\bar{e}}^{ij}$.
In a FN setup, this coupling arises from a non-renormalizable operator with a minimal number of flavons. It can be UV completed with a tree-level chain of heavy fermions and flavons with a single leptoquark vertex, as illustrated on the left in Figure~\ref{fig:loop_correction_S1_yuk}. 
However, the same operator can also be generated at higher order by including SM fermions in the FN chain $\bar{u}^i \to Q^k \to L^l \to \bar{e}^j$, as shown on the right in Figure~\ref{fig:loop_correction_S1_yuk}. The first and last connections include additional Higgs insertions that are tied together to form a loop, and the $Q^k \to L^l$ connection involves a leptoquark interaction. Thus, the higher-order contribution to $\Delta_{\bar{u}\bar{e}}^{ij}$ is:
\begin{equation}
\Delta_{\bar{u}\bar{e}}^{ij}\Big|_{\mathrm{loop}} \sim \frac{1}{16\pi^2} \left( Y_{Q\bar{u}}^{T} \cdot \Delta_{QL}^{*} \cdot Y_{L\bar{e}} \right)^{ij}.
\label{eq:loop_contribution}
\end{equation}
Demanding this contribution to be smaller than the tree-level contribution, and assuming the absence of any artificial cancellations, leads to a lower bound on the Yukawa coupling $\Delta_{\bar{u}\bar{e}}^{ij}$ and an upper bound on the entries of $\Delta^*_{QL}$. This bound begets a set of consistency conditions on the wrinkles:
\begin{equation}
\begin{aligned}
\label{eq:exconsistency}
\big|\Delta_{\bar{u}\bar{e}}^{ij}\big| & \gtrsim 
\frac{1}{16\pi^2} \Big|\big( Y_{Q\bar{u}}^{T} \cdot \Delta_{QL}^{*} \cdot Y_{L\bar{e}} \big)^{ij}\Big|,\\
\qquad \qquad \implies 
\lambda^{\omega_{\bar{u}\bar{e}}^{ij} + \scharges{\bar{u}_i}{\bar{e}_j}}
& \gtrsim \frac{1}{16\pi^2} \big|Y_{Q\bar{u}}^{T}\big|^{ik} \lambda^{\omega_{QL}^{kl} + \scharges{Q_k}{L_l}} \big|Y_{L\bar{e}}\big|^{lj}\ ,
\end{aligned}
\end{equation}
where there is an implicit summation over the indices $k$ and $l$ above. While the SM Yukawas on the right hand side of this relation may also contain wrinkles, it is the IR value of the coupling 
that appears, which is fit to the SM masses and mixing angles. 

Similar \textit{consistency conditions} were proposed in Refs.~\cite{Feldmann:2006jk,Bordone:2019uzc,Bordone:2020lnb}, neglecting the loop factor. Other similar constraints (including the loop factor) have been considered as naturalness constraints on models of flavorful new physics \cite{Arnold:2013cva}. 
We settle for the weaker constraint, including the loop factor, as a concrete, irreducible bound.\footnote{RG evolution of the leptoquark couplings also does not change the above set of bounds, as long as one imposes the consistency conditions at the matching scale of order $M$. The structure of the one loop Yukawa RGEs involves the same higher order operators as appearing in our consistency condition. Thus, imposing the consistency condition at the matching scale ensures that running is a small effect and can be neglected. Consequently, RG evolution to scales below the matching scale ensures that the consistency condition (inequality) holds at all such scales. } 
Note that there are also other higher order contributions to the spurions, such as those from higher-dimensional operators with the Higgs replaced by its vacuum expectation value, but they will be smaller than the one in Eq.~\eqref{eq:loop_contribution}, since $v^2 / M^2 < 1/16\pi^2$. 

More generally, a complete set of consistency conditions can be derived by again considering the Yukawas as spurions under $G_{\textrm{flavor}}$.   
In the absence of any additional symmetries, contributions similar to Eq.~\eqref{eq:loop_contribution} arise from any combination of Yukawa couplings that transform in the same representation of $G_{\textrm{flavor}}$. The complete list of leading consistency conditions for all of the Yukawa couplings in the SM extended with the $S_1$ leptoquark are listed in Appendix~\ref{app:consistency}.

These inequalities must be satisfied for any wrinkled FN setup involving additional flavor spurions, and they impose non-trivial constraints on the size of the wrinkles introduced in Eq.~\eqref{eq:wrinkledef}.\footnote{The consistency conditions, as written, hold neglecting $\mathcal{O}(1)$ couplings; there may be small deviations from including them.} 
The details of these constraints depend on the particular charge assignment of the SM fermions, but once these are fixed, a degree of predictiveness is returned to the FN Ansatz, even in the presence of wrinkles.
As pointed out in Refs.~\cite{Feldmann:2006jk, Bordone:2019uzc, Bordone:2020lnb}, these consistency conditions are trivially satisfied in a vanilla FN setup without wrinkles, as a result of the triangle inequality. 

As an example of how this bound works with nonzero wrinkles, consider the charge assignment in Table~\ref{tab:gen_charges} with $q_0 = 0$, $l_0 = -1$, $X = +1$, $Y = -1$ and all other sign choices being positive. Assuming no wrinkles in the SM Yukawas, the bound on $\omega_{\bar{u}\bar{e}}^{33}$ from Eq.~\eqref{eq:exconsistency} becomes 
\begin{equation}
\begin{aligned}
\omega^{33}_{\bar{u}\bar{e}} & ~\lesssim~
\sum_{k,l} \Big( \big| [Q_k] + [\bar{u}_3]\big| + \big| [Q_k] + [L_l] \big| + \omega_{QL}^{kl} + \big|[L_l] + [\bar{e}_3]\big| \Big)\\[-0.5em]
& \qquad \qquad 
+ \log_{\lambda}\!\frac{1}{16\pi^2} - \big|[\bar{e}_3] + [\bar{u}_3]\big| \\[0.25em]
& ~\lesssim~ 2 + \omega_{QL}^{33} + \log_{\lambda}\!\frac{1}{16\pi^2} ,
\label{eq:exwrinkle}
\end{aligned}
\end{equation}
where in the last line we have assumed that $k = l = 3$ is the largest entry in $\omega_{QL}^{kl}$, which is typically the case. We see that, at least for this consistency condition, up to five wrinkles on $\Delta_{\bar{u}\bar{e}}^{33}$ are allowed, even without extra wrinkles on $\Delta_{QL}^{33}$.

A similar argument for general couplings, again using the triangle inequality, makes it clear that if all $\omega_{\psi\bar{\chi}}^{ij} \geq 0$, a {\em sufficient} condition on the wrinkles is that they are all greater than a loop factor: 
\begin{equation}
\left( W_{\psi \bar{\chi}} \right)^{ij} \gtrsim \frac{1}{16\pi^2} .
    \label{eq:radiativebound}
\end{equation}
Note that in this equation, we have assumed a mild separation of scales so that the logarithms in the loop contribution can be neglected along with other $\mathcal{O}(1)$ factors in the loop calculation. 
In this work, we focus on the bound in Eq.~\eqref{eq:radiativebound} and leave further studies of more accurate lower bounds on wrinkles for future work. 
As shown in Eq.~\eqref{eq:exwrinkle}, this bound may be overly restrictive, but it provides a useful shortcut for employing wrinkles in an EFT without having to manually check all the consistency conditions.

\subsection{UV Completions}
\label{subsec:uv_completions}

We now turn to UV completions of the wrinkles introduced in Eq.~\eqref{eq:wrinkledef}. Our goal is not to provide an exhaustive or detailed list of examples, but demonstrate a proof of principle of potential ways these wrinkles can arise from more complicated UV completions. 

\subsubsection{Missing Heavy Fermions \label{sec:heavyFermions}}

As a first concrete realization of the idea sketched in Figure~\ref{fig:wrinkle_cartoon}, 
we consider a situation where one of the heavy fermions with a particular horizontal charge does not exist in the spectrum. Instead, the chain leading to the effective operator can only be completed by including additional fermions and scalars, causing additional suppression.

\begin{figure}[t]
\centering
\includegraphics[height=3.5cm]{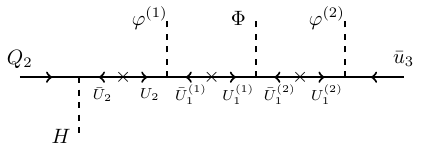}
\caption{
An explicit realization of a ``wrinkled'' FN chain, where the heavy quark with horizontal charge $+1$ is replaced by two heavy quarks, along with additional flavons, transforming under additional symmetries.}
\label{fig:uv_wrinkle}
\end{figure}

To illustrate this mechanism, we consider the example in Figure~\ref{fig:fn_chain_example} and replace a single heavy vector-like pair of fermions $U_1$, $\bar{U}_1$ with two sets of vector-like pairs, which we will denote by $U_1^{(1)}$, $\bar{U}_1^{(1)}$ and $U_1^{(2)}$, $\bar{U}_1^{(2)}$.  
These are assumed to have the same SM and horizontal charges as $U_1$, $\bar{U}_1$, but also transform as conjugate pairs under new symmetry groups, $G_1$ and $G_2$, respectively. 
To be explicit, we will take $G_1 = \lab{SU}(N_1)$ and $G_2 = \lab{SU}(N_2)$ to be two different continuous, non-Abelian groups, but the following construction works for arbitrary (continuous or discrete) groups as well, with straightforward modifications.
To complete the chain diagram, we must also introduce new flavons, which we take to be in the representations,
\begin{equation}
\varphi^{(1)}: (\mathbf{N_1}, \mathbf{1})_{-1}, \qquad
\varphi^{(2)}: (\mathbf{1}, \mathbf{\overline{N}_2})_{-1}, \qquad
\Phi^{(1,2)}: (\mathbf{\overline{N}_1}, \mathbf{N_2})_0, 
\end{equation}
where the parentheses indicate the $\lab{SU}(N_1) \times \lab{SU}(N_2)$ representation, and the subscript is the horizontal charge.
These allow us to construct the diagram shown in Figure~\ref{fig:uv_wrinkle}, where both of the extra heavy fermion pairs are traversed between $Q_2$ and $\bar{u}_3$. The charge assignments forbid the couplings $\varphi^{(1)} U^{\phantom{(}}_2 \bar{U}_1^{(2)}$ and $\varphi^{(2)}U_1^{(1)} \bar{u}_3$, so that this diagram is the leading effective operator containing $H Q_2 \bar{u}_3$.

Assuming all the scalars acquire vevs $\sim \langle \varphi \rangle$ and that the new fermions have vector-like masses $\sim M$, this replaces the $\lambda^2$ suppression inferred from the horizontal charges with a $\lambda^3$ suppression. In other words, this leads to a ``wrinkle'', $W^{23}_{Q\bar{u}} \sim \lambda$.

This construction can be extended to include arbitrarily many wrinkles in place of a single heavy fermion. For example, $W^{23}_{Q\bar{u}} \sim \lambda^2$ is obtained by introducing additional mirror quarks, $U_1^{(3)}$, $\bar{U}_1^{(3)}$, replacing $\varphi^{(2)}$ with a bi-fundamental $\Phi^{(2,3)}$ transforming as $(\mathbf{1}, \mathbf{\overline{N}_2}, \mathbf{N_3})_0$, closing the chain with $\varphi^{(3)}$, which transforms as a $(\mathbf{1}, \mathbf{1}, \mathbf{\overline{N}_3})_{-1}$; further wrinkles are obtained for additional mirror quarks.
In these types of examples, the Higgs and chiral fermions of the SM are neutral under the new symmetries, so as to be compatible with the general arguments in Ref.~\cite{Leurer:1992wg}. 

Note that with this mechanism, we see an example of the correlation between wrinkles in different chains. We constructed this wrinkle in the context of the $Q_2$ and $\bar{u}_3$ chain, but since we have removed $U_1$, $\bar{U}_1$ from the spectrum, the wrinkle necessarily appears in any chain involving them. For instance, assuming heavy up-like quarks are responsible for all of the up-type Yukawa couplings, it would also appear in the $Q_1 H \bar{u}_3$ operator.

\subsubsection{Extra Abelian Symmetries}

Another concrete example in which wrinkles can appear in an effective theory with the FN Ansatz is realized by considering additional Abelian symmetries in the UV, under which the SM fermions are charged. 
In particular, we can consider gauging the non-anomalous combinations of baryon number, $B$, and the individual lepton numbers, $L_e$, $L_{\mu}$, and $L_{\tau}$, as is frequently done in model-building for various flavor anomalies~\cite{Altmannshofer:2019xda}. 
These symmetries are preserved by the SM Yukawa couplings, but generically violated by neutrino masses and additional Yukawa couplings between SM fermions and new BSM fields, such as leptoquarks. 
For concreteness, we again work with the $S_1$ leptoquark and assume it is neutral under the new symmetry; therefore
the flavor spurion must absorb the remaining $\lab{U}(1)$ charge.
This means that additional flavons charged under the extra symmetries also must be included in order to complete the leptoquark Yukawa couplings. 
The usual flavon, with $\lab{U}(1)_H$ charge $-1$, is still present, since it is required to complete the SM Yukawa couplings.\footnote{In the presence of neutrino masses, the extra flavons may also be required to generate the PMNS matrix structure, depending on the additional symmetries we impose.}

In contrast to the UV completions discussed in \S\ref{sec:heavyFermions}, where the wrinkles are always additional suppression factors, the wrinkles that result from these extra symmetries can naturally
either suppress or enhance the size of the flavor spurions. Another distinction is that we have not removed any fermions of particular charges from the UV spectrum in this case: we allow fermions with all required quantum numbers to exist. 

Just like other UV models, additional symmetries and the flavons charged under them can generate a correlated pattern of wrinkles for the different chains. The details of those correlations depend on whether the new symmetries are flavor universal or flavor specific; we will discuss examples of both cases. In order to maintain the predictivity of our example, we also assume additional symmetries are spontaneously broken at similar scales to the $\lab{U}(1)_H$ symmetry. 

The flavor universal case is simpler, but also less flexible because of interdependence between different chains. 
For example, assuming $\lab{U}(1)_{B-L}$ is a symmetry of the theory, we can construct the leptoquark Yukawa spurions by introducing an additional flavon, $\tilde{\varphi}$, which we take to have $B-L$ charge $1/3$.
The new flavon will not change any of the SM chains, since they respect the $B-L$ symmetry, but the leptoquark chains can be different from the usual FN scenario. 
For instance, if $\tilde{\varphi}$ has no $\lab{U}(1)_H$ charge, all the leptoquark Yukawas will become smaller by $\lambda^2$, since the external fermions all have $B-L$ charge difference $\pm 2/3$ without the new flavon. If $\tilde{\varphi}$ also has $\lab{U}(1)_H$ charge $\geq 1$, then the pattern of leptoquark chains becomes more intricate. Since each leptoquark chain must contain exactly two copies of the new $B-L$ flavon and the remaining difference in horizontal charge requires insertions of the original flavon, whether a given chain becomes shorter or longer depends on the details of the assigned horizontal charges.

\begin{figure}
\centering
\includegraphics[width=\linewidth]{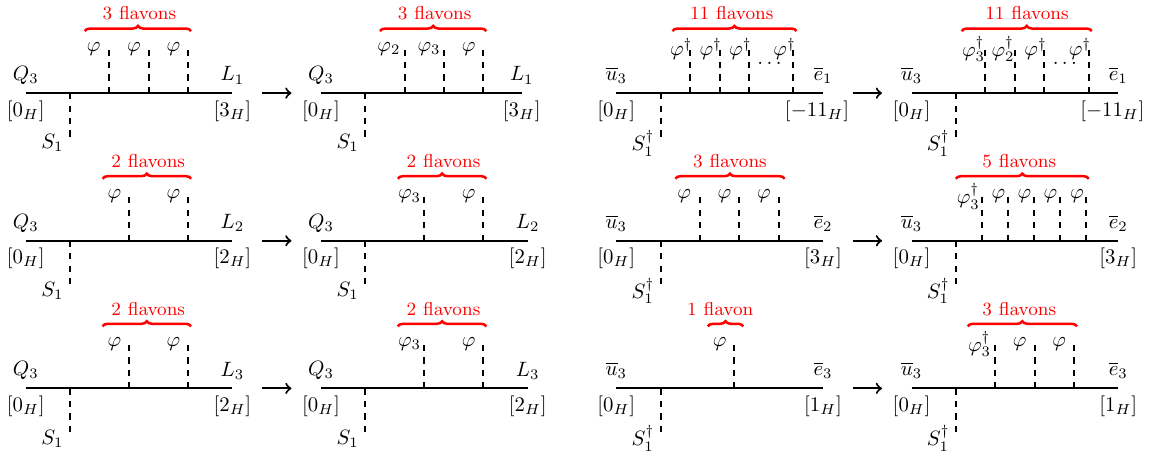}
\caption{
Chains before and after adding flavons charged under a $\lab{U}(1)_{B - 3L_e}$ symmetry to generate wrinkles. The horizontal charges correspond to Table~\ref{tab:gen_charges} with $q_0 = 0$, $l_0 = 2$, and $Y = 1$. We observe that the new $\lab{U}(1)_{B - 3L_e}$ symmetry and its flavons modify the prediction of the model for some of the leptoquark couplings in the IR.}
\label{fig:ExExtraFlavons}
\end{figure}

We have somewhat more freedom in the flavor specific case. As an example, consider introducing a new $\lab{U}(1)_{B - 3L_e}$ symmetry. 
Now both the PMNS matrix and chains for the leptoquark Yukawa couplings require new flavons charged under both $\lab{U}(1)_H$ and $\lab{U}(1)_{B-3L_e}$. We introduce two additional flavons: $\varphi_2$ is necessary to generate PMNS matrix entries of the correct size, and $\varphi_3$ is necessary to complete the leptoquark chains while respecting the additional symmetries. These flavons have $B - 3L_e$ charges
\begin{equation}
[\varphi] = 0 \qquad [\varphi_2] = 3 \qquad 
[\varphi_3] = -1/3.
\label{eq:exExtraFlavons}
\end{equation}
Each also carries $\lab{U}(1)_H$ charge $-1$. Including these extra symmetries and flavons charged under them creates wrinkles by changing the required number of vev insertions for the leptoquark couplings compared to the spurion size we would naively expect with only these $\lab{U}(1)_H$ charges. For example, if we consider only the couplings to the third generation leptons, we make the right-handed $\mu$ and $\tau$ couplings smaller while leaving the right-handed $e$ coupling and the left-handed couplings unaffected. This is shown in Figure~\ref{fig:ExExtraFlavons}. Nonetheless, despite the additional freedom in the flavor specific case, it is still challenging to obtain certain patterns of wrinkles, such as those constrained by the triangle inequality.

Finally, we comment on a few modifications to the examples above. First, we note that it is possible to modify this approach by charging the leptoquark under $\lab{U}(1)_H$ instead of/in addition to additional flavon(s). Similar to the $B-L$ charged flavons, this is another mechanism to add wrinkles to the leptoquark couplings without affecting the SM couplings. 
In principle, we can also charge the leptoquark under the additional symmetries we discussed in this section, but note that we are not always guaranteed a charge assignment which makes all of the couplings invariant. 
Second, we note that like the previous case, other modifications such as using discrete Abelian symmetries also behave similarly. 
However, we can not replace these Abelian symmetries with non-Abelian ones \cite{Leurer:1992wg}, because we are charging the SM fermions under the new symmetry. 
This is in contrast to the previous case, where only internal fermions are charged under new non-Abelian symmetries.

While we have provided two different ways in which wrinkles could be generated, we have not exhausted the possibilities. These are only examples, and there are undoubtedly many more options for generating wrinkles, which would be interesting for future work. Since the details of a particular model are not the central point of this paper, we now move to discussing a full example in the IR.

\section{Dessert: \texorpdfstring{$B \to K\bar{\nu}\nu$}{B to K nu nu} in a Wrinkled Setup}
\label{sec:bknunu}

To demonstrate the ideas of the previous sections with a specific example, in this section we study the phenomenology of the $S_1$ leptoquark introduced in Eq.~\eqref{eq:S1lagrangian} with particular flavor Ans\"atze in detail. 
Such Ans\"atze correlate the contribution of $S_1$ to different observables.
As mentioned in the previous section, the inclusion of wrinkles in a FN Ansatz can change the relative sizes of predictions for different flavor observables. This could allow a model to accommodate a significant excess over the SM in one observable, while suppressing other observables 
that would otherwise be too constraining.\footnote{Signals of leptoquarks in all flavor experiments can also be suppressed by choosing $q_0$ and $l_0$ (defined in Table~\ref{tab:gen_charges}) such that the quarks' and leptons' charges are very far apart, but this requires an unnaturally large separation of charges. This choice also does not permit the explanation of any discrepancies in flavor experiments because it suppresses leptoquark contribution to all observables.
}

As an illustration, we will focus on constructing a model that can give rise to a large signal in the semi-leptonic decay $\bknn$. $\bknnbr$ is an interesting test case for several reasons.
Assuming the vanilla FN Ansatz, the mass range preferred for new physics near the current experimental sensitivity  is in the few TeV range, and small hints of flavorful new physics may have already been detected~\cite{Belle-II:2021rof}. 
Like all flavor-changing neutral currents, the $b\to s\bar{\nu}\nu$ transition is greatly suppressed in the SM. 
It is also relatively clean theoretically, with the uncertainties in the hadronic form factors and from perturbative effects well under control~\cite{Buchalla:1993wq,Misiak:1999yg,Buchalla:1998ba, Ball:2004ye, Ball:2004rg, Khodjamirian:2010vf, Brod:2010hi, Bouchard:2013eph, Horgan:2013hoa}. 
This situation, along with the prospect of observing the decay at the Belle~II experiment in the near future, make it an intriguing probe of BSM physics~\cite{Altmannshofer:2009ma, Buras:2014fpa, Blake:2016olu}. 
We use this specific observable as a testbed of various ideas introduced in the previous section; similar studies can be carried out for any other flavorful anomalies that may emerge in experimental data.

\subsection{\texorpdfstring{$B\to K\bar{\nu}\nu$}{B to K nu nu} in the SM and Beyond}

In order to understand how various FN Ans\"atze contribute to $\bknnbr$, we first need to discuss the SM and leptoquark contributions, as well as experimental bounds.
Typically, $\bknnbr$ is parameterized in terms of the Wilson coefficients $C_{R}^{ij}$ and $C_{L}^{ij}$, which are defined implicitly in the effective Hamiltonian governing $b \to s\bar{\nu}\nu$ transitions 
\begin{equation}
\mathcal{H}_{\textrm{eff}} = -\frac{4 G_F}{\sqrt{2}} V_{tb}^{\phantom{*}} V_{ts}^* \big(C_L^{ij} \mathcal{O}_L^{ij} + C_R^{ij} \mathcal{O}_R^{ij}\big) + \hc
\label{eq:Hbknn}
\end{equation}
where
\begin{equation}
\mathcal{O}^{ij}_L = \frac{\alpha_{\mathrm{em}}}{2\pi}
\big( s_L^{\dagger} \bar{\sigma}^{\mu} b_L \big)
\big( \nu_j^{\dagger} \bar{\sigma}_{\mu} \nu_i \big),
\qquad
\mathcal{O}^{ij}_R = \frac{\alpha_{\mathrm{em}}}{2\pi}
\big( s_R^{\dagger} \sigma^{\mu} b_R  \big) 
\big( \nu_j^{\dagger} \bar{\sigma}_{\mu} \nu_i \big),
\label{eq:defOL}
\end{equation}
and $i,j=e,\mu,\tau$ are neutrino flavor indices.

\begin{figure}
\centering
\includegraphics[height=3.5cm]{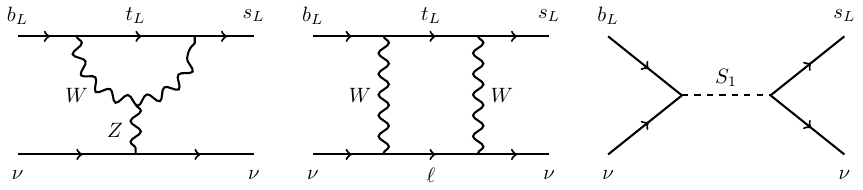}
\caption{Example Feynman diagrams leading to $b \to s\bar{\nu}\nu$ transitions in the SM extended with an $S_1$ leptoquark. The left (center) diagram show the leading one loop SM contributions with the penguin (box) topology, while the right diagram illustrates the tree-level leptoquark contribution. The $Z$ in the left diagram could also connect to the top line instead.}
\label{fig:btosnunu_diags}
\end{figure}

In the SM (and in the $S_1$ leptoquark model we consider below), only $C_L$ is non-zero. The leading contribution to the SM value of the Wilson coefficient arises from diagrams such as those in Figure~\ref{fig:btosnunu_diags}. Also including NLO QCD corrections \cite{Buchalla:1993wq,Misiak:1999yg,Buchalla:1998ba} and two-loop electroweak contributions \cite{Brod:2010hi}, the SM Wilson coefficient is 
\begin{equation}
\label{eq:CLSM}
C_L^{\,ij,\,\textrm{SM}} = (-6.353 \pm 0.074)\,\delta_{ij},
\end{equation}
where $\delta_{ij}$ captures the fact that the SM contributions are lepton flavor conserving.
This leads to a prediction for the branching ratio~\cite{Buras:2014fpa,Blake:2016olu},
\begin{equation}
\bknnbr \Big|_{\mathrm{SM}} = (0.46 \pm 0.05) \times 10^{-5},
\label{eq:bknnSM}
\end{equation}
where we are inclusive to neutrino flavor. 

This process has been searched for at Belle and BaBar by tagging the second $B$ meson in either a hadronic or semileptonic decay~\cite{Belle:2013tnz, BaBar:2013npw, Belle:2017oht}.
Similar searches exist for $\mathrm{BR}(B\rightarrow K^* \bar{\nu}\nu)$, e.g. see Refs.~\cite{Belle:2013tnz, BaBar:2013npw}. Each of these channels leads to the same qualitative conclusions; thus, for the rest of this work we will focus on $\bknnbr$ measurements, for simplicity.
A combination of these results yields a 90\% C.L. upper limit on the branching ratio of
\begin{equation}
\label{eq:bknnbr_limit}
\bknnbr < 1.6 \times 10^{-5} .
\end{equation}
Recently, Belle~II has searched for the same decay using an inclusive tagging technique, which allows them to partially compensate for their smaller dataset and larger backgrounds~\cite{Belle-II:2021rof}. Though not yet statistically significant, a combination of these results (assuming their uncertainties are uncorrelated) leads to a best fit value of 
\begin{equation}
\label{eq:bknnbr_best}
\bknnbr = (1.1 \pm 0.4) \times 10^{-5},
\end{equation}
which leaves room for a BSM contribution on top of the SM prediction in Eq.~\eqref{eq:bknnSM}. The uncertainties in all of these estimates---both the tagged and inclusive searches---are predominantly statistical, and are expected to improve and become comparable to the theoretical uncertainty in Eq.~\eqref{eq:bknnSM} with the forthcoming full Belle~II dataset \cite{Belle-II:2018jsg}. 
Therefore, while it remains to be seen if any signals of new physics exist in this channel, it provides an interesting application of our wrinkled FN setup.

The $S_1$ leptoquark contributes to $b \to s\bar{\nu}\nu$ transitions via the tree-level diagram shown on the right in Figure~\ref{fig:btosnunu_diags}. It generates a Wilson coefficient 
\begin{equation}
C_L^{ij} \propto \frac{v^2}{m_{S_1}^2} \Delta_{QL}^{3i} \Delta_{QL}^{2j\,*}
\end{equation}
for the effective theory of Eq.~\eqref{eq:Hbknn}. 
Since this is the same operator as generated in the SM, it is convenient to capture these effects by considering the ratio:
\begin{equation}
R_K^{\nu\nu} \equiv \frac{\bknnbr}{\bknnbr\big|}_{\mathrm{SM}}.
\label{eq:RKdef}
\end{equation}
The contribution from $S_1$ is given by \cite{Bauer:2015knc,Becirevic:2016oho} (see also Refs.~\cite{Buras:2014fpa,Dorsner:2016wpm}) 
\begin{equation}
R_K^{\nu\nu} = 1 - y \Re \left[ \frac{(\Delta_{QL}^{3i}\Delta_{QL}^{2i\,*})}{V_{tb}^{\phantom{*}} V_{ts}^*} \right] + \frac{3 y^2}{4}  \frac{(\Delta_{QL}^{3i}\Delta_{QL}^{3i\,*}) (\Delta_{QL}^{2j}\Delta_{QL}^{2j\,*})}{\big| V_{tb}^{\phantom{*}} V_{ts}^*\big|^2} ,
\label{eq:RKSI}
\end{equation}
with a sum over repeated lepton indices in each term, and 
\begin{equation}
y \equiv - \frac{2\pi v^2}{6 C_L^{\mathrm{SM}} \alpha_{\mathrm{em}} m_{S_1}^2} \simeq \bigg(\frac{1.2 \, \mathrm{TeV}}{m_{S_1}}\bigg)^2.
\label{eq:defyCL}
\end{equation}

In terms of $R_K^{\nu\nu}$, the 90\% C.L. limit and 68\% C.L. preferred values of the branching ratio in Eqs.~\eqref{eq:bknnbr_limit} and \eqref{eq:bknnbr_best} translate to
\begin{equation}
R_K^{\nu\nu} < 3.4, \qquad
R_K^{\nu\nu} \in [1.5, 3.3],
\end{equation}
respectively. The interpretation of these bounds in the context of the leptoquark depends on the assumptions made about the hierarchies in $\Delta_{QL}^{ij}$, to which we now turn.


\begin{table}[t!]
\centering
\renewcommand\arraystretch{1.8}
\resizebox{\columnwidth}{!}{
\begin{tabular}{c|c|c|c}
\hline 
Observable & $S_1$ Yukawa Couplings & Experimental Result & Future Bounds\\
\hline 
$\textrm{BR}(B^+ \to K^+\bar{\nu}\nu)$ & 
$\Delta_{QL}^{3i} \times (\Delta_{QL}^{2j})^*$ & 
$(1.1 \pm 0.4) \times 10^{-5}$ ~ \cite{Belle-II:2021rof} & - \\
\hline \hline
electron EDM & 
$(V^{*} \Delta_{QL})^{31} \times (\Delta_{\bar{u}\bar{e}}^{31})^*$  & 
$ < 4.1 \times 10^{-30}~e\,$cm ~ \cite{Roussy:2022cmp} & $< 10^{-31}~e~$cm \cite{Ho:2020ucd, Fitch:2020jil} \\
\hline
$\textrm{BR}(\mu \to e\gamma)$ &
$\begin{array}{c} (V^{*} \Delta_{QL})^{32} \times \Delta_{\bar{u}\bar{e}}^{31}{}^{\phantom{*}} \\ \Delta_{\bar{u}\bar{e}}^{32\, *} \times (V^{*} \Delta_{QL})^{31\, *} \end{array}$ & 
$ < 4.2 \times 10^{-13}$ ~ \cite{MEG:2016leq} & $< 6 \times 10^{-14}$~\cite{Baldini:2018nnn} \\
\hline
$\textrm{CR}(\mu \to e)_N$ &
$(V^* \Delta_{QL})^{11\,*} \times (V^* \Delta_{QL})^{12}$
& $< 7.0 \times 10^{-13}$~\cite{SINDRUMII:2006dvw} & $ <2.5\times 10^{-18}$~\cite{Bartoszek:2014mya, Abusalma:2018xem} \\
\hline
$\textrm{BR}(\tau \to \mu\gamma)$ &
$\begin{array}{c} (V^{*} \Delta_{QL})^{33} \times \Delta_{\bar{u}\bar{e}}^{32}{}^{\phantom{*}} \\ \Delta_{\bar{u}\bar{e}}^{33\, *} \times (V^{*} \Delta_{QL})^{32\, *} \end{array}$ & 
$< 4.2 \times 10^{-8}$~\cite{Belle:2021ysv} & $<6.9\times 10^{-9}$~\cite{Belle-II:2022cgf, Banerjee:2022vdd} \\
\hline
$\textrm{BR}(K^{+} \to \pi^+ \bar{\nu} \nu)$  &
$\Delta_{QL}^{2k} \times (\Delta_{QL}^{1k})^{*}$ & 
$< 1.88 \times 10^{-10}$~\cite{NA62:2021zjw} & $(8.4 \pm 0.4) \times 10^{-11}$~\cite{NA62:2020upd} \\
\hline
%
%
$\Delta m_{B_s}$& 
$(\Delta_{QL}\, \Delta_{QL}^\dagger)^{32}$ & 
$\Delta C_{B_{s}}\leq0.09$ ~\cite{Cerri:2018ypt} & $\Delta C_{B_{s}}\leq0.026$ ~\cite{Cerri:2018ypt}  \\
\hline
\end{tabular} 
}
\caption{Here we show the experimental results for $\bknnbr$ and a few other constraining observables; we also show the predominant $S_1$ Yukawa couplings contributing to each. Note that for $B$-mixing, we use the experimental uncertainty on the quantity $C_{B_s}$ as defined in Eq.~\eqref{eq:utfit_ratio}. For $K^+ \to \pi^+\nu\bar{\nu}$, the future bound corresponds to reaching a $5\%$ experimental uncertainty on the SM branching ratio~\cite{Buras:2015qea}. The muon to electron conversion rate in nuclei, $\textrm{CR}(\mu \to e)_N$, gets contributions from both dipole and four-fermion operators; 
we show the Yukawas entering the four-fermion operator that is dominant in the FN Ansatz (associated with a left-handed vector current) here, while the complete set is given in Appendix~\ref{app:obs}. The current (future) bound listed for it is on the conversion rate in a gold (aluminum) nucleus.}
\label{tab:mainobs}
\end{table}

\subsection{Constraints with Different Flavor Ans\"atze}
\label{subsec:constraints}

In addition to $b \to s\bar{\nu}\nu$ transitions discussed above, the $S_1$ leptoquark can contribute to a number of flavor-changing processes or precision observables that are constrained by experiments. 
These include electric and magnetic dipole moments of SM particles, LFV decays, leptonic and semi-leptonic meson decays, flavor-violating decays of gauge bosons, and neutral meson mixing. 
Some of the most powerful observables, and their dependence on the leptoquark Yukawa couplings are summarized in Table~\ref{tab:mainobs}.\footnote{For simplicity, we work with flavor basis neutrinos, so no dependence on the PMNS matrix appears.}
As is apparent from the table, the observables depend on numerous different combinations of the leptoquark couplings.
More details about the observables, including the dependence on the leptoquark couplings and references to more complete treatments in the literature, are given in Appendix~\ref{app:obs}. 

Because the contributions to various observables are correlated, we need to pick a particular Ansatz and study it in order to understand these constraints. In the rest of this section, we study these constraints in the context of three different flavor Ans\"atze: flavor anarchy, vanilla FN, and FN with wrinkles. In particular, we explore how adding wrinkles can alleviate constraints while maintaining consistency with $\bknnbr$ measurements.

\begin{figure}[t]
\centering
\includegraphics[width=0.95\linewidth]{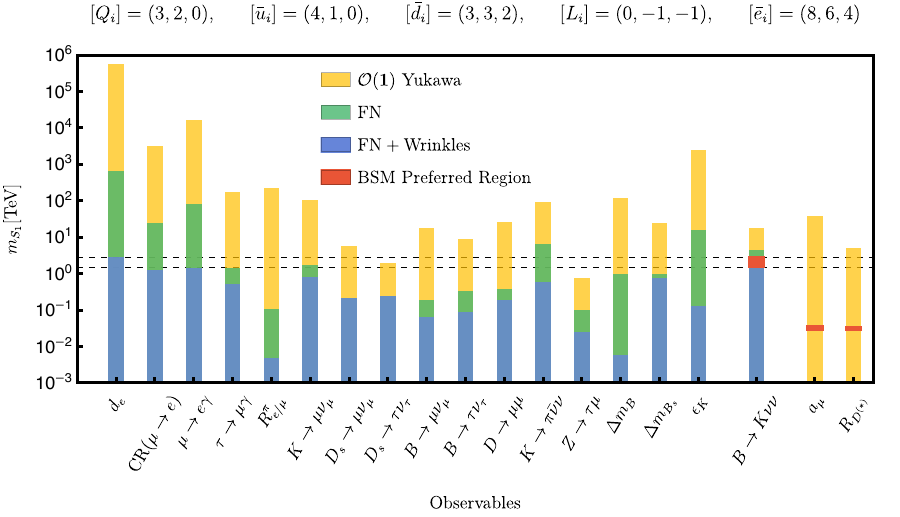}
\caption{
The leptoquark mass range probed by various observables if the Yukawa couplings of the leptoquark are either $\mathcal{O}(1)$ (yellow), follow the vanilla FN setup in Eq.~\eqref{eq:lambdasBknunu} (green), or the same FN setup plus the wrinkles from Eq.~\eqref{eq:wrinklesLQ} (blue). 
The preferred range for explaining some existing anomalies are shown in red, assuming the wrinkled setup. 
The undetermined $\mathcal{O}(1)$ factors in the Yukawas (folded in $r_{ij}$ in Eq.~\eqref{eq:fn_eft}) can further affect the leptoquark contribution and slightly change the mass range probed by each observable. 
We see that in our wrinkled setup, the mass range that explains the current discrepancy in $\bknnbr$ measurement (between the horizontal dashed lines) can also be probed by the LFV processes $\mu \rightarrow e \gamma$ and $\mathrm{CR}(\mu \to e)$, and the electron EDM in near future measurements.}
\label{fig:observable_bounds}
\end{figure}

Without any assumptions about the underlying structure, a minimal assumption is that all elements of $\Delta_{QL}$ and $\Delta_{\bar{u}\bar{e}}$ are $\mathcal{O}(1)$. This assumption is commonly referred to as ``flavor anarchy''. Under this assumption, the mass of the leptoquark consistent with the $\bknnbr$ measurements is $m_{S_1} \in (9,~18)$~TeV.
On the other hand, measurements of the electron EDM and other flavor-changing processes constrain the mass of the leptoquark to be above $\sim 10^5~\textrm{TeV}$. 
The resulting limits for some of the observables considered are shown as yellow bars in Figure~\ref{fig:observable_bounds}.
To calculate these ranges for observables that are already measured experimentally, we demand the leptoquark contribution to be within one standard deviation of the measured value, while for others we use the reported upper bounds from Ref.~\cite{ParticleDataGroup:2022pth}.\footnote{The exceptions to this are $R_D$ and $a_\mu$, where we take the maximum leptoquark mass consistent to within $3 \sigma$ and $4\sigma$, respectively, of the experimental measurement for the anarchic coupling case, and use the $2 \sigma$ ellipse for the preferred mass range in the wrinkled case.
} For the  electron EDM, a CP-odd observable, we assume a purely imaginary coupling to show the maximum reach of the experimental results. 

It is clear that without any flavor texture on the leptoquark Yukawas, observables such as the electron EDM, LFV decays, or meson-mixing parameters rule out the leptoquark mass range relevant for $\bknnbr$.\footnote{Our model can also contribute to $a_\mu$ at one loop to explain the observed anomaly \cite{Muong-2:2006rrc,Muong-2:2021ojo}, although recent lattice calculations~\cite{Borsanyi:2020mff,Wang:2022lkq,Ce:2022kxy,ExtendedTwistedMass:2022jpw,FermilabLattice:2022izv,Bazavov:2023has,Blum:2023qou} and measurements~\cite{CMD-3:2023alj} hint toward a smaller discrepancy with the experimental data. However, other observables already rule out the leptoquark mass range that has a large enough contribution to $a_\mu$. See Refs.~\cite{Stockinger:2006zn,Blanke:2007db,Pospelov:2008zw,Feruglio:2008ht,Dermisek:2013gta,Agrawal:2014ufa,Calibbi:2014yha,Bauer:2015knc,Dorsner:2016wpm,Calibbi:2018rzv,Saad:2020ihm,Calibbi:2020emz,Calibbi:2021qto,FileviezPerez:2021lkq,Nakai:2021mha,Lopez-Ibanez:2021yzu} for other solutions to this anomaly, including attempts at embedding the solution in a FN construction.} 
We have also checked the contribution of our setup to many other similar observables (electron and tau MDM, $\tau \rightarrow e \gamma$, $K \rightarrow e \nu$, various other $D$ meson decays, $D_s \rightarrow e \nu$, $B \rightarrow e \nu$, $\pi \rightarrow e e$, $\pi \rightarrow \mu e$), but find that the constraints they place are not as competitive for our model.

Thus we are led to consider embedding the $S_1$ leptoquark in a FN model of flavor. This has the benefit of not only alleviating some of the experimental constraints discussed above, but also relating it to the SM flavor puzzle.

As discussed in \S\ref{sec:fn_intro}, aside from the general shifts in the lepton and quark horizontal charges, there are only a handful of possible charge assignments that give rise to the correct pattern of SM masses and mixing angles. 
For concreteness, we choose horizontal charges from Table~\ref{tab:gen_charges} with $q_0 = 0$, $l_0 = -1$, and $X = -Y = -1$. This yields 
\begin{equation}
\begin{gathered}
([Q_1],\, [Q_2],\, [Q_3]) = (3, 2, 0), \qquad
([\bar{u}_1],\, [\bar{u}_2],\, [\bar{u}_3]) = (4, 1, 0), \qquad
([\bar{d}_1],\, [\bar{d}_2],\, [\bar{d}_3]) = (3, 3, 2), \\[0.5em]
([L_1],\, [L_2],\, [L_3]) = (0, -1, -1), \qquad
([\bar{e}_1],\, [\bar{e}_2],\, [\bar{e}_3]) = (8, 6, 4) .
\end{gathered}
\label{eq:chargesmodel}
\end{equation}
With these charge assignments, the FN Ansatz for the leptoquark couplings is:
\begin{equation}
\Delta_{QL} ~\sim~ 
\begin{pmatrix}
\lambda^3 & \lambda^2 & \lambda^2 \\
\lambda^2 & \lambda & \lambda \\
1 & \lambda & \lambda 
\end{pmatrix},
\qquad
\Delta_{\bar{u}\bar{e}} ~\sim~
\begin{pmatrix}
\lambda^{12} & \lambda^{10} & \lambda^8 \\
\lambda^9 & \lambda^7 & \lambda^5 \\
\lambda^8 & \lambda^6 & \lambda^4
\end{pmatrix}.
\label{eq:lambdasBknunu}
\end{equation}
The resulting bounds, neglecting $\mathcal{O}(1)$ Yukawa factors, are shown as the green bars in Figure~\ref{fig:observable_bounds}.
Compared to the anarchic Ansatz, the bounds on the leptoquark mass are significantly relaxed.

Nevertheless, it is clear that the mass range consistent with the $\bknnbr$ measurements at Belle~II is still excluded by other observables under the FN Ansatz. We have checked that---while the exact bounds for different observables can change significantly---this conclusion remains unchanged for the other possible charge assignments enumerated in Table~\ref{tab:gen_charges}.
If any deviation from SM is observed in $\bknnbr$, the $S_1$ leptoquark embedded in a vanilla FN model cannot explain the anomaly while respecting bounds from other measurements.

Adding wrinkles to the FN Ansatz as discussed in \S\ref{sec:wrinkles} can ameliorate the tension with these observables. Using the scaling of the observables with the leptoquark Yukawas shown in Table~\ref{tab:mainobs} as a guide, we add the following wrinkles (as defined in Eq.~\eqref{eq:wrinkledef}) to the leptoquark Yukawa matrices:  
\begin{equation}
    W_{\bar{u}\bar{e}}^{ij} = 
    \lambda^3, \qquad
    W_{QL} = \left( \begin{matrix}
        \lambda^3 & \lambda^3 & \lambda^3 \\
        \lambda^3 & 1 & 1 \\
        \lambda^3 & 1 & 1
    \end{matrix}
    \right).
    \label{eq:wrinklesLQ}
\end{equation}
This is the largest number of wrinkles we can add to suppress the leptoquark contribution to the most constraining observables (especially electron EDM, $\mu \rightarrow e \gamma$, $\tau \rightarrow \mu \gamma$, and meson mixing observables), while retaining consistency with the na\"{i}ve constraint $\omega \gtrsim \lambda^3 \sim 1/16\pi^2$ from \S\ref{subsec:radiative} and leaving the contribution to $\bknnbr$ mostly intact. 
Further suppression with additional powers of $\lambda$ may be possible, but must be carefully checked with all of the consistency conditions in Appendix~\ref{app:consistency}.

It is worth emphasizing that it is not obvious how to get the pattern of wrinkles in Eq.~\eqref{eq:wrinklesLQ} from the example UV completions discussed in \S\ref{subsec:uv_completions}. Nevertheless, we can treat them consistently in an effective field theory approach, and leave the model-building to future work. 
Note also that with the additional suppression of the right-handed Yukawa couplings, the phenomenology of this model resembles that of the RPV down squark as discussed in \S\ref{sec:FNplusBSM}.

The contribution of this wrinkled FN setup to various observables is shown by blue bars in Figure~\ref{fig:observable_bounds}. 
We find that the set of wrinkles from Eq.~\eqref{eq:wrinklesLQ} sufficiently suppresses the contribution to other observables, so that they are all compatible with the mass range of interest for $\bknnbr$. 
In particular, bounds from meson mixing observables and leptonic meson decays are circumvented. 
Within this wrinkled setup, the viable leptoquark mass range that can account for a signal in $\bknnbr$ is slightly above the current direct search bounds at the LHC (see Refs.~\cite{Diaz:2017lit, Schmaltz:2018nls}) and could be detected in future searches  at the LHC or future hadron \cite{Allanach:2017bta, Allanach:2019zfr, Bandyopadhyay:2020wfv, Hiller:2021pul} or lepton \cite{Huang:2021biu, Asadi:2021gah, Bandyopadhyay:2021pld, Qian:2021ihf, Parashar:2022wrd, MuonCollider:2022xlm} colliders. 

There are several observables which probe a similar mass range to $\bknnbr$ which will see significant improvement in experimental measurements soon. In particular, these observables include $\mu \to e\gamma$, $\mathrm{CR}(\mu \to e)$, and electron EDM, though the precise mass range depends on $\mathcal{O}(1)$ Yukawa couplings in the UV completion.
As a result, they could be the smoking gun signal of an FN-like $S_1$ leptoquark solution to any future excess observed in $\bknnbr$. Since the experimental precision on both of these (and several other) observables is expected to improve significantly in the near future, we will dedicate the next subsection to discussing potential discovery prospects for this wrinkled FN scenario.

\subsection{Predictions for Future Measurements}
\label{subsec:predictions}

We have already seen that adding wrinkles to a FN Ansatz allows for greater flexibility in simultaneously accommodating experimental deviations from the SM while satisfying constraints from other observables and explaining the observed pattern of SM masses and mixing angles.
As we will now emphasize, despite this added flexibility, these choices still make concrete predictions for other observables, which can be tested in future experiments. The importance of these tests lies in being able to probe indirect information about the underlying UV model which is hidden in the charge assignments and wrinkles in the IR.

Several upcoming experiments will provide concrete tests of our wrinkled Ansatz. When assuming the wrinkled FN Ansatz from Eq.~\eqref{eq:wrinklesLQ} for the leptoquark Yukawa couplings, several classes of observables--- including LFV processes, the electron EDM, meson-mixing measurements, and the decay $K \rightarrow \mu \nu$---have a present sensitivity to roughly the same mass scale as $\bknnbr$. 
Moreover, the mass reach of many of these observables is expected to improve significantly with forthcoming experimental data. 
Since we have suppressed our model contribution to these observables as far as possible while satisfying the bound from the consistency condition in Eq.~\eqref{eq:radiativebound}, these correlated signals allow for a definitive test of these types of wrinkled models within the FN mechanism.

\begin{figure}[t]
\centering
\includegraphics[width=0.45\linewidth]{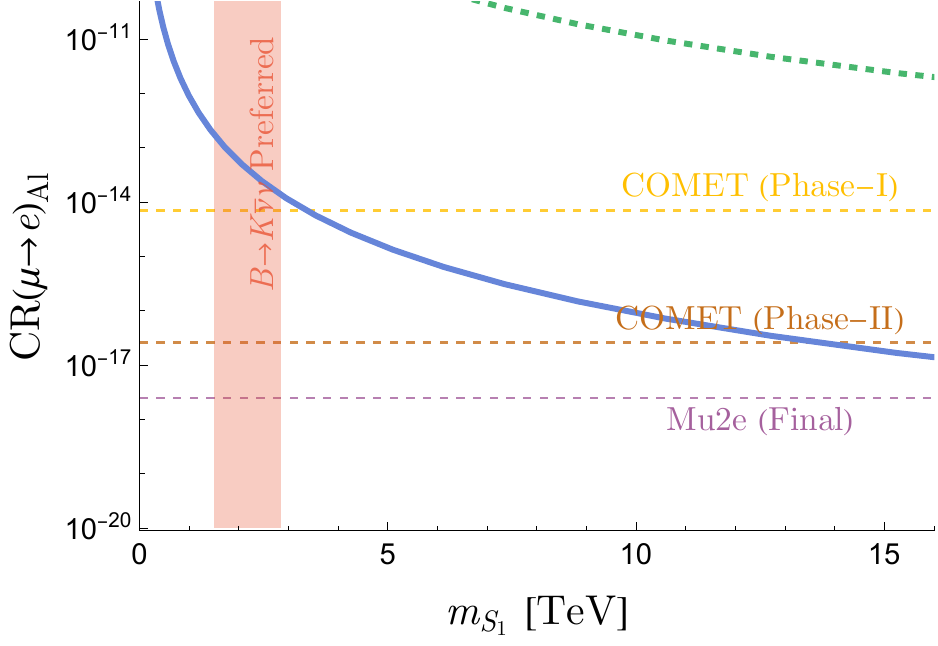}
\quad
\includegraphics[width=0.45\linewidth]{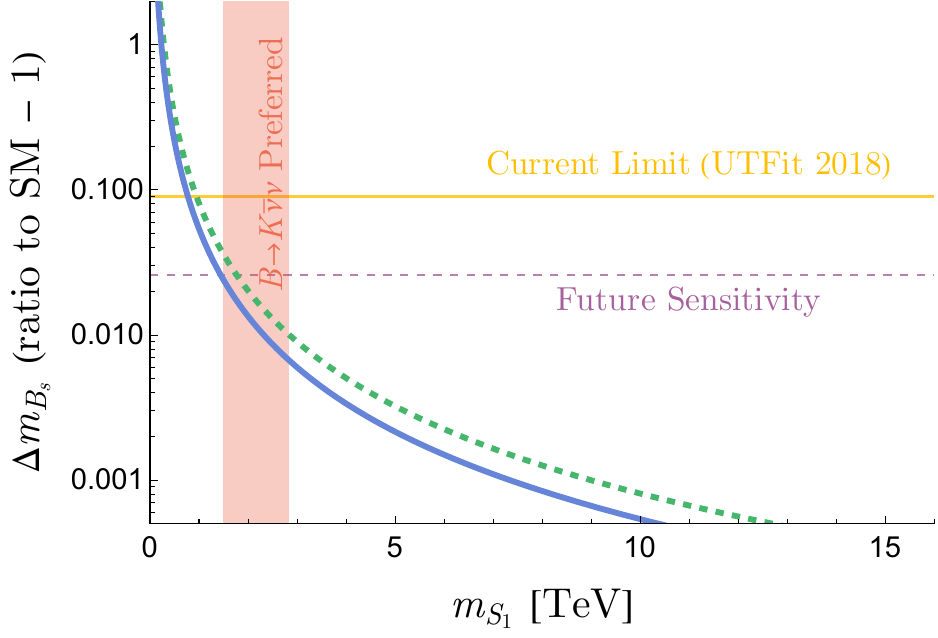} \\
\includegraphics[width=0.45\linewidth]{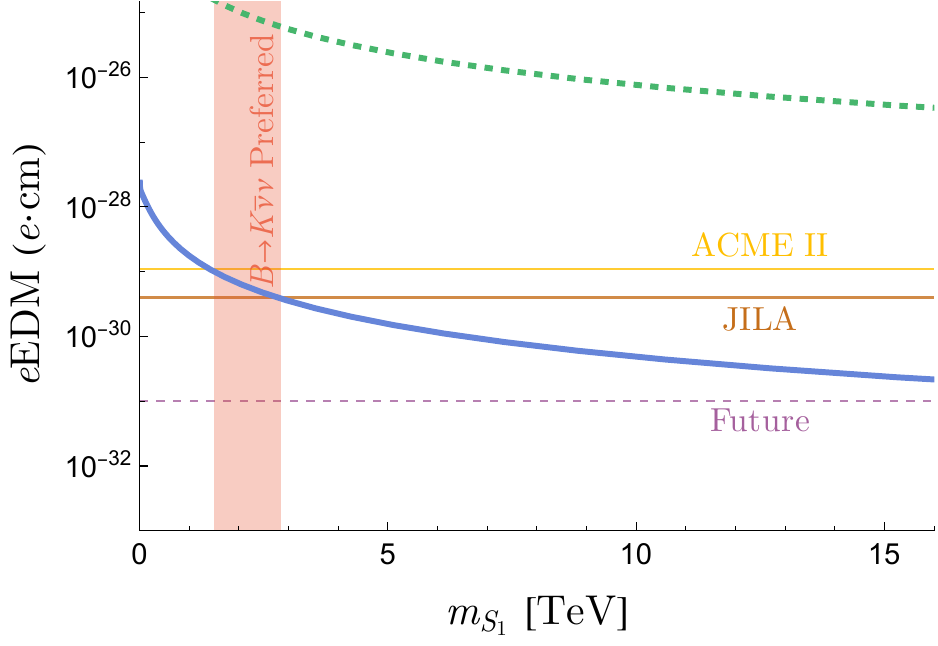}
\quad
\includegraphics[width=0.45\linewidth]{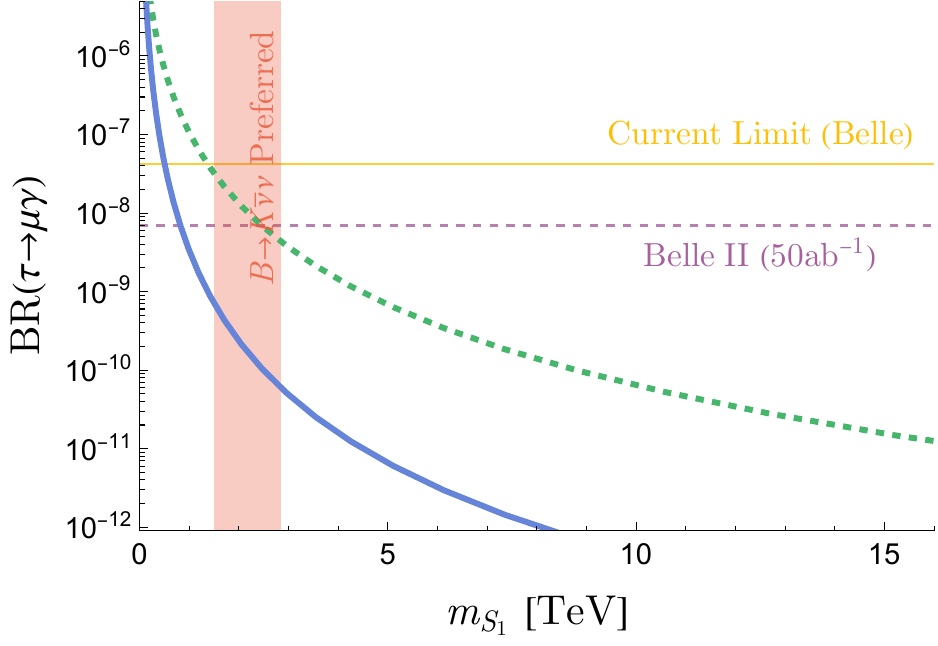}
\caption{
Predictions for the $S_1$ leptoquark contributions to precision observables with the wrinkled (blue, solid) and vanilla (green, dashed) FN Ans\"atze described in \S\ref{subsec:constraints}. We show the $\mu \to e$ conversion rate in an aluminum nucleus (top left), the electron EDM (bottom left), the relative new physics contribution to $\Delta m_{B_s}$ (top right), and $\textrm{BR}(\tau\to\mu\gamma)$ (bottom right), using solid (dashed) lines for current (future) experimental bounds or sensitivity.
We do not show the best current bounds on $\mu \to e$ conversion rate, $<7 \times 10^{-13}$, from SINDRUM~II \cite{SINDRUMII:2006dvw} since it was made with a different nucleus (gold).
The red 
band indicates the mass range of interest for $\bknnbr$, as in Figure~\ref{fig:observable_bounds}. }\label{fig:predicted_obs}
\end{figure}

At the moment, the strongest bound on LFV processes involving muons is the 90\%~C.L. limit, $\textrm{BR}(\mu \to e\gamma) < 4.2 \times 10^{-13}$ set by the MEG experiment~\cite{MEG:2016leq}.
In the future, however, the most powerful probes of this model will come from searches for $\mu \to e$ conversion in atomic nuclei. 
As discussed in more detail in Appendix~\ref{app:obs}, the conversion rate depends not only on the dipole operator relevant for $\mu \to e \gamma$ and $\mu \to 3e$ decays, but also on four-fermion operators including the first generation quarks generated by integrating out the leptoquark. 
Future prospects for detecting $\mu \rightarrow e$ conversion include the COMET experiment, which will set a limit on the conversion rate of $7 \times 10^{-15}$ ($2.6 \times 10^{-17}$) in Phase-I (Phase-II)~\cite{Adamov:2018vin, Angelique:2018svf}, and at Mu2e, which aims at a final sensitivity of $2.5 \times 10^{-18}$~\cite{Bartoszek:2014mya, Abusalma:2018xem}\footnote{This sensitivity might be achievable at Mu2e-II, a proposed upgrade of Mu2e using the PIP-II accelerator at Fermilab, potentially with a target material other than aluminum~\cite{Mu2e-II:2022blh, CGroup:2022tli}.}, both in aluminum nuclei.
For more discussion on current and forthcoming searches for LFV, see  Refs.~\cite{Ellis:2016yje, Homiller:2022iax, Baldini:2018uhj, Artuso:2022ouk, Mu2e-II:2022blh, CGroup:2022tli}.

In the top left panel of Figure~\ref{fig:predicted_obs}, we show the predicted $\mu \to e$ conversion rate in aluminum nuclei as a function of the leptoquark mass, with the wrinkled FN Ansatz taken for the Yukawa couplings. The $\bknnbr$-preferred region discussed in \S\ref{subsec:constraints} is highlighted in red, while the dashed horizontal lines show the future sensitivities for the conversion rate. 
We see that even Phase-I of the COMET experiment will be sensitive to the mass range preferred by $\bknn$ measurements, while Mu2e will decisively test all of the relevant parameter space predicted by this model of flavor.

For the electron EDM, the bounds from 
the ACME~II and JILA experiments~\cite{ACME:2018yjb, Roussy:2022cmp} are at the level $d_e < 1.1 \times 10^{-29}$ and $4.1\times 10^{-30}\,e\textrm{ cm}$, respectively. For anarchic flavor couplings, this excludes masses up to $\sim 10^5\,\textrm{TeV}$. A vanilla FN Ansatz relaxes this constraint to $\sim 10^2\,\textrm{TeV}$, and with the additional wrinkles invoked in Eq.~\eqref{eq:wrinklesLQ}, this bound weakens to $m_{S_1} \gtrsim 2.7\,\textrm{TeV}$. 
Random factors of $\mathcal{O}(1)$, neglected throughout our calculations, can slightly affect the reach on $m_{S_1}$. 
The fact that this is the same mass range as favored by $\bknnbr$ measurements, and that the reach in $m_{S_1}$ scales faster with improvements to electron EDM measurements compared to other observables, underscores the importance of future electron EDM experiments in probing our model. In the coming years, experimental advances and new technologies promise to increase the sensitivity of EDM experiments by an order of magnitude or more~\cite{Ho:2020ucd, Fitch:2020jil, Alarcon:2022ero}. 
In the lower-left panel of Figure~\ref{fig:predicted_obs}, we show the predicted value of the electron EDM as a function of the leptoquark mass, alongside current bounds and a projected constraint of $10^{-31}\,e\textrm{ cm}$, assuming an $\mathcal{O}(1)$ CP-violating phase.
As is clear from the figure, future EDM experiments will decisively test this model, up to scales $m_{S_1} \sim \mathcal{O}(10)\,\textrm{TeV}$.

For the meson mixing observables, we focus in particular on the neutral $B_s$ meson mass difference, $\Delta m_{B_s}$,  
whose matrix element is directly related to the $\bknn$ process for the $S_1$ leptoquark.
To understand the current sensitivity to new physics of $B_s-\bar{B}_s$ mixing, we follow the UTFit analysis~\cite{UTfit:2007eik,Cerri:2018ypt,Ferrari:2023slj} and compute the quantity $C_{B_s}$, defined as 
\begin{equation}
C_{B_s} e^{2i\phi_{B_s}} \equiv \myfrac[2pt]{\langle B_s| \mathcal{H}_{\textrm{mix}}^{\textsc{SM}+\textsc{NP}}|\bar{B}_s \rangle}{\langle B_s| \mathcal{H}_{\textrm{mix}}^{\textsc{SM}}|\bar{B}_s \rangle}, \label{eq:utfit_ratio}
\end{equation}
where $\mathcal{H}_{\textrm{mix}}$ includes the four-fermion operators responsible for $\Delta F = 2$ transitions, as defined in Appendix~\ref{App:Meson_mixing}.
The SM is defined as the point $C_{B_s} = 1$, $\phi_{B_s} = 0$, and the allowed size of the new physics contribution is determined by a global fit to the flavor sector, with the range determined primarily by the uncertainties on the input parameters, such as the CKM matrix elements. To be conservative, we consider only the absolute value of the matrix elements above, and avoid making any assumptions about the relative phase between the SM and leptoquark contributions, which is constrained by $\phi_{B_s}$. 

The resulting current and future sensitivities (where we assume the current central value is at the SM, for consistency with future projections) are shown on the top right in Figure~\ref{fig:predicted_obs}. The projected future sensitivity of $\Delta C_{B_s} = 0.026$ is taken from Ref.~\cite{Cerri:2018ypt}, based on projections of HL-LHC results and Belle~II results with $50\,\textrm{ab}^{-1}$ integrated luminosity. We see that the improved sensitivity will start to probe the leptoquark mass range preferred by the $\bknnbr$ measurements. It is also worth emphasizing that these projections do not account for potential improvements in lattice inputs, and thus could be quite conservative. A statistically significant signal in any of the aforementioned channels would also warrant a much more careful analysis of these $B_s$-mixing constraints and projections, including phase information that depends in more detail on the flavor Ansatz, which could improve sensitivity even further.

A number of additional flavor-changing or flavor-violating decays will be probed with increasing sensitivity at Belle~II. 
A notable example is the LFV decay $\tau \to \mu \gamma$, for which the current bound set by Belle is $\textrm{BR}(\tau \to \mu\gamma) < 4.2 \times 10^{-8}$~\cite{Belle:2021ysv}.
Belle~II is projected to improve this bound to $6.9\times 10^{-9}$~\cite{Belle-II:2022cgf, Banerjee:2022vdd}.
In the lower-right panel of Figure~\ref{fig:predicted_obs}, we show the predicted branching ratio of $\tau \to \mu \gamma$ as a function of mass. We see that, for the mass range preferred by the $\bknnbr$ measurements, the addition of wrinkles in our flavor Ansatz suppresses what would otherwise be a predicted signal from assuming the FN mechanism.

Finally, the $K \to \pi \nu\bar{\nu}$ decays, which would rule out the preferred mass range for $\bknn$ without wrinkles, have a sensitivity $\sim 1\,\mathrm{TeV}$ in the wrinkled FN Ansatz.
The $K^+ \to \pi^+\nu\bar{\nu}$ decay was only recently measured (with a significance of $3.4\sigma$) at the NA62 experiment~\cite{NA62:2021zjw}. A $10$ -- $20\%$ precision on this branching ratio is necessary to start excluding $m_{S_1} \sim 2$ -- $3\,\textrm{TeV}$, and the requirement for $K_L \to \pi^0 \nu\bar{\nu}$ is similar.
Both of these may be achievable with future runs at NA62, or at future experiments planned at the NA62 hall at CERN~\cite{NA62:2020upd, Goudzovski:2022vbt} and at J-PARC~\cite{Aoki:2021cqa}, and would be an interesting complementary probe of the same physics considered here.

The preceding discussion demonstrates that all of these powerful, forthcoming measurements could have a similar sensitivity to new mass scales for an appropriate choice of wrinkles.
Exactly which search channel is ideal depends on the precise pattern of charges and wrinkles in the IR.
However, the expectation that we will probe these other correlated signals is relatively robust since the wrinkles in Eq.~\eqref{eq:wrinklesLQ} were chosen to saturate the bound in Eq.~\eqref{eq:radiativebound} without diminishing the $\bknn$ signal. While this enhancement to $\bknn$ was only for illustration, and not a fit to a true, significant deviation from the SM, it reveals that for some motivated UV models of flavor, upcoming experiments can simultaneously test explanations for the SM flavor puzzle.

\section{Digestifs}
\label{sec:conclusion}

When new physics is embedded in the FN mechanism, the FN Ansatz determines the size of both the SM and new physics couplings. In this paper, we have put forward a systematic extension of this Ansatz which can change the expected scaling of the new physics and SM couplings. 
These changes, referred to as wrinkles, deviate from the FN pattern that is dictated by the horizontal symmetry charges.  
Wrinkles allow us to demand consistency with other experimental measurements and searches: modifying the relative size of couplings restores some theories that would otherwise be unfeasible due to the correlations between different observables from the FN Ansatz. Therefore, they vastly increase the FN mechanism's versatility in accommodating solutions to flavor anomalies.
However, owing to radiative corrections, we have also argued that wrinkles can not give rise to arbitrarily large deviations from vanilla FN predictions. There are consistency conditions which must be obeyed by the size of the new wrinkled Yukawas. 

While the primary purpose of wrinkles is to give a consistent IR description for various flavor observables, we have also explored how they can be UV completed by various different models. Specifically, in this paper we have given some simple schematic examples of possible UV realizations. In future work, it would also be interesting to understand more about what patterns of wrinkles can be realistically realized in the UV and the various models that can be used to realize them.

Throughout this work, we focused on the phenomenological example of the $S_1$ leptoquark. 
We discussed the implementation in the IR when the leptoquark is embedded in a FN model. 
We also provided a detailed example of how an enhancement of the leptoquark contribution to $\bknnbr$ can consistently respect other experimental bounds, but only if wrinkles are invoked. 
This wrinkled setup also motivates future measurements, since several signals would be on the verge of discovery in this model, even when the number of wrinkles is enlarged to saturate the simplest consistency condition. 
In particular, we showed predictions for the most sensitive upcoming probes, namely $\mu \to e$ conversion and the electron EDM. 

While we limited our exploration to a specific example with the $S_1$ leptoquark in this paper, it would be interesting to explore how wrinkles can be applied more broadly. 
For instance, in our example we fixed the horizontal charges of the SM particles, but there are many other possible choices that reliably yield the SM masses and mixing angles. 
One could explore how changing the charges affects the correlations and hence the allowed wrinkled Ansatz, and see which observables remain correlated to the same mass scale more generally.
It would also be intriguing to include other flavor spurions or to add wrinkles to the SM couplings in addition to the new physics couplings. 
Moreover, it would be useful to do a broad methodical study on the effect of $\mathcal{O}(1)$ numbers in different spurions to explore naturalness in these types of models; see Refs.~\cite{Fedele:2020fvh,Aloni:2021wzk, Cornella:2023zme} for previous studies of naturalness in such models.

Aside from the flexibility permitted by wrinkles, it is worthwhile to emphasize a separate point about FN models in general: there is more than one
charge assignment that can naturally generate the observed SM masses and mixings, beyond just the 
overall shift in the quark and lepton charges. 
In particular, we find that the charges of first generation fermions can be either larger than or smaller than other two generations. 
This is in contrast to a criterion in Ref.~\cite{Froggatt:1978nt}, where it was demanded that charges increase monotonically between generations. 
However, this general FN charge assignment is still not anomaly free and requires some cancellation mechanism, such as Green-Schwarz.

With a number of precision flavor experiments gathering data in the near future that could probe the underlying mechanisms for the flavor structure of the SM, it is the right moment to think about sophisticated UV flavor structures beyond the vanilla FN setup.  
Wrinkles---a systematic deviation from the vanilla FN prediction for the relationship between different couplings---are one such example that significantly increase the versatility of FN constructions in confronting potential signs of flavorful new physics. We encourage their use in embedding solutions to anomalous signals in UV complete models of flavor.

\subsection*{Acknowledgments}

We thank Daniel Aloni, Wolfgang Altmannshofer, Avital Dery, Darius Faroughy, Seth Koren, Graham Kribs, Clara Murgui, Matthew Reece, Matthew Strassler, and Lian-Tao Wang for helpful discussions. 
The work of PA is supported in part by the U.S. Department of Energy under Grant Number DE-SC0011640. 
AB, KF and SH are supported in part by the DOE grant DE-SC0013607. 
KF and SH are also supported in part by the Alfred P.~Sloan Foundation Grant No.~G-2019-12504, and KF is also supported in part by the NASA ATP Grant NNX16AI12G. 
The work of AP is supported in part by the US National Science Foundation Grant PHY2210533 and the Simons Foundation Grant No. 623940. 
PA thanks Mainz Institute for Theoretical Physics (MITP) of the Cluster of Excellence PRISMA$^+$ (Project ID 39083149) and KF thanks the Aspen Center for Physics (which is supported by NSF grant PHY-2210452) for their hospitality during the completion of this work.

\appendix

\clearpage
\section{Full Set of Consistency Conditions}
\label{app:consistency}

Here we list the full set of consistency conditions that arise for the Yukawa couplings of the $S_1$ leptoquark model embedded in an FN setup. They arise from considering the representation of the Yukawas under the SM flavor symmetry group, $G_{\textrm{flavor}}$ (see Eq.~\eqref{eq:Gflavor}), and constructing the other combinations of Yukawas that transform in the same way. Each combination produces a one-loop contribution via a diagram analogous to Figure~\ref{fig:loop_correction_S1_yuk}. The representations of $Y_{Q\bar{u}}$, $Y_{Q\bar{d}}$, $Y_{L\bar{e}}$, $\Delta_{QL}$, and $\Delta_{\bar{u}\bar{e}}$ are listed in Eqs.~\eqref{eq:GflavorSM} and \eqref{eq:GflavorLQ}. We find
\begin{equation}
\label{eq:consistency-app}
\begin{split}
\Big|\Delta_{\bar{u}\bar{e}}^{ij}\Big| 
& \geq \frac{1}{16\pi^2} 
\Big|\big(\Delta_{\bar{u}\bar{e}} \cdot \Delta_{\bar{u}\bar{e}}^\dagger \cdot \Delta_{\bar{u}\bar{e}}  \big)^{ij}\Big|, \\
\Big|\Delta_{\bar{u}\bar{e}}^{ij}\Big| 
& \geq \frac{1}{16\pi^2} 
\Big|\big(  \Delta_{\bar{u}\bar{e}}  \cdot Y_{L\bar{e}}^{\dagger} \cdot Y_{L\bar{e} } \big)^{ij}\Big|, \\
\Big|\Delta_{\bar{u}\bar{e}}^{ij}\Big| 
& \geq \frac{1}{16\pi^2} 
\Big|\big( Y_{Q\bar{u}}^{T} \cdot Y_{Q\bar{u}}^{*} \cdot \Delta_{\bar{u}\bar{e}} \big)^{ij}\Big|, \\
\Big|\Delta_{\bar{u}\bar{e}}^{ij}\Big| 
& \geq \frac{1}{16\pi^2} 
\Big|\big( Y_{Q\bar{u}}^{T} \cdot \Delta_{QL}^{*} \cdot Y_{L\bar{e}} \big)^{ij}\Big|, 
\\[1em]
\Big|Y_{Q\bar{d}}^{ij}\Big| 
& \geq \frac{1}{16\pi^2} 
\Big|\big( Y_{Q\bar{d}}  \cdot Y_{Q\bar{d}}^\dagger \cdot Y_{Q\bar{d}} \big)^{ij}\Big|, \\
\Big|Y_{Q\bar{d}}^{ij}\Big| 
& \geq \frac{1}{16\pi^2} 
\Big|\big( Y_{Q\bar{u}}  \cdot Y_{Q\bar{u}}^\dagger \cdot Y_{Q\bar{d}} \big)^{ij}\Big|, \\
\Big|Y_{Q\bar{d}}^{ij}\Big| 
& \geq \frac{1}{16\pi^2} 
\Big|\big( \Delta_{QL}  \cdot \Delta_{QL}^\dagger \cdot Y_{Q\bar{d}} \big)^{ij}\Big|. 
\\[1em]
\Big|Y_{L\bar{e}}^{ij}\Big| 
& \geq \frac{1}{16\pi^2} 
\Big|\big( Y_{L\bar{e}}  \cdot Y_{L\bar{e}}^\dagger \cdot Y_{L\bar{e}} \big)^{ij}\Big|, \\
\Big|Y_{L\bar{e}}^{ij}\Big| 
& \geq \frac{1}{16\pi^2} 
\Big|\big( \Delta_{QL}^{T} \cdot \Delta_{QL}^{*} \cdot Y_{L\bar{e}}\big)^{ij}\Big|, \\
\Big|Y_{L\bar{e}}^{ij}\Big| 
& \geq \frac{1}{16\pi^2} 
\Big|\big(Y_{L\bar{e}} \cdot  \Delta_{\bar{u}\bar{e}}^{\dagger} \cdot \Delta_{\bar{u}\bar{e}} \big)^{ij}\Big|, \\
\Big|Y_{L\bar{e}}^{ij}\Big| 
& \geq \frac{1}{16\pi^2} 
\Big|\big( \Delta_{QL}^{T} \cdot Y_{Q\bar{u}}^{*} \cdot \Delta_{\bar{u}\bar{e}}\big)^{ij}\Big|, 
\end{split}
\qquad
\begin{split}
\Big|\Delta_{QL}^{ij}\Big| 
& \geq \frac{1}{16\pi^2} 
\Big|\big(\Delta_{QL} \cdot \Delta_{QL}^\dagger \cdot \Delta_{QL} \big)^{ij}\Big|, \\
\Big|\Delta_{QL}^{ij}\Big| 
& \geq \frac{1}{16\pi^2} 
\Big|\big( Y_{Q\bar{d}} \cdot Y^{\dagger}_{Q\bar{d} } \cdot \Delta_{QL} \big)^{ij}\Big|, \\
\Big|\Delta_{QL}^{ij}\Big| 
& \geq \frac{1}{16\pi^2} 
\Big|\big( Y_{Q\bar{u}} \cdot Y^{\dagger}_{Q\bar{u} } \cdot \Delta_{QL} \big)^{ij}\Big|, \\
\Big|\Delta_{QL}^{ij}\Big| 
& \geq \frac{1}{16\pi^2} 
\Big|\big( \Delta_{QL} \cdot Y_{L\bar{e}}^{*} \cdot Y^{T}_{L\bar{e} } \big)^{ij}\Big|, \\
\Big|\Delta_{QL}^{ij}\Big| 
& \geq \frac{1}{16\pi^2} 
\Big|\big( Y_{Q\bar{u}} \cdot \Delta_{\bar{u}\bar{e}}^{*} \cdot Y^{T}_{L\bar{e}} \big)^{ij}\Big|,
\\[1em]
\Big|Y_{Q\bar{u}}^{ij}\Big| 
& \geq \frac{1}{16\pi^2} 
\Big|\big( Y_{Q\bar{u}}  \cdot Y_{Q\bar{u}}^\dagger \cdot Y_{Q\bar{u}} \big)^{ij}\Big|, \\
\Big|Y_{Q\bar{u}}^{ij}\Big| 
& \geq \frac{1}{16\pi^2} 
\Big|\big( \Delta_{QL} \cdot \Delta_{QL}^\dagger \cdot Y_{Q\bar{u}} \big)^{ij}\Big|, \\
\Big|Y_{Q\bar{u}}^{ij}\Big| 
& \geq \frac{1}{16\pi^2} 
\Big|\big( Y_{Q\bar{u}} \cdot \Delta_{\bar{u}\bar{e}}^{*} \cdot \Delta_{\bar{u}\bar{e}}^{T} \big)^{ij}\Big|, \\
\Big|Y_{Q\bar{u}}^{ij}\Big| 
& \geq \frac{1}{16\pi^2}
\Big|\big( Y_{Q\bar{d}} \cdot Y^\dagger_{Q\bar{d}} \cdot Y_{Q\bar{u}}\big)^{ij}\Big|, \\
\Big|Y_{Q\bar{u}}^{ij}\Big| 
& \geq \frac{1}{16\pi^2} 
\Big|\big( \Delta_{QL} \cdot Y^{*}_{L\bar{e}} \cdot \Delta_{\bar{u}\bar{e}}^{T}\big)^{ij}\Big|,
\end{split}
\end{equation}
We could also consider additional consistency conditions with more 
Yukawa couplings on the right-hand side, but those will be sub-dominant to those listed above.
The consistency conditions listed above are specific to the spurions we have considered, but this procedure generalizes to arbitrary new 
spurions under $G_{\textrm{flavor}}$.

\clearpage
\section{Calculation of Other Observables}
\label{app:obs}

In this appendix we review the contributions of the $S_1$ leptoquark to various flavor observables. The emphasis is on the dependence on the flavor spurions, $\Delta_{QL}$ and $\Delta_{\bar{u}\bar{e}}$, with many details left to the references. In what follows, $V$ is the CKM matrix and $v$ is the SM Higgs vev. We use the CKM parameters as determined in Ref.~\cite{Ferrari:2023slj} while the remainder of our inputs are taken from the PDG~\cite{ParticleDataGroup:2022pth}. Furthermore, we work with a set of operators where the neutrinos are left in the flavor basis as the processes we consider have either a final state neutrino of a specific flavor, or a sum over all possible final state neutrinos, which can be done in any basis. Therefore, we do not include explicit factors of the PMNS matrix in the expressions for the Wilson coefficients.
We assume the leptoquark Yukawas are given in the IR and neglect the running effects. 
These calculations are used in \S\ref{sec:bknunu} to identify the most relevant constraints and the wrinkles which are useful for evading them. 
We also employ mostly four-component spinor notation in this appendix for consistency with the majority of the references.

\subsection{Dipole Moments}

\begin{figure}[!htb]
    \centering
    \includegraphics[height=3.5cm]{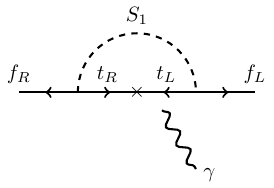}
    \caption{Feynman diagram (in two-component notation) for the $S_1$ leptoquark contribution to the dipole operators of charged fermions, including $(g-2)_{\mu}$. The largest contribution arises from the top quark in the loop. The photon can attach to either internal line in the loop.}
    \label{fig:mu_g2_diagrams}
\end{figure}

First we calculate the contribution of $S_1$ to the electric and magnetic dipole moments of SM particles. After integrating out the leptoquark, the one loop diagrams of Figure~\ref{fig:mu_g2_diagrams} can give rise to the effective operators
\begin{equation}
\mathcal{L} \supset c^R_{ij} \bar{f}_i \sigma^{\mu\nu} P_R f_j F_{\mu\nu} + \mathrm{h.c.,}
    \label{eq:LFVlagrangian}
\end{equation}
where $f_{i,j}$ are SM fermions, $F_{\mu\nu}$ is the electromagnetic field strength, and $c_{ij}^R$ is the corresponding Wilson coefficient. 
By matching the diagrams in Figure~\ref{fig:mu_g2_diagrams} to this operator, we can calculate $c_{ij}^R$ values in our setup. 
See Refs.~\cite{Hisano:1995cp,Ellis:2008zy,Ellis:2016yje,Crivellin:2018qmi,Parikh:2018huy,Aloni:2021wzk,Nakai:2016atk} for details of the calculation. 
Following the notation of Ref.~\cite{Aloni:2021wzk}, we have 
\begin{multline}
\label{eq:cR_UV_Yukawa}
c^R_{ij} = \sum_{\bar{q}}  \frac{e}{64 \pi^2 m_{S_1}^2}
\bigg[ 	m_{\bar{q}} (V^{*} \Delta_{QL})^{\bar{q} i\,*} \Delta^{\bar{q} j\,*}_{\bar{u}\bar{e}} \Big( Q_{S_1} A(r) - Q_{\bar{q}} B(r)	\Big)	 \\
+  \left(	m_i 	\Delta^{\bar{q}i}_{\bar{u}\bar{e}} \Delta^{\bar{q}j\,*}_{\bar{u}\bar{e}} + m_j 	(V^{*} \Delta_{QL})^{ \bar{q} i\, *} (V^{*} \Delta_{QL})^{\bar{q} j} \right)   \Big( Q_{S_1} \bar{A}(r) - Q_{\bar{q}} \bar{B}(r)		\Big)    \bigg],
\end{multline}
where the sum is over all possible up-type anti-quarks $\bar{q}$ that can go in the loop, $Q$ is the electric charge, $m_{i,j}$ are the masses of the external leptons, $r=m_{\bar{q}}^2/m_{S_1}^2$, and the loop functions are defined in the appendix of Ref.~\cite{Aloni:2021wzk}.

In terms of these Wilson coefficients, the electric and magnetic dipole moments can be written as
\begin{equation}
    \label{eq:dipolecalcs}
    d_f = 2 ~\mathrm{Im}\,c_{ff}^R \qquad a_f = \frac{4m_f}{e} ~\mathrm{Re}\, c_{ff}^R.
\end{equation}
Note that because the two fermions in the operator Eq.~\eqref{eq:LFVlagrangian} have opposite chirality, all the contributions in Eq.~\eqref{eq:cR_UV_Yukawa} are proportional to the external fermion or internal quark mass. As a result, unless the Yukawas are very suppressed, the $S_1$ contribution to EDMs and MDMs are dominated by diagrams with the top 
quark in the loop, which are proportional to $m_t$.

\subsection{Lepton Flavor Violating Observables}

The Lagrangian from Eq.~\eqref{eq:LFVlagrangian} also contributes to LFV decays as~\cite{Hisano:1995cp,Ellis:2016yje,Aloni:2021wzk}
\begin{equation}
\mathrm{BR} \left( \ell \rightarrow \ell' \gamma \right) = \frac{48\pi^2}{G_F^2 m_\ell^2} \left( |c_{\ell\ell'}^R|^2 + |c_{\ell'\ell}^R|^2 \right).
    \label{eq:lfvappx}
\end{equation}
Similar to the previous section, dominant contributions to $c_{\ell'\ell}^R$ come from diagrams with the heaviest quarks in the loop. More concretely, we find that in the limit $m_{\ell},m_{\ell^{\prime}}\ll m_{S_{1}}$ 
\begin{align}
    c^R_{\ell\ell'} \approx \frac{e m_{q}}{16\pi^2 m_{S_1}^2 }\left[\ln(\frac{m_{S_1}^2}{m_q^2})-\frac{7}{4}\right] (V^{*} \Delta_{QL})^{q\ell\, *} \Delta^{q\ell'\,*}_{\bar{u}\bar{e}}.
\end{align}
These dipole operators also contribute to the well-constrained LFV processes $\mu \to 3 e$ and $\mu \to e$ conversion in nuclei. 
In our leptoquark model, the dipole operator is the only contribution to $\mu \to 3e$, so these branching ratios are directly correlated:
\begin{equation}
\textrm{BR}(\mu \to 3e) = \frac{\alpha}{3\pi} \Big( \log \frac{m_{\mu}^2}{m_e^2} - \frac{11}{4}\Big)\, \textrm{BR}(\mu \to e\gamma) \simeq \frac{1}{162}\textrm{BR}(\mu \to e\gamma) .
\end{equation}

The $\mu-e$ conversion process in nuclei, however, also receives contributions from four-fermion operators coupling the muon and electron to quarks. 
The effective Hamiltonian for this process can be written~\cite{Okada:1999zk, Kuno:1999jp}:
\begin{equation}
\begin{aligned}
\mathcal{H} & \supset \frac{G_F}{\sqrt{2}} \sum_{q=u,d,s}
\bigg[ 
\big(c_{LS}^{(q)} \bar{e}P_R\mu + c_{RS}^{(q)} \bar{e}P_L\mu\big) \bar{q}q
+ \big( c_{LP}^{(q)} \bar{e}P_R \mu + c_{RP}^{(q)} \bar{e}P_L \mu\big) \bar{q}\gamma_5 q \\
& \quad \qquad
+ \big( c_{LV}^{(q)} \bar{e}\gamma^{\mu}P_L\mu + c_{RV}^{(q)}\bar{e}\gamma^{\mu}P_R \mu\big) \bar{q}\gamma_{\mu} q
+ \big( c_{LA}^{(q)} \bar{e}\gamma^{\mu}P_L\mu + c_{RA}^{(q)}\bar{e}\gamma^{\mu}P_R \mu\big) \bar{q}\gamma_{\mu}\gamma_5 q \\[0.25em]
& \quad \qquad 
+ \frac{1}{2}\big( c_{LT}^{(q)} \bar{e}\sigma^{\mu\nu} P_R \mu + c_{RT}^{(q)} \bar{e}\sigma^{\mu\nu} P_L \mu\big) \bar{q}\sigma_{\mu\nu} q
+ \hc 
\bigg]
\end{aligned}
\end{equation}
where for the $S_1$ leptoquark, 
\begin{equation}
\begin{aligned}
c_{LS}^{(u)} = +c_{LP}^{(u)} = -c_{LT}^{(u)} & =
-\frac{1}{2} \frac{v^2}{m_{S_1}^2} (V^* \Delta_{QL})^{11\,*} \,\Delta_{\bar{u}\bar{e}}^{12\,*} \\
c_{RS}^{(u)} = -c_{RP}^{(u)} = -c_{RT}^{(u)} & = 
- \frac{1}{2} \frac{v^2}{m_{S_1}^2} (V^* \Delta_{QL})^{12}\, \Delta_{\bar{u}\bar{e}}^{11} \\
c_{LV}^{(u)} = -c_{LA}^{(u)} & = -\frac{1}{2}\frac{v^2}{m_{S_1}^2} (V^* \Delta_{QL})^{12}\, (V^* \Delta_{QL})^{11\,*}  \\
c_{RV}^{(u)} = c_{RA}^{(u)} & = -\frac{1}{2}\frac{v^2}{m_{S_1}^2} \Delta_{\bar{u}\bar{e}}^{11}\, \Delta_{\bar{u}\bar{e}}^{12\,*}
\end{aligned}
\end{equation}
The conversion rate is then computed by evaluating the overlap integrals of the fermion wave-function and nucleon densities. This has been performed in Ref.~\cite{Kitano:2002mt}, assuming the coherent conversion process (where the initial and final state nucleus are the same) dominates. We use the average values of their overlap integrals for the different nuclei (Al and Au).

\subsection{Leptonic Meson Decays}

\subsubsection{\texorpdfstring{$P \rightarrow \ell \nu$}{P to l nu}}

The EFT for a generic meson decaying to a neutrino and a charged lepton is \cite{Fajfer:2012jt,Bordone:2020lnb,Bauer:2015knc}

\begin{eqnarray}
\label{eq:Hleptonic}
\mathcal{H}_{\mathrm{eff}} &=& \frac{4 G_F V_{ud}}{\sqrt{2}} \left[  C^V_{L,ud\ell\nu} \left(  \bar{u}_L \gamma^\mu d_L \right) \left(  \bar{\ell}_L \gamma_\mu \nu_L \right) + C^V_{R,ud\ell\nu} \left(  \bar{u}_R \gamma^\mu d_R \right) \left(  \bar{\ell}_L \gamma_\mu \nu_L \right)    \right. \\
& + & \left.  C^S_{L,ud\ell\nu} \left(  \bar{u}_R d_L \right) \left(  \bar{\ell}_R   \nu_L \right) + C^S_{R,ud\ell\nu} \left(  \bar{u}_L d_R \right) \left(  \bar{\ell}_R  \nu_L \right)  \right] + \mathrm{h.c.}, \nonumber 
\end{eqnarray}
where $u$ ($d$) labels the involved up-type (down-type) quark.
In the SM, these decays are mediated by a $W$ exchange and the overall normalization is chosen such that $C^V_L=1$, with other Wilson coefficients set to zero.

For the $S_1$ leptoquark, we can show that at the leptoquark mass scale
\begin{align}
\label{eq:CVST}
C^V_{L,ud\ell\nu} & =  \frac{\Delta_{QL}^{d\nu}(V^{*} \Delta_{QL})^{u\ell\,*}}  {V_{ud}} \frac{v^2}{4 m_{S_1}^2}, \\
C^S_{L,ud\ell\nu} &= \frac{\Delta_{QL}^{d\nu} \Delta_{\bar{u}\bar{e}}^{u\ell\,} }{V_{ud}} \frac{v^2}{4 m_{S_1}^2}, \nonumber 
\end{align}
In our model there are no couplings to RH down-type quarks, so $C^S_{R,ud\ell\nu} = C^V_{R,ud\ell\nu}=0$. 

The meson branching ratio to $\ell\nu$ is given by 
\begin{multline}
\mathrm{BR} \left( P_{ud}^{-} \rightarrow \ell \nu \right) = \tau_{P} \frac{m_P f_P^2 G_F^2 |V_{ud}|^2}{8\pi} m_\ell^2 \left(   1- \frac{m_\ell^2}{m_P^2}    \right)^2  \\ 
\times  \Big| (C^V_{L,ud\ell\nu} - C^V_{R,ud\ell\nu}) + \frac{m_P^2}{m_\ell (m_u+m_d)} (C^S_{R,ud\ell\nu} - C^S_{L,ud\ell\nu} ) \Big|^2,
\label{eq:decaylmeson}
\end{multline}
where $\tau_P$ is the meson lifetime, $m_P$ is the meson mass, $f_P$ is the meson decay constant, $m_{\ell}$ is the final state lepton's mass, and $m_u$ ($m_d$) is the mass of the up-type (down-type) valence quark of the meson. 
This equation has been used to calculate the contribution of our model to various leptonic meson decays in the main text.

\subsubsection{\texorpdfstring{$P \rightarrow \ell \ell^{\prime} $}{P to l lprime} and \texorpdfstring{$P \rightarrow \nu \nu'$}{P to n nprime}}

The Hamiltonian describing a meson $P$ decaying to charged leptons $l$ and $l'$ is \cite{Becirevic:2016zri,Dorsner:2016wpm}
\begin{eqnarray}
    \label{eq:HPll}
    \mathcal{H}_{\mathrm{eff}} \supset \frac{4G_F}{\sqrt{2}} \lambda_{\textsc{CKM}} \left[
    \sum_{X=S,P,9,10} C_X^{qq';\ell\ell'} \mathcal{O}_X^{qq';\ell\ell'} + C_{X'}^{qq';\ell\ell'} \mathcal{O}_{X'}^{qq';\ell\ell'}
    + \hc
    \right] .
\end{eqnarray}
Here, $\lambda_{\textsc{CKM}}$ is a combination of two CKM entries involving the valence quarks of the meson, $C_X$ are Wilson coefficients, and their associated operators are
\begin{eqnarray}
    \label{eq:OperatorPll}
    \mathcal{O}_S^{qq';\ell\ell'} = \frac{\alpha_{\mathrm{em}}}{4\pi} (\bar{q} P_R q')(\bar{\ell} \ell') & ~ &     \mathcal{O}_P^{qq';\ell\ell'} = \frac{\alpha_{\mathrm{em}}}{4\pi} (\bar{q} P_R q')(\bar{\ell} \gamma^5 \ell') \\[0.5em] 
    \mathcal{O}_{9}^{qq';\ell\ell'} = \frac{\alpha_{\mathrm{em}}}{4\pi} (\bar{q} \gamma^\mu P_L q')(\bar{\ell} \gamma_\mu \ell') & ~ &    \mathcal{O}_{10}^{qq';\ell\ell'} = \frac{\alpha_{\mathrm{em}}}{4\pi} (\bar{q} \gamma^\mu P_L q')(\bar{\ell} \gamma_\mu \gamma^5 \ell'), \nonumber
\end{eqnarray}
The operators with a prime on the subscript are obtained by the replacement $P_{L/R} \rightarrow P_{R/L}$. 

At tree level, our leptoquark only gives rise to decays of $D$ and $\pi$ via $t$-channel diagrams, while decays of $K$, $~B$, and $B_s$ take place at one-loop level and are suppressed.  For the tree-level decays, the Wilson coefficients above can be calculated as a function of the leptoquark Yukawa couplings \cite{Dorsner:2016wpm}
\begin{equation}
\begin{aligned}
\label{eq:WCsPllLQ}
C_9^{qq';\ell\ell'} = 
-C_{10}^{qq';\ell\ell'} & = 
-\frac{v^2 \pi }{2\alpha_{\mathrm{em}} \lambda_{\textsc{CKM}} m_{S_1}^2 } (V^{*} \Delta_{QL})_{q'\ell'} (V  \Delta_{QL})_{q\ell}^{*} \\
C_{9'}^{qq';\ell\ell'} = 
C_{10'}^{qq';\ell\ell'} & = 
-\frac{v^2 \pi }{2\alpha_{\mathrm{em}} \lambda_{\textsc{CKM}} m_{S_1}^2 } (\Delta_{\bar{u}\bar{e}})_{q'\ell'}^{*} (\Delta_{\bar{u}\bar{e}})_{q\ell} \\  
C_{S}^{qq';\ell\ell'} = 
C_{P}^{qq';\ell\ell'} & = 
-\frac{v^2 \pi }{2\alpha_{\mathrm{em}} \lambda_{\textsc{CKM}} m_{S_1}^2 } (\Delta_{\bar{u}\bar{e}})_{q'\ell'}^{*} (V^{*} \Delta_{QL})_{q\ell}^{*} \\  
C_{S'}^{qq';\ell\ell'} =
- C_{P'}^{qq';\ell\ell'} & = 
-\frac{v^2 \pi }{2\alpha_{\mathrm{em}} \lambda_{\textsc{CKM}} m_{S_1}^2 } (V^{*} \Delta_{QL})_{q'\ell'}  (\Delta_{\bar{u}\bar{e}})_{q\ell}.  
\end{aligned}
\end{equation}
For $D$ and $\pi$ mesons decays we set $\lambda_{\textsc{CKM}} = V^{*}_{q'b} V_{qb}$ with $q$, $q'$ referring to the valence quarks of the meson.

In terms of the Wilson coefficients above, the BR of the meson to $\ell^-$ and $\ell'^+$ is given by \cite{Becirevic:2016zri,Dorsner:2016wpm} 
\begin{equation}
\begin{aligned}
\label{eq:Pllprime}
BR (P \rightarrow \ell^- \ell'^{+}) & = \tau_P f_P^2 m_P^3 \frac{\alpha_{\mathrm{em}}^2 G_F^2}{64\pi^3} \lambda_{\textsc{CKM}}^2 \sqrt{\left(1- \frac{(m_1-m_2)^2}{m_P^2}\right)\left(1- \frac{(m_1+m_2)^2}{m_P^2}\right)} \\
& \mkern-36mu \times  \left[  \left( 1- \frac{(m_1+m_2)^2}{m_P^2}\right) \Big| (C_9 - C_{9'}) \frac{m_1 - m_2}{m_P}  +  \frac{m_P}{m_{q'} + m_q} (C_S -C_{S'}) \Big|^2 \right. \\
& \mkern-36mu +  \left. \left( 1- \frac{(m_1-m_2)^2}{m_P^2}\right) \Big| (C_{10} - C_{10'}) \frac{m_1 + m_2}{m_P}  +  \frac{m_P}{m_{q'} + m_q} (C_P -C_{P'}) \Big|^2   \right],
\end{aligned}
\end{equation}
where $\tau_P$ is the meson lifetime, $m_P$ is the meson mass, and $m_1$ ($m_2$) is the mass of the $\ell$ ($\ell'$) lepton.

We can use Eq.~\eqref{eq:Pllprime} to calculate meson decay to a pair of neutrinos too. 
For that, we should set $m_1=m_2=0$ and only keep couplings to LH fermions in the SM. 
Doing that, we find zero contribution for the $S_1$ leptoquark.

\subsection{Semi-leptonic Meson Decays \label{app:RD}}

Next we compute the leptoquark contribution to semi-leptonic meson decays. We ignore constraints from $B\rightarrow K^{(*)} \ell \ell$, since the $S_1$ leptoquark only contributes at loop-level, which is subdominant for leptoquark masses above a few TeV \cite{Buras:2014fpa}.
Instead, we study the more sensitive observables $B\rightarrow D^{(*)} l \nu$ and $K \to \pi \nu \bar{\nu}$, which receive contributions at tree-level. 

\subsubsection{\texorpdfstring{$R_{D^{(*)}}$}{RD}}

$B\rightarrow D^{(*)} l \nu$ proceeds at tree-level via the exchange of the $W$ and the leptoquark \cite{Fajfer:2012jt,Sakaki:2013bfa,Freytsis:2015qca,Bauer:2015knc,Dorsner:2016wpm,Cai:2017wry}. This and other leptoquark models have generated significant interest in the context of $B\rightarrow D^{(*)} l \nu$ because some evidence of a lepton flavor non-universal BSM contribution in this channel, captured by the ratio
\begin{equation}
    R_{D^{(*)}} \equiv \myfrac[3pt]{\mathrm{BR}\left(B\rightarrow D^{(*)} \tau \nu \right)}{\mathrm{BR}\left(B\rightarrow D^{(*)} \ell \nu\right)},
    \label{eq:RDdef}
\end{equation}
has been detected in various experiments \cite{BaBar:2007hvx,Belle:2010tvu,BaBar:2012obs,BaBar:2013mob,LHCb:2015gmp,Belle:2015qfa,Abdesselam:2016xqt} ($\ell=e,\mu$).

When computing the decay rate, integrating the heavy mediators out allows us to work with a set of dimension-6 operators given by 
\begin{equation}
    \mathcal{H}_{\rm eff} = \frac{4G_{F}V_{cb}}{\sqrt{2}}\bigg(\mathcal{O}^{V}_{LL} + \sum_{\substack{X = S,V,T \\ M = L,R}} C^{X}_{ML}\mathcal{O}^{X}_{ML}\bigg)
    \label{eq:HeffRD}
\end{equation}
where
\begin{equation}
    \mathcal{O}^{S}_{ML} \equiv (\bar{c}P_{M}b)(\bar{\tau}P_{L}\nu) \quad
    \mathcal{O}^{V}_{ML} \equiv (\bar{c}\gamma^{\mu}P_{M}b)(\bar{\tau}\gamma_{\mu}P_{L}\nu) \quad
    \mathcal{O}^{T}_{ML} \equiv (\bar{c}\sigma^{\mu\nu}P_{M}b)(\bar{\tau}\sigma_{\mu\nu}P_{L}\nu)
\end{equation}
Note that we have split apart the contributions to the vector operator such that the Wilson coefficients only capture leptoquark contributions.

For the process of interest, the helicity amplitude we wish to compute is
\begin{equation}
    -i \mathcal{M} = \langle\ell(p_{\ell},\lambda_{\ell}),\bar{\nu}_{\ell}(p_{\nu}),D^{(*)}(p_{\mu},\epsilon(\lambda_{M}))|\mathcal{H}_{eff}|B(p_{B})\rangle .
\end{equation}
Each of these operators can be split apart into the constituent quark and lepton bilinears, which allows us to split apart the total amplitude into a product of hadronic and leptonic amplitudes. Details of the calculation can be found in~\cite{Tanaka:2012nw,Asadi:2018sym}. The leptonic amplitudes, which are generically functions of various angles, are identical for both $D$ and $D^{*}$, while the hadronic amplitudes, which are functions of $q^{2}$, vary and are determined by the specific helicity of the $D^{(*)}$ meson. The leptonic amplitudes can be found in multiple references, including~\cite{Tanaka:2012nw}; the expressions for the relevant hadronic functions are taken from~\cite{Sakaki:2013bfa,Bardhan:2016uhr}.\footnote{The correct sign of $h_{T_{3}}(w)$ is in~Ref.~\cite{Sakaki:2013bfa}.}

To compute the differential decay rate, we use 
\begin{equation}
    \frac{\dd\Gamma}{\dd q^2\, \dd \cos \theta} = \frac{1}{2 m_B} \sum_\ell \Big|\mathcal{M}(q^2, \cos \theta)\Big|^2 \frac{\sqrt{(m_B + m_D)^2 - q^2}\sqrt{(m_B - m_D)^2 - q^2}}{256 \pi^{3} m_B^2}\left(1 - \frac{m_{\tau}^2}{q^2}\right)
\end{equation}
where we sum over neutrinos in the final state.
Performing the angular integral over the leptonic functions first, we recover Eqs.~(B.6) and (B.8) in Ref.~\cite{Asadi:2018sym} for the differential decay rates of $B\to D\tau\nu$ and $B\to D^{*}\tau\nu$ respectively. This is the result for a $\tau$ in the final state, but making the replacement $m_{\tau} \to m_{\ell}$ gives us the expression for decays involving any of the SM leptons. The total decay rate can then be obtained by performing the $q^{2}$ integral over the interval $[m_{\ell}^{2}, (m_{B} - m_{D})^{2}]$.

The expressions from (B.6) and (B.8) in Ref.~\cite{Asadi:2018sym} are given in terms of the Wilson coefficients defined in Eq.~\eqref{eq:HeffRD}, therefore, the last ingredient required to complete this computation is the set of pertinent Wilson coefficients for the leptoquark model. They are given by 
\begin{equation}
\begin{aligned}
C^{S}_{LL} & = -\frac{v^{2}}{4m_{LQ}^{2}} \frac{\Delta_{QL}^{3j} \,\Delta_{\bar{u}\bar{e}}^{23}}{V_{cb}} \\ 
C^{V}_{LL} & = \frac{v^{2}}{4m_{LQ}^{2}} \frac{\Delta_{QL}^{3j} (V^{*}\Delta_{QL})^{23\,*}}{V_{cb}} \\ 
C^{T}_{LL} & = \frac{v^{2}}{16m_{LQ}^{2}} \frac{\Delta_{QL}^{3j}\, \Delta_{\bar{u}\bar{e}}^{23}}{V_{cb}}
\end{aligned}
\end{equation}

\subsubsection{\texorpdfstring{$K \to \pi \nu\bar{\nu}$}{K to pi nu nu}}

The decays $K^+ \to \pi^+ \nu\bar{\nu}$ and $K_L \to \pi^0 \nu\bar{\nu}$ can be described with an effective Hamiltonian very similar to Eq.~\eqref{eq:Hbknn}~\cite{Buras:2004uu, Altmannshofer:2009ma}:
\begin{equation}
\mathcal{H}_{\textrm{eff}}  = -\frac{4 G_F}{\sqrt{2}} \bigg[
\mathcal{H}_{\textrm{eff}}^{(c)} + V_{td}^* V_{ts}^{\phantom{*}} (C_L^{K\nu} \mathcal{O}_L^{K\nu} + C_R^{K\nu} \mathcal{O}_R^{K\nu}) + \hc
\bigg]
\end{equation}
where 
\begin{equation}
\mathcal{O}_{L(R)}^{K\nu} = \frac{\alpha_{\mathrm{em}}}{4\pi}(\bar{d}\gamma^{\mu} P_{L(R)} s)(\bar{\nu}\gamma_{\mu}(1-\gamma_5)\nu) ,
\end{equation}
and $\mathcal{H}^{(c)}_{\textrm{eff}}$ includes operators that encode physics below the weak scale.
The branching ratios for $K^+ \to \pi^+\nu\bar{\nu}$ and $K_L \to \pi^0\nu\bar{\nu}$ are then written as
\begin{equation}
\begin{aligned}
\textrm{BR}(K^+ \to \pi^+\nu\bar{\nu}) & = \kappa_+\,\bigg[ 
\Big( \frac{\textrm{Im}(\lambda_t X^{K\nu})}{\lambda^5}\Big)^2 + \Big(-P_{(u,c)} + \frac{\textrm{Re}(\lambda_t X^{K\nu})}{\lambda^5}\Big)^2 \bigg] \\[0.5em]
\textrm{BR}(K_L \to \pi^0 \nu\bar{\nu}) & = \kappa_L\, \Big( \frac{\textrm{Im}(\lambda_t X^K)}{\lambda^5}\Big)^2 
\end{aligned}
\end{equation}
where $X^{K\nu} = -\sin^2\theta_W (C_L^{K\nu} + C_R^{K\nu})$, $\lambda_t = V_{td}^* V_{ts}^{\phantom{*}}$ and $\lambda = 0.2255$ is the Wolfenstein parameter of the CKM matrix. 
The $\kappa$-factors encode input from hadronic matrix elements. Following Ref.~\cite{Altmannshofer:2009ma}, we take $\kappa_+ = (5.27 \pm 0.03) \times 10^{-11}$ and $\kappa_L = (2.27 \pm 0.01) \times 10^{-10}$. The quantity $P_{(u,c)} = 0.41 \pm 0.05$ encodes contributions from charm and light-quark loops. 
These two decays are related via the Grossman-Nir bound \cite{Grossman:1997sk}.

The SM Wilson coefficient $C_L^{K\nu\,\textsc{SM}}$ is the same as Eq.~\eqref{eq:CLSM}, while the leptoquark contribution is
\begin{equation}
C_L^{K\nu} = \frac{v^2}{m_{S_1}^2} \frac{\pi}{2 \alpha_{\textrm{em}}}\frac{\Delta_{QL}^{2k} \Delta_{QL}^{1k\,*}}{\lambda_t}
\end{equation}

We set a constraint on the leptoquark mass by demanding that the total predicted branching ratio (including the SM contribution) be less than the $2\sigma$ upper limit of the measured branching ratio in Ref.~\cite{NA62:2021zjw}: $\textrm{BR}(K^+ \to \pi^+\nu\bar{\nu}) < 1.88 \times 10^{-10}$.
The analogous limit for $K_L$ decays set by the KOTO experiment, $\textrm{BR}(K_L \to \pi^0\nu\bar{\nu}) < 4.9 \times 10^{-9}$~\cite{KOTO:2020prk} is not yet competitive in the context of this model.

\subsection{\texorpdfstring{$Z \rightarrow \ell \ell'$}{Z to l lprime}}

Virtual corrections involving SM fermions and the $S_1$ leptoquark can also contribute to lepton flavor universality violating decays of the SM gauge bosons. 
The strongest bound on the leptoquark comes from measurements of the $Z \to \ell \ell'$ decays, which are constrained by ATLAS~\cite{ATLAS:2014vur}. 
Constraints on $Z$ decays can be cast as bounds on anomalous couplings of the $Z$ boson, $\delta g$, where 
\begin{equation}
\label{eq:Zanomcoup_lag}
\mathcal{L} \supset \frac{g}{\cos\theta_W} \sum_{f,i,j} \bar{f}_i \gamma^{\mu}
\Big[ (\delta_{ij} g^{f_L}_{\textsc{SM}} + \delta g^{f_L}_{ij}) P_L + (\delta_{ij} g^{f_R}_{\textsc{SM}} + \delta g^{f_R}_{ij})P_R \Big] f_j Z_{\mu},
\end{equation}
with $g^{f_L}_{\textsc{SM}} = T_3^f - Q_f \sin^2\theta_W$ and $g^{f_R}_{\textsc{SM}} = -Q_f \sin^2\theta_W$ being the left- and right-handed fermion couplings to the $Z$ boson in the SM.

The $S_1$ leptoquark contributions to these anomalous couplings have been worked out in Refs.~\cite{Becirevic:2017jtw, Arnan:2019olv}. In particular Ref.~\cite{Arnan:2019olv} includes additional finite terms that are numerically important. The $S_1$ leptoquark contributions to the charged lepton couplings of the $Z$ is
\begin{equation}
\begin{aligned}
\label{eq:Zanomcoup_LQ}
\delta g^{\ell\,L(R)}_{ij} & = 
\frac{N_c}{16\pi^2} w_{L(R)}^{tj}(w_{L(R)}^{ti})^*
\bigg[ 
\big(g^{u_{L(R)}}_{\textsc{SM}} - g^{u_{R(L)}}_{\textsc{SM}}\big) \frac{x_t(x_t - 1 - \log x_t)}{(x_t - 1)^2} + \frac{x_Z}{12} F_{L(R)}(x_t)
\bigg] \\
& \qquad + \frac{N_c}{48\pi^2}x_Z \sum_{k = u,c} 
w_{L(R)}^{kj} (w_{L(R)}^{ki})^* \bigg[
g^{u_L(R)}_{\textsc{SM}}\big(\log x_Z - i \pi - \frac{1}{6}\big) + \frac{1}{6}g^{\ell_{L(R)}}_{\textsc{SM}}
\bigg]
\end{aligned}
\end{equation}
where $x_Z = m_Z^2 / m_{S_1}^2$, $x_t = m_t^2/m_{S_1}^2$, $w_L^{ij} = (V^* \Delta_{QL})^{ij}$, $w_R^{ij} = \Delta_{\bar{u}\bar{e}}^{ij}$, and 
$F_{L(R)}(x)$ are loop functions, which can be found in Ref.~\cite{Arnan:2019olv}.

Ref.~\cite{Efrati:2015eaa} sets bounds on combinations of these anomalous couplings with a variety of flavor Ans\"atze, by combining the LFV decay bounds with LEP data at the $Z$-pole~\cite{ALEPH:2005ab}. To extract a constraint on the $S_1$ leptoquark, we simply demand that the anomalous couplings computed above satisfy their bounds assuming generic LFV coupling, which limits 
\begin{equation}
\sqrt{|\delta g^{\ell_L}_{12}|^2 + |\delta g^{\ell_R}_{12}|^2} < 1.2 \times 10^{-3}, \qquad
\sqrt{|\delta g^{\ell_L}_{23}|^2 + |\delta g^{\ell_R}_{23}|^2} < 4.8 \times 10^{-3}
.
\end{equation}
The $e\mu$ bound is most constraining for the anarchic and vanilla FN flavor Ans\"atze, while the $\mu\tau$ bound is strongest with the additional wrinkles from Eq.~\eqref{eq:wrinklesLQ}.

\subsection{Meson Mixing \label{App:Meson_mixing}}
The leptoquark $S_1$ also contributes at the one-loop level to operators in the SM that are responsible for meson mixing. In particular for the down type quarks, the important operator for meson mixing is the dimension-six, four-quark bilinear 
\begin{equation}
\begin{aligned}
\mathcal{H}_{\mathrm{mix}} \supset C^{ij}_{\mathrm{mix}}\,(\bar{d}^{i}_{L}\gamma^\mu d^j_{L})\,(\bar{d}^{i}_{L}\gamma^\mu d^j_{L}).
\label{eq:Hmix}
\end{aligned}
\end{equation}
The associated Wilson coefficient for this operator generated by the $S_1$ leptoquark is \cite{Cai:2017wry}
\begin{align}
\label{eq:Cmix_leptoquark}
C^{ij}_{\mathrm{mix}} = \frac{1}{128\pi^2 m_{S_1}^2}\sum_{k=1}^{3}\left[ (\Delta_{QL}^{ik\ *})\Delta_{QL}^{jk}\right]^2,
\end{align}
where the sum above is over all neutrino flavors. Several experimental quantities of interest can then be derived from this; for instance (in the limit of negligible CP violating phases) the mass difference $\Delta m$ between the mass eigenstates of the oscillating meson is given by 
\begin{equation}
\begin{aligned}
    \Delta m &=\frac{\mel{P}{\mathcal{H}_{\text{mix}}}{\bar{P}}}{m_P}\ =\frac{C^{ij}_{\mathrm{mix}}}{m_{P}} \mel{P}{(\bar{d}^{i}_{L}\gamma^\mu d^j_
    {L})\ (\bar{d}^{i}_{L}\gamma^\mu d^j_
    {L})}{\bar{P}}\ .
\end{aligned}
\end{equation}
Here, $P$ denotes the meson whose constituent down-type quarks are in the $i,j$ generation. The non-perturbative hadronic matrix element above is
\begin{align}
\bra{P}(\bar{d}^{i}_{L}\gamma^\mu d^j_
    {L})\ (\bar{d}^{i}_{L}\gamma^\mu d^j_
    {L}) \ket{\bar{P}} = \frac{2}{3} f_{P}^2 m_{P}^2 B_{P},
\end{align}
where $f_P$ is the meson decay constant and $B_P$ is the meson bag factor, which can be extracted from lattice computations~\cite{Juttner:2007sn, Dowdall:2019bea, Grinstein:2015nya}.

In order to reduce uncertainties from the hadronic matrix elements, we find it advantageous to compare ratios of the matrix elements of the mixing operator (as given in Eq.~\eqref{eq:utfit_ratio}). We define 
\begin{align}
    C_{B_q}e^{2i\phi_{B_q}}=\myfrac[2pt]{\mel{B_q}{\mathcal{H}_{\mathrm{mix}}^{\textsc{SM}+\textsc{NP}}}{\bar{B}_q}}{\mel{B_q}{\mathcal{H}_{\mathrm{mix}}^{\textsc{SM}}}{\bar{B}_q}}\ ,
\end{align}
where $q = d, s$ and by definition in the SM, $C_{B_q}=1$ and $ \phi_{B_q}=0$. By  definition, the $C_{B_q}$ are free from the non-perturbative matrix elements and depend only on perturbative, short-distance Wilson coefficients. The aforementioned ratio is experimentally determined by the UTFit collaboration~\cite{UTfit:2007eik,Cerri:2018ypt,Ferrari:2023slj}, and can be understood as a short-distance proxy for the mass difference $\Delta m$. In principle, there can be intricate interplay between the phases of leptoquark couplings, leading to interference with the SM contributions in this ratio. In this work, we avoid making any assumptions on the underlying complex phases of the leptoquark couplings in $C_{B_q}$, and simply compute the absolute value of $C_{B_q}$.

Additional CP violation from BSM physics is also strongly constrained by other meson mixing measurements, especially in the Kaon system. The quantity of interest is $\epsilon_K$, which, following standard assumptions (see e.g.~\cite{Lenz:2010gu}), is given by 
\begin{equation}
\begin{aligned}
\epsilon_K = \frac{1}{4}\myfrac[2pt]{\mel{K_0}{\mathcal{H}_{\text{mix}}}{\bar{K}_0}}{\mel{\bar{K}_{0}}{\mathcal{H}_{\text{mix}}}{K_0}}
-\frac{1}{4}\ .
\end{aligned}
\end{equation}
To account for $\epsilon_K$, which is much more constraining than the Kaon mass difference, we define
\begin{equation}
C_{\epsilon_K} = \myfrac[2pt]{\Im \langle K^0 | \mathcal{H}_{\mathrm{mix}}^{\textsc{SM}+\textsc{NP}} | \bar{K}^0 \rangle}{\Im \langle K^0 | \mathcal{H}_{\mathrm{mix}}^{\textsc{SM}} | \bar{K}^0 \rangle},
\end{equation}
where again $C_{\epsilon_K} = 1$ in the SM. 

For all of these quantities, we compute the leptoquark contributions using Eq.~\eqref{eq:Cmix_leptoquark}. We compare to the SM matrix elements, which are computed following Refs.~\cite{Buchalla:1995vs, Buras:2005xt, Lenz:2010gu}, including the scale-independent, short-distance QCD corrections. Then we set constraints using the latest results from UTFit~\cite{Ferrari:2023slj}.

We do not consider effects of the $S_1$ leptoquark on mixing in mesons with up-type quarks such as the $D^0$, primarily due large hadronic undertainties \cite{Golowich:2007ka, Bazavov:2017weg} in current SM predictions that make it difficult to glean any information from new physics contributions. 

\end{spacing}

\clearpage
\addcontentsline{toc}{section}{References}
{\small
\bibliographystyle{utphys}
\bibliography{main}

\providecommand{\href}[2]{#2}\begingroup\raggedright\begin{thebibliography}{100}

\bibitem{Bjorken:1964gz}
J.~D. Bjorken and S.~L. Glashow, ``{Elementary Particles and SU(4)},''
  \href{http://dx.doi.org/10.1016/0031-9163(64)90433-0}{{\em Phys. Lett.}
  {\bfseries 11} (1964) 255--257}.

\bibitem{Glashow:1970gm}
S.~L. Glashow, J.~Iliopoulos, and L.~Maiani, ``{Weak Interactions with
  Lepton-Hadron Symmetry},''
  \href{http://dx.doi.org/10.1103/PhysRevD.2.1285}{{\em Phys. Rev. D}
  {\bfseries 2} (1970) 1285--1292}.

\bibitem{Campbell:1981rg}
B.~A. Campbell and P.~J. O'Donnell, ``{Mass of the Top Quark and Induced Decay
  and Neutral Mixing of B Mesons},''
  \href{http://dx.doi.org/10.1103/PhysRevD.25.1989}{{\em Phys. Rev. D}
  {\bfseries 25} (1982) 1989}.

\bibitem{Shifman:1987nd}
M.~A. Shifman, ``{Theoretical Status of Weak Decays},''
  \href{http://dx.doi.org/10.1016/0920-5632(88)90190-9}{{\em Nucl. Phys. B
  Proc. Suppl.} {\bfseries 3} (1988) 289}.

\bibitem{Ellis:1987mm}
J.~R. Ellis, J.~S. Hagelin, S.~Rudaz, and D.~D. Wu, ``{Implications of recent
  measurements of $B$ meson mixing and $\epsilon^\prime / \epsilon_K$},''
  \href{http://dx.doi.org/10.1016/0550-3213(88)90625-6}{{\em Nucl. Phys. B}
  {\bfseries 304} (1988) 205--235}.

\bibitem{Martin:1989ux}
A.~D. Martin, ``{Top and Bottom Physics: The K-M Matrix and CP Violation},''
  \href{http://dx.doi.org/10.1088/0954-3899/15/8/008}{{\em J. Phys. G}
  {\bfseries 15} (1989) 1073}.

\bibitem{CDF:1995wbb}
{\bfseries CDF} Collaboration, F.~Abe {\em et~al.}, ``{Observation of top quark
  production in $\bar{p}p$ collisions},''
  \href{http://dx.doi.org/10.1103/PhysRevLett.74.2626}{{\em Phys. Rev. Lett.}
  {\bfseries 74} (1995) 2626--2631},
  \href{http://arxiv.org/abs/hep-ex/9503002}{{\ttfamily arXiv:hep-ex/9503002}}.

\bibitem{D0:1995jca}
{\bfseries D0} Collaboration, S.~Abachi {\em et~al.}, ``{Observation of the top
  quark},'' \href{http://dx.doi.org/10.1103/PhysRevLett.74.2632}{{\em Phys.
  Rev. Lett.} {\bfseries 74} (1995) 2632--2637},
  \href{http://arxiv.org/abs/hep-ex/9503003}{{\ttfamily arXiv:hep-ex/9503003}}.

\bibitem{EuropeanStrategyforParticlePhysicsPreparatoryGroup:2019qin}
R.~K. Ellis {\em et~al.}, ``{Physics Briefing Book}: {Input for the European
  Strategy for Particle Physics Update 2020},''
  \href{http://arxiv.org/abs/1910.11775}{{\ttfamily arXiv:1910.11775
  [hep-ex]}}.

\bibitem{Artuso:2022ouk}
M.~Artuso {\em et~al.}, ``{Report of the Frontier For Rare Processes and
  Precision Measurements},'' \href{http://arxiv.org/abs/2210.04765}{{\ttfamily
  arXiv:2210.04765 [hep-ex]}}.

\bibitem{Froggatt:1978nt}
C.~D. Froggatt and H.~B. Nielsen, ``{Hierarchy of Quark Masses, Cabibbo Angles
  and CP Violation},''
  \href{http://dx.doi.org/10.1016/0550-3213(79)90316-X}{{\em Nucl. Phys. B}
  {\bfseries 147} (1979) 277--298}.

\bibitem{Leurer:1992wg}
M.~Leurer, Y.~Nir, and N.~Seiberg, ``{Mass matrix models},''
  \href{http://dx.doi.org/10.1016/0550-3213(93)90112-3}{{\em Nucl. Phys. B}
  {\bfseries 398} (1993) 319--342},
  \href{http://arxiv.org/abs/hep-ph/9212278}{{\ttfamily arXiv:hep-ph/9212278}}.

\bibitem{Leurer:1993gy}
M.~Leurer, Y.~Nir, and N.~Seiberg, ``{Mass matrix models: The Sequel},''
  \href{http://dx.doi.org/10.1016/0550-3213(94)90074-4}{{\em Nucl. Phys. B}
  {\bfseries 420} (1994) 468--504},
  \href{http://arxiv.org/abs/hep-ph/9310320}{{\ttfamily arXiv:hep-ph/9310320}}.

\bibitem{Pouliot:1993zm}
P.~Pouliot and N.~Seiberg, ``{(S)quark masses and nonAbelian horizontal
  symmetries},'' \href{http://dx.doi.org/10.1016/0370-2693(93)91801-S}{{\em
  Phys. Lett. B} {\bfseries 318} (1993) 169--173},
  \href{http://arxiv.org/abs/hep-ph/9308363}{{\ttfamily arXiv:hep-ph/9308363}}.

\bibitem{Arkani-Hamed:1999ylh}
N.~Arkani-Hamed and M.~Schmaltz, ``{Hierarchies without symmetries from extra
  dimensions},'' \href{http://dx.doi.org/10.1103/PhysRevD.61.033005}{{\em Phys.
  Rev. D} {\bfseries 61} (2000) 033005},
  \href{http://arxiv.org/abs/hep-ph/9903417}{{\ttfamily arXiv:hep-ph/9903417}}.

\bibitem{Gherghetta:2000qt}
T.~Gherghetta and A.~Pomarol, ``{Bulk fields and supersymmetry in a slice of
  AdS},'' \href{http://dx.doi.org/10.1016/S0550-3213(00)00392-8}{{\em Nucl.
  Phys. B} {\bfseries 586} (2000) 141--162},
  \href{http://arxiv.org/abs/hep-ph/0003129}{{\ttfamily arXiv:hep-ph/0003129}}.

\bibitem{Huber:2000ie}
S.~J. Huber and Q.~Shafi, ``{Fermion masses, mixings and proton decay in a
  Randall-Sundrum model},''
  \href{http://dx.doi.org/10.1016/S0370-2693(00)01399-X}{{\em Phys. Lett. B}
  {\bfseries 498} (2001) 256--262},
  \href{http://arxiv.org/abs/hep-ph/0010195}{{\ttfamily arXiv:hep-ph/0010195}}.

\bibitem{Kaplan:2001ga}
D.~E. Kaplan and T.~M.~P. Tait, ``{New tools for fermion masses from extra
  dimensions},'' \href{http://dx.doi.org/10.1088/1126-6708/2001/11/051}{{\em
  JHEP} {\bfseries 11} (2001) 051},
  \href{http://arxiv.org/abs/hep-ph/0110126}{{\ttfamily arXiv:hep-ph/0110126}}.

\bibitem{Ahmed:2019zxm}
A.~Ahmed, A.~Carmona, J.~Castellano~Ruiz, Y.~Chung, and M.~Neubert,
  ``{Dynamical origin of fermion bulk masses in a warped extra dimension},''
  \href{http://dx.doi.org/10.1007/JHEP08(2019)045}{{\em JHEP} {\bfseries 08}
  (2019) 045}, \href{http://arxiv.org/abs/1905.09833}{{\ttfamily
  arXiv:1905.09833 [hep-ph]}}.

\bibitem{Nelson:2000sn}
A.~E. Nelson and M.~J. Strassler, ``{Suppressing flavor anarchy},''
  \href{http://dx.doi.org/10.1088/1126-6708/2000/09/030}{{\em JHEP} {\bfseries
  09} (2000) 030}, \href{http://arxiv.org/abs/hep-ph/0006251}{{\ttfamily
  arXiv:hep-ph/0006251}}.

\bibitem{Weinberg:1972ws}
S.~Weinberg, ``{Electromagnetic and weak masses},''
  \href{http://dx.doi.org/10.1103/PhysRevLett.29.388}{{\em Phys. Rev. Lett.}
  {\bfseries 29} (1972) 388--392}.

\bibitem{Georgi:1972hy}
H.~Georgi and S.~L. Glashow, ``{Attempts to calculate the electron mass},''
  \href{http://dx.doi.org/10.1103/PhysRevD.7.2457}{{\em Phys. Rev. D}
  {\bfseries 7} (1973) 2457--2463}.

\bibitem{Barr:1976bk}
S.~M. Barr and A.~Zee, ``{A New Approach to the electron-Muon Mass Ratio},''
  \href{http://dx.doi.org/10.1103/PhysRevD.15.2652}{{\em Phys. Rev. D}
  {\bfseries 15} (1977) 2652}.

\bibitem{Balakrishna:1988ks}
B.~S. Balakrishna, A.~L. Kagan, and R.~N. Mohapatra, ``{Quark Mixings and Mass
  Hierarchy From Radiative Corrections},''
  \href{http://dx.doi.org/10.1016/0370-2693(88)91676-0}{{\em Phys. Lett. B}
  {\bfseries 205} (1988) 345--352}.

\bibitem{Babu:2009fd}
K.~S. Babu, \href{http://dx.doi.org/10.1142/9789812838360_0002}{``{TASI
  Lectures on Flavor Physics},''} in {\em {Theoretical Advanced Study Institute
  in Elementary Particle Physics}: {The Dawn of the LHC Era}}, pp.~49--123.
\newblock 2010.
\newblock \href{http://arxiv.org/abs/0910.2948}{{\ttfamily arXiv:0910.2948
  [hep-ph]}}.

\bibitem{Feruglio:2015jfa}
F.~Feruglio, ``{Pieces of the Flavour Puzzle},''
  \href{http://dx.doi.org/10.1140/epjc/s10052-015-3576-5}{{\em Eur. Phys. J. C}
  {\bfseries 75} no.~8, (2015) 373},
  \href{http://arxiv.org/abs/1503.04071}{{\ttfamily arXiv:1503.04071
  [hep-ph]}}.

\bibitem{Feruglio:2019ybq}
F.~Feruglio and A.~Romanino, ``{Lepton flavor symmetries},''
  \href{http://dx.doi.org/10.1103/RevModPhys.93.015007}{{\em Rev. Mod. Phys.}
  {\bfseries 93} no.~1, (2021) 015007},
  \href{http://arxiv.org/abs/1912.06028}{{\ttfamily arXiv:1912.06028
  [hep-ph]}}.

\bibitem{Smolkovic:2019jow}
A.~Smolkovi\v{c}, M.~Tammaro, and J.~Zupan, ``{Anomaly free Froggatt-Nielsen
  models of flavor},'' \href{http://dx.doi.org/10.1007/JHEP10(2019)188}{{\em
  JHEP} {\bfseries 10} (2019) 188},
  \href{http://arxiv.org/abs/1907.10063}{{\ttfamily arXiv:1907.10063
  [hep-ph]}}. [Erratum: JHEP 02, 033 (2022)].

\bibitem{DAmbrosio:2002vsn}
G.~D'Ambrosio, G.~F. Giudice, G.~Isidori, and A.~Strumia, ``{Minimal flavor
  violation: An Effective field theory approach},''
  \href{http://dx.doi.org/10.1016/S0550-3213(02)00836-2}{{\em Nucl. Phys. B}
  {\bfseries 645} (2002) 155--187},
  \href{http://arxiv.org/abs/hep-ph/0207036}{{\ttfamily arXiv:hep-ph/0207036}}.

\bibitem{Egana-Ugrinovic:2018znw}
D.~Egana-Ugrinovic, S.~Homiller, and P.~Meade, ``{Aligned and Spontaneous
  Flavor Violation},''
  \href{http://dx.doi.org/10.1103/PhysRevLett.123.031802}{{\em Phys. Rev.
  Lett.} {\bfseries 123} no.~3, (2019) 031802},
  \href{http://arxiv.org/abs/1811.00017}{{\ttfamily arXiv:1811.00017
  [hep-ph]}}.

\bibitem{Nir:1993mx}
Y.~Nir and N.~Seiberg, ``{Should squarks be degenerate?},''
  \href{http://dx.doi.org/10.1016/0370-2693(93)90942-B}{{\em Phys. Lett. B}
  {\bfseries 309} (1993) 337--343},
  \href{http://arxiv.org/abs/hep-ph/9304307}{{\ttfamily arXiv:hep-ph/9304307}}.

\bibitem{Feldmann:2006jk}
T.~Feldmann and T.~Mannel, ``{Minimal Flavour Violation and Beyond},''
  \href{http://dx.doi.org/10.1088/1126-6708/2007/02/067}{{\em JHEP} {\bfseries
  02} (2007) 067}, \href{http://arxiv.org/abs/hep-ph/0611095}{{\ttfamily
  arXiv:hep-ph/0611095}}.

\bibitem{Bordone:2019uzc}
M.~Bordone, O.~Cat\`a, and T.~Feldmann, ``{Effective Theory Approach to New
  Physics with Flavour: General Framework and a Leptoquark Example},''
  \href{http://dx.doi.org/10.1007/JHEP01(2020)067}{{\em JHEP} {\bfseries 01}
  (2020) 067}, \href{http://arxiv.org/abs/1910.02641}{{\ttfamily
  arXiv:1910.02641 [hep-ph]}}.

\bibitem{Bordone:2020lnb}
M.~Bordone, O.~Cat\`a, T.~Feldmann, and R.~Mandal, ``{Constraining flavour
  patterns of scalar leptoquarks in the effective field theory},''
  \href{http://dx.doi.org/10.1007/JHEP03(2021)122}{{\em JHEP} {\bfseries 03}
  (2021) 122}, \href{http://arxiv.org/abs/2010.03297}{{\ttfamily
  arXiv:2010.03297 [hep-ph]}}.

\bibitem{Dorsner:2016wpm}
I.~Dor\v{s}ner, S.~Fajfer, A.~Greljo, J.~F. Kamenik, and N.~Ko\v{s}nik,
  ``{Physics of leptoquarks in precision experiments and at particle
  colliders},'' \href{http://dx.doi.org/10.1016/j.physrep.2016.06.001}{{\em
  Phys. Rept.} {\bfseries 641} (2016) 1--68},
  \href{http://arxiv.org/abs/1603.04993}{{\ttfamily arXiv:1603.04993
  [hep-ph]}}.

\bibitem{Baver:1994hh}
E.~Baver and M.~Leurer, ``{Naturally light leptoquarks},''
  \href{http://dx.doi.org/10.1103/PhysRevD.51.260}{{\em Phys. Rev. D}
  {\bfseries 51} (1995) 260--264},
  \href{http://arxiv.org/abs/hep-ph/9407324}{{\ttfamily arXiv:hep-ph/9407324}}.

\bibitem{Belle-II:2021rof}
{\bfseries Belle-II} Collaboration, F.~Abudin\'en {\em et~al.}, ``{Search for
  $B^+ \to K^+\nu\bar{\nu}$ Decays Using an Inclusive Tagging Method at Belle
  II},'' \href{http://dx.doi.org/10.1103/PhysRevLett.127.181802}{{\em Phys.
  Rev. Lett.} {\bfseries 127} no.~18, (2021) 181802},
  \href{http://arxiv.org/abs/2104.12624}{{\ttfamily arXiv:2104.12624
  [hep-ex]}}.

\bibitem{Wolfenstein:1983yz}
L.~Wolfenstein, ``{Parametrization of the Kobayashi-Maskawa Matrix},''
  \href{http://dx.doi.org/10.1103/PhysRevLett.51.1945}{{\em Phys. Rev. Lett.}
  {\bfseries 51} (1983) 1945}.

\bibitem{Cabibbo:1963yz}
N.~Cabibbo, ``{Unitary Symmetry and Leptonic Decays},''
  \href{http://dx.doi.org/10.1103/PhysRevLett.10.531}{{\em Phys. Rev. Lett.}
  {\bfseries 10} (1963) 531--533}.

\bibitem{Kobayashi:1973fv}
M.~Kobayashi and T.~Maskawa, ``{CP Violation in the Renormalizable Theory of
  Weak Interaction},'' \href{http://dx.doi.org/10.1143/PTP.49.652}{{\em Prog.
  Theor. Phys.} {\bfseries 49} (1973) 652--657}.

\bibitem{Pontecorvo:1957qd}
B.~Pontecorvo, ``{Inverse beta processes and nonconservation of lepton
  charge},'' {\em Zh. Eksp. Teor. Fiz.} {\bfseries 34} (1957) 247.

\bibitem{Maki:1962mu}
Z.~Maki, M.~Nakagawa, and S.~Sakata, ``{Remarks on the unified model of
  elementary particles},'' \href{http://dx.doi.org/10.1143/PTP.28.870}{{\em
  Prog. Theor. Phys.} {\bfseries 28} (1962) 870--880}.

\bibitem{Aloni:2021wzk}
D.~Aloni, P.~Asadi, Y.~Nakai, M.~Reece, and M.~Suzuki, ``{Spontaneous CP
  violation and horizontal symmetry in the MSSM: toward lepton flavor
  naturalness},'' \href{http://dx.doi.org/10.1007/JHEP09(2021)031}{{\em JHEP}
  {\bfseries 09} (2021) 031}, \href{http://arxiv.org/abs/2104.02679}{{\ttfamily
  arXiv:2104.02679 [hep-ph]}}.

\bibitem{Cornella:2023zme}
C.~Cornella, D.~Curtin, E.~T. Neil, and J.~O. Thompson, ``{Mapping and Probing
  Froggatt-Nielsen Solutions to the Quark Flavor Puzzle},''
  \href{http://arxiv.org/abs/2306.08026}{{\ttfamily arXiv:2306.08026
  [hep-ph]}}.

\bibitem{Nir:1996am}
Y.~Nir and R.~Rattazzi, ``{Solving the supersymmetric CP problem with Abelian
  horizontal symmetries},''
  \href{http://dx.doi.org/10.1016/0370-2693(96)00571-0}{{\em Phys. Lett. B}
  {\bfseries 382} (1996) 363--368},
  \href{http://arxiv.org/abs/hep-ph/9603233}{{\ttfamily arXiv:hep-ph/9603233}}.

\bibitem{Nakai:2021mha}
Y.~Nakai, M.~Reece, and M.~Suzuki, ``{Supersymmetric alignment models for (g -
  2)$_{\mu}$},'' \href{http://dx.doi.org/10.1007/JHEP10(2021)068}{{\em JHEP}
  {\bfseries 10} (2021) 068}, \href{http://arxiv.org/abs/2107.10268}{{\ttfamily
  arXiv:2107.10268 [hep-ph]}}.

\bibitem{Fedele:2020fvh}
M.~Fedele, A.~Mastroddi, and M.~Valli, ``{Minimal Froggatt-Nielsen textures},''
  \href{http://dx.doi.org/10.1007/JHEP03(2021)135}{{\em JHEP} {\bfseries 03}
  (2021) 135}, \href{http://arxiv.org/abs/2009.05587}{{\ttfamily
  arXiv:2009.05587 [hep-ph]}}.

\bibitem{Banks:2010zn}
T.~Banks and N.~Seiberg, ``{Symmetries and Strings in Field Theory and
  Gravity},'' \href{http://dx.doi.org/10.1103/PhysRevD.83.084019}{{\em Phys.
  Rev. D} {\bfseries 83} (2011) 084019},
  \href{http://arxiv.org/abs/1011.5120}{{\ttfamily arXiv:1011.5120 [hep-th]}}.

\bibitem{Wilczek:1982rv}
F.~Wilczek, ``{Axions and Family Symmetry Breaking},''
  \href{http://dx.doi.org/10.1103/PhysRevLett.49.1549}{{\em Phys. Rev. Lett.}
  {\bfseries 49} (1982) 1549--1552}.

\bibitem{Ema:2016ops}
Y.~Ema, K.~Hamaguchi, T.~Moroi, and K.~Nakayama, ``{Flaxion: a minimal
  extension to solve puzzles in the standard model},''
  \href{http://dx.doi.org/10.1007/JHEP01(2017)096}{{\em JHEP} {\bfseries 01}
  (2017) 096}, \href{http://arxiv.org/abs/1612.05492}{{\ttfamily
  arXiv:1612.05492 [hep-ph]}}.

\bibitem{Calibbi:2016hwq}
L.~Calibbi, F.~Goertz, D.~Redigolo, R.~Ziegler, and J.~Zupan, ``{Minimal axion
  model from flavor},''
  \href{http://dx.doi.org/10.1103/PhysRevD.95.095009}{{\em Phys. Rev. D}
  {\bfseries 95} no.~9, (2017) 095009},
  \href{http://arxiv.org/abs/1612.08040}{{\ttfamily arXiv:1612.08040
  [hep-ph]}}.

\bibitem{Bonnefoy:2019lsn}
Q.~Bonnefoy, E.~Dudas, and S.~Pokorski, ``{Chiral Froggatt-Nielsen models,
  gauge anomalies and flavourful axions},''
  \href{http://dx.doi.org/10.1007/JHEP01(2020)191}{{\em JHEP} {\bfseries 01}
  (2020) 191}, \href{http://arxiv.org/abs/1909.05336}{{\ttfamily
  arXiv:1909.05336 [hep-ph]}}.

\bibitem{Allanach:2018vjg}
B.~C. Allanach, J.~Davighi, and S.~Melville, ``{An Anomaly-free Atlas: charting
  the space of flavour-dependent gauged $U(1)$ extensions of the Standard
  Model},'' \href{http://dx.doi.org/10.1007/JHEP02(2019)082}{{\em JHEP}
  {\bfseries 02} (2019) 082}, \href{http://arxiv.org/abs/1812.04602}{{\ttfamily
  arXiv:1812.04602 [hep-ph]}}. [Erratum: JHEP 08, 064 (2019)].

\bibitem{Costa:2019zzy}
D.~B. Costa, B.~A. Dobrescu, and P.~J. Fox, ``{General Solution to the U(1)
  Anomaly Equations},''
  \href{http://dx.doi.org/10.1103/PhysRevLett.123.151601}{{\em Phys. Rev.
  Lett.} {\bfseries 123} no.~15, (2019) 151601},
  \href{http://arxiv.org/abs/1905.13729}{{\ttfamily arXiv:1905.13729
  [hep-th]}}.

\bibitem{Green:1984sg}
M.~B. Green and J.~H. Schwarz, ``{Anomaly Cancellation in Supersymmetric D=10
  Gauge Theory and Superstring Theory},''
  \href{http://dx.doi.org/10.1016/0370-2693(84)91565-X}{{\em Phys. Lett. B}
  {\bfseries 149} (1984) 117--122}.

\bibitem{Barbier:2004ez}
R.~Barbier {\em et~al.}, ``{R-parity violating supersymmetry},''
  \href{http://dx.doi.org/10.1016/j.physrep.2005.08.006}{{\em Phys. Rept.}
  {\bfseries 420} (2005) 1--202},
  \href{http://arxiv.org/abs/hep-ph/0406039}{{\ttfamily arXiv:hep-ph/0406039}}.

\bibitem{Arnold:2013cva}
J.~M. Arnold, B.~Fornal, and M.~B. Wise, ``{Phenomenology of scalar
  leptoquarks},'' \href{http://dx.doi.org/10.1103/PhysRevD.88.035009}{{\em
  Phys. Rev. D} {\bfseries 88} (2013) 035009},
  \href{http://arxiv.org/abs/1304.6119}{{\ttfamily arXiv:1304.6119 [hep-ph]}}.

\bibitem{Altmannshofer:2019xda}
W.~Altmannshofer, J.~Davighi, and M.~Nardecchia, ``{Gauging the accidental
  symmetries of the standard model, and implications for the flavor
  anomalies},'' \href{http://dx.doi.org/10.1103/PhysRevD.101.015004}{{\em Phys.
  Rev. D} {\bfseries 101} no.~1, (2020) 015004},
  \href{http://arxiv.org/abs/1909.02021}{{\ttfamily arXiv:1909.02021
  [hep-ph]}}.

\bibitem{Buchalla:1993wq}
G.~Buchalla and A.~J. Buras, ``{The rare decays $K^+ \to \pi^+ \nu \bar \nu$
  and $K_L \to \mu^+ \mu^-$ beyond leading logarithms},''
  \href{http://dx.doi.org/10.1016/0550-3213(94)90496-0}{{\em Nucl. Phys. B}
  {\bfseries 412} (1994) 106--142},
  \href{http://arxiv.org/abs/hep-ph/9308272}{{\ttfamily arXiv:hep-ph/9308272}}.

\bibitem{Misiak:1999yg}
M.~Misiak and J.~Urban, ``{QCD corrections to FCNC decays mediated by Z
  penguins and W boxes},''
  \href{http://dx.doi.org/10.1016/S0370-2693(99)00150-1}{{\em Phys. Lett. B}
  {\bfseries 451} (1999) 161--169},
  \href{http://arxiv.org/abs/hep-ph/9901278}{{\ttfamily arXiv:hep-ph/9901278}}.

\bibitem{Buchalla:1998ba}
G.~Buchalla and A.~J. Buras, ``{The rare decays $K\to \pi \nu\bar\nu$, $B \to X
  \nu\bar\nu$ and $B \to l^+ l^-$: An Update},''
  \href{http://dx.doi.org/10.1016/S0550-3213(99)00149-2}{{\em Nucl. Phys. B}
  {\bfseries 548} (1999) 309--327},
  \href{http://arxiv.org/abs/hep-ph/9901288}{{\ttfamily arXiv:hep-ph/9901288}}.

\bibitem{Ball:2004ye}
P.~Ball and R.~Zwicky, ``{New results on $B \to \pi, K, \eta$ decay formfactors
  from light-cone sum rules},''
  \href{http://dx.doi.org/10.1103/PhysRevD.71.014015}{{\em Phys. Rev. D}
  {\bfseries 71} (2005) 014015},
  \href{http://arxiv.org/abs/hep-ph/0406232}{{\ttfamily arXiv:hep-ph/0406232}}.

\bibitem{Ball:2004rg}
P.~Ball and R.~Zwicky, ``{$B_{d,s} \to \rho, \omega, K^*, \phi$ decay
  form-factors from light-cone sum rules revisited},''
  \href{http://dx.doi.org/10.1103/PhysRevD.71.014029}{{\em Phys. Rev. D}
  {\bfseries 71} (2005) 014029},
  \href{http://arxiv.org/abs/hep-ph/0412079}{{\ttfamily arXiv:hep-ph/0412079}}.

\bibitem{Khodjamirian:2010vf}
A.~Khodjamirian, T.~Mannel, A.~A. Pivovarov, and Y.~M. Wang, ``{Charm-loop
  effect in $B \to K^{(*)} \ell^{+} \ell^{-}$ and $B\to K^*\gamma$},''
  \href{http://dx.doi.org/10.1007/JHEP09(2010)089}{{\em JHEP} {\bfseries 09}
  (2010) 089}, \href{http://arxiv.org/abs/1006.4945}{{\ttfamily arXiv:1006.4945
  [hep-ph]}}.

\bibitem{Brod:2010hi}
J.~Brod, M.~Gorbahn, and E.~Stamou, ``{Two-Loop Electroweak Corrections for the
  $K \to \pi \nu \bar{\nu}$ Decays},''
  \href{http://dx.doi.org/10.1103/PhysRevD.83.034030}{{\em Phys. Rev. D}
  {\bfseries 83} (2011) 034030},
  \href{http://arxiv.org/abs/1009.0947}{{\ttfamily arXiv:1009.0947 [hep-ph]}}.

\bibitem{Bouchard:2013eph}
{\bfseries HPQCD} Collaboration, C.~Bouchard, G.~P. Lepage, C.~Monahan, H.~Na,
  and J.~Shigemitsu, ``{Rare decay $B \to K \ell^+ \ell^-$ form factors from
  lattice QCD},'' \href{http://dx.doi.org/10.1103/PhysRevD.88.054509}{{\em
  Phys. Rev. D} {\bfseries 88} no.~5, (2013) 054509},
  \href{http://arxiv.org/abs/1306.2384}{{\ttfamily arXiv:1306.2384 [hep-lat]}}.
  [Erratum: Phys.Rev.D 88, 079901 (2013)].

\bibitem{Horgan:2013hoa}
R.~R. Horgan, Z.~Liu, S.~Meinel, and M.~Wingate, ``{Lattice QCD calculation of
  form factors describing the rare decays $B \to K^* \ell^+ \ell^-$ and $B_s
  \to \phi \ell^+ \ell^-$},''
  \href{http://dx.doi.org/10.1103/PhysRevD.89.094501}{{\em Phys. Rev. D}
  {\bfseries 89} no.~9, (2014) 094501},
  \href{http://arxiv.org/abs/1310.3722}{{\ttfamily arXiv:1310.3722 [hep-lat]}}.

\bibitem{Altmannshofer:2009ma}
W.~Altmannshofer, A.~J. Buras, D.~M. Straub, and M.~Wick, ``{New strategies for
  New Physics search in $B \to K^{*} \nu \bar{\nu}$, $B \to K \nu \bar{\nu}$
  and $B \to X_{s} \nu \bar{\nu}$ decays},''
  \href{http://dx.doi.org/10.1088/1126-6708/2009/04/022}{{\em JHEP} {\bfseries
  04} (2009) 022}, \href{http://arxiv.org/abs/0902.0160}{{\ttfamily
  arXiv:0902.0160 [hep-ph]}}.

\bibitem{Buras:2014fpa}
A.~J. Buras, J.~Girrbach-Noe, C.~Niehoff, and D.~M. Straub, ``{$ B\to
  {K}^{\left(\ast \right)}\nu \overline{\nu} $ decays in the Standard Model and
  beyond},'' \href{http://dx.doi.org/10.1007/JHEP02(2015)184}{{\em JHEP}
  {\bfseries 02} (2015) 184}, \href{http://arxiv.org/abs/1409.4557}{{\ttfamily
  arXiv:1409.4557 [hep-ph]}}.

\bibitem{Blake:2016olu}
T.~Blake, G.~Lanfranchi, and D.~M. Straub, ``{Rare $B$ Decays as Tests of the
  Standard Model},'' \href{http://dx.doi.org/10.1016/j.ppnp.2016.10.001}{{\em
  Prog. Part. Nucl. Phys.} {\bfseries 92} (2017) 50--91},
  \href{http://arxiv.org/abs/1606.00916}{{\ttfamily arXiv:1606.00916
  [hep-ph]}}.

\bibitem{Belle:2013tnz}
{\bfseries Belle} Collaboration, O.~Lutz {\em et~al.}, ``{Search for $B \to
  h^{(*)} \nu \bar{\nu}$ with the full Belle $\Upsilon(4S)$ data sample},''
  \href{http://dx.doi.org/10.1103/PhysRevD.87.111103}{{\em Phys. Rev. D}
  {\bfseries 87} no.~11, (2013) 111103},
  \href{http://arxiv.org/abs/1303.3719}{{\ttfamily arXiv:1303.3719 [hep-ex]}}.

\bibitem{BaBar:2013npw}
{\bfseries BaBar} Collaboration, J.~P. Lees {\em et~al.}, ``{Search for $B \to
  K^{(*)} \nu \overline \nu$ and invisible quarkonium decays},''
  \href{http://dx.doi.org/10.1103/PhysRevD.87.112005}{{\em Phys. Rev. D}
  {\bfseries 87} no.~11, (2013) 112005},
  \href{http://arxiv.org/abs/1303.7465}{{\ttfamily arXiv:1303.7465 [hep-ex]}}.

\bibitem{Belle:2017oht}
{\bfseries Belle} Collaboration, J.~Grygier {\em et~al.}, ``{Search for
  $\boldsymbol{B\to h\nu\bar{\nu}}$ decays with semileptonic tagging at
  Belle},'' \href{http://dx.doi.org/10.1103/PhysRevD.96.091101}{{\em Phys. Rev.
  D} {\bfseries 96} no.~9, (2017) 091101},
  \href{http://arxiv.org/abs/1702.03224}{{\ttfamily arXiv:1702.03224
  [hep-ex]}}. [Addendum: Phys.Rev.D 97, 099902 (2018)].

\bibitem{Belle-II:2018jsg}
{\bfseries Belle-II} Collaboration, W.~Altmannshofer {\em et~al.}, ``{The Belle
  II Physics Book},'' \href{http://dx.doi.org/10.1093/ptep/ptz106}{{\em PTEP}
  {\bfseries 2019} no.~12, (2019) 123C01},
  \href{http://arxiv.org/abs/1808.10567}{{\ttfamily arXiv:1808.10567
  [hep-ex]}}. [Erratum: PTEP 2020, 029201 (2020)].

\bibitem{Bauer:2015knc}
M.~Bauer and M.~Neubert, ``{Minimal Leptoquark Explanation for the
  $R_{D^{(*)}}$ , $R_K$ , and $(g-2)_\mu$ Anomalies},''
  \href{http://dx.doi.org/10.1103/PhysRevLett.116.141802}{{\em Phys. Rev.
  Lett.} {\bfseries 116} no.~14, (2016) 141802},
  \href{http://arxiv.org/abs/1511.01900}{{\ttfamily arXiv:1511.01900
  [hep-ph]}}.

\bibitem{Becirevic:2016oho}
D.~Be\v{c}irevi\'c, N.~Ko\v{s}nik, O.~Sumensari, and R.~Zukanovich~Funchal,
  ``{Palatable Leptoquark Scenarios for Lepton Flavor Violation in Exclusive
  $b\to s\ell_1\ell_2$ modes},''
  \href{http://dx.doi.org/10.1007/JHEP11(2016)035}{{\em JHEP} {\bfseries 11}
  (2016) 035}, \href{http://arxiv.org/abs/1608.07583}{{\ttfamily
  arXiv:1608.07583 [hep-ph]}}.

\bibitem{Roussy:2022cmp}
T.~S. Roussy {\em et~al.}, ``{A new bound on the electron's electric dipole
  moment},'' \href{http://dx.doi.org/10.1126/science.adg4084}{{\em Science}
  {\bfseries 381} (2023) 46}, \href{http://arxiv.org/abs/2212.11841}{{\ttfamily
  arXiv:2212.11841 [physics.atom-ph]}}.

\bibitem{Ho:2020ucd}
C.~J. Ho, J.~A. Devlin, I.~M. Rabey, P.~Yzombard, J.~Lim, S.~C. Wright, N.~J.
  Fitch, E.~A. Hinds, M.~R. Tarbutt, and B.~E. Sauer, ``{New techniques for a
  measurement of the electron\textquoteright{}s electric dipole moment},''
  \href{http://dx.doi.org/10.1088/1367-2630/ab83d2}{{\em New J. Phys.}
  {\bfseries 22} no.~5, (2020) 053031},
  \href{http://arxiv.org/abs/2002.02332}{{\ttfamily arXiv:2002.02332
  [physics.atom-ph]}}.

\bibitem{Fitch:2020jil}
N.~J. Fitch, J.~Lim, E.~A. Hinds, B.~E. Sauer, and M.~R. Tarbutt, ``{Methods
  for measuring the electron\textquoteright{}s electric dipole moment using
  ultracold YbF molecules},''
  \href{http://dx.doi.org/10.1088/2058-9565/abc931}{{\em Quantum Sci. Technol.}
  {\bfseries 6} no.~1, (2021) 014006},
  \href{http://arxiv.org/abs/2009.00346}{{\ttfamily arXiv:2009.00346
  [physics.atom-ph]}}.

\bibitem{MEG:2016leq}
{\bfseries MEG} Collaboration, A.~M. Baldini {\em et~al.}, ``{Search for the
  lepton flavour violating decay $\mu ^+ \rightarrow \mathrm {e}^+ \gamma $
  with the full dataset of the MEG experiment},''
  \href{http://dx.doi.org/10.1140/epjc/s10052-016-4271-x}{{\em Eur. Phys. J. C}
  {\bfseries 76} no.~8, (2016) 434},
  \href{http://arxiv.org/abs/1605.05081}{{\ttfamily arXiv:1605.05081
  [hep-ex]}}.

\bibitem{Baldini:2018nnn}
{\bfseries MEG II} Collaboration, A.~Baldini {\em et~al.}, ``{The design of the
  MEG II experiment},''
  \href{http://dx.doi.org/10.1140/epjc/s10052-018-5845-6}{{\em Eur. Phys. J. C}
  {\bfseries 78} no.~5, (2018) 380},
  \href{http://arxiv.org/abs/1801.04688}{{\ttfamily arXiv:1801.04688
  [physics.ins-det]}}.

\bibitem{SINDRUMII:2006dvw}
{\bfseries SINDRUM II} Collaboration, W.~H. Bertl {\em et~al.}, ``{A Search for
  muon to electron conversion in muonic gold},''
  \href{http://dx.doi.org/10.1140/epjc/s2006-02582-x}{{\em Eur. Phys. J. C}
  {\bfseries 47} (2006) 337--346}.

\bibitem{Bartoszek:2014mya}
{\bfseries Mu2e} Collaboration, L.~Bartoszek {\em et~al.}, ``{Mu2e Technical
  Design Report},'' \href{http://arxiv.org/abs/1501.05241}{{\ttfamily
  arXiv:1501.05241 [physics.ins-det]}}.

\bibitem{Abusalma:2018xem}
{\bfseries Mu2e} Collaboration, F.~Abusalma {\em et~al.}, ``{Expression of
  Interest for Evolution of the Mu2e Experiment},''
  \href{http://arxiv.org/abs/1802.02599}{{\ttfamily arXiv:1802.02599
  [physics.ins-det]}}.

\bibitem{Belle:2021ysv}
{\bfseries Belle} Collaboration, A.~Abdesselam {\em et~al.}, ``{Search for
  lepton-flavor-violating tau-lepton decays to $\ell\gamma$ at Belle},''
  \href{http://dx.doi.org/10.1007/JHEP10(2021)019}{{\em JHEP} {\bfseries 10}
  (2021) 19}, \href{http://arxiv.org/abs/2103.12994}{{\ttfamily
  arXiv:2103.12994 [hep-ex]}}.

\bibitem{Belle-II:2022cgf}
{\bfseries Belle-II} Collaboration, L.~Aggarwal {\em et~al.}, ``{Snowmass White
  Paper: Belle II physics reach and plans for the next decade and beyond},''
  \href{http://arxiv.org/abs/2207.06307}{{\ttfamily arXiv:2207.06307
  [hep-ex]}}.

\bibitem{Banerjee:2022vdd}
S.~Banerjee, ``{Searches for Lepton Flavor Violation in Tau Decays at Belle
  II},'' \href{http://dx.doi.org/10.3390/universe8090480}{{\em Universe}
  {\bfseries 8} no.~9, (2022) 480},
  \href{http://arxiv.org/abs/2209.11639}{{\ttfamily arXiv:2209.11639
  [hep-ex]}}.

\bibitem{NA62:2021zjw}
{\bfseries NA62} Collaboration, E.~Cortina~Gil {\em et~al.}, ``{Measurement of
  the very rare K$^{+}$\textrightarrow{}$ {\pi}^{+}\nu \overline{\nu} $
  decay},'' \href{http://dx.doi.org/10.1007/JHEP06(2021)093}{{\em JHEP}
  {\bfseries 06} (2021) 093}, \href{http://arxiv.org/abs/2103.15389}{{\ttfamily
  arXiv:2103.15389 [hep-ex]}}.

\bibitem{NA62:2020upd}
{\bfseries NA62, KLEVER} Collaboration, ``{Rare decays at the CERN
  high-intensity kaon beam facility},''
  \href{http://arxiv.org/abs/2009.10941}{{\ttfamily arXiv:2009.10941
  [hep-ex]}}.

\bibitem{Cerri:2018ypt}
A.~Cerri {\em et~al.}, ``{Report from Working Group 4}: {Opportunities in
  Flavour Physics at the HL-LHC and HE-LHC},''
  \href{http://dx.doi.org/10.23731/CYRM-2019-007.867}{{\em CERN Yellow Rep.
  Monogr.} {\bfseries 7} (2019) 867--1158},
  \href{http://arxiv.org/abs/1812.07638}{{\ttfamily arXiv:1812.07638
  [hep-ph]}}.

\bibitem{Buras:2015qea}
A.~J. Buras, D.~Buttazzo, J.~Girrbach-Noe, and R.~Knegjens, ``{$ {K}^{+}\to
  {\pi}^{+}\nu \overline{\nu} $ and $ {K}_L\to {\pi}^0\nu \overline{\nu} $ in
  the Standard Model: status and perspectives},''
  \href{http://dx.doi.org/10.1007/JHEP11(2015)033}{{\em JHEP} {\bfseries 11}
  (2015) 033}, \href{http://arxiv.org/abs/1503.02693}{{\ttfamily
  arXiv:1503.02693 [hep-ph]}}.

\bibitem{ParticleDataGroup:2022pth}
{\bfseries Particle Data Group} Collaboration, R.~L. Workman {\em et~al.},
  ``{Review of Particle Physics},''
  \href{http://dx.doi.org/10.1093/ptep/ptac097}{{\em PTEP} {\bfseries 2022}
  (2022) 083C01}.

\bibitem{Muong-2:2006rrc}
{\bfseries Muon g-2} Collaboration, G.~W. Bennett {\em et~al.}, ``{Final Report
  of the Muon E821 Anomalous Magnetic Moment Measurement at BNL},''
  \href{http://dx.doi.org/10.1103/PhysRevD.73.072003}{{\em Phys. Rev. D}
  {\bfseries 73} (2006) 072003},
  \href{http://arxiv.org/abs/hep-ex/0602035}{{\ttfamily arXiv:hep-ex/0602035}}.

\bibitem{Muong-2:2021ojo}
{\bfseries Muon g-2} Collaboration, B.~Abi {\em et~al.}, ``{Measurement of the
  Positive Muon Anomalous Magnetic Moment to 0.46 ppm},''
  \href{http://dx.doi.org/10.1103/PhysRevLett.126.141801}{{\em Phys. Rev.
  Lett.} {\bfseries 126} no.~14, (2021) 141801},
  \href{http://arxiv.org/abs/2104.03281}{{\ttfamily arXiv:2104.03281
  [hep-ex]}}.

\bibitem{Borsanyi:2020mff}
S.~Borsanyi {\em et~al.}, ``{Leading hadronic contribution to the muon magnetic
  moment from lattice QCD},''
  \href{http://dx.doi.org/10.1038/s41586-021-03418-1}{{\em Nature} {\bfseries
  593} no.~7857, (2021) 51--55},
  \href{http://arxiv.org/abs/2002.12347}{{\ttfamily arXiv:2002.12347
  [hep-lat]}}.

\bibitem{Wang:2022lkq}
{\bfseries chiQCD} Collaboration, G.~Wang, T.~Draper, K.-F. Liu, and Y.-B.
  Yang, ``{Muon g-2 with overlap valence fermions},''
  \href{http://arxiv.org/abs/2204.01280}{{\ttfamily arXiv:2204.01280
  [hep-lat]}}.

\bibitem{Ce:2022kxy}
M.~C\`e {\em et~al.}, ``{Window observable for the hadronic vacuum polarization
  contribution to the muon g-2 from lattice QCD},''
  \href{http://dx.doi.org/10.1103/PhysRevD.106.114502}{{\em Phys. Rev. D}
  {\bfseries 106} no.~11, (2022) 114502},
  \href{http://arxiv.org/abs/2206.06582}{{\ttfamily arXiv:2206.06582
  [hep-lat]}}.

\bibitem{ExtendedTwistedMass:2022jpw}
{\bfseries Extended Twisted Mass} Collaboration, C.~Alexandrou {\em et~al.},
  ``{Lattice calculation of the short and intermediate time-distance hadronic
  vacuum polarization contributions to the muon magnetic moment using
  twisted-mass fermions},''
  \href{http://dx.doi.org/10.1103/PhysRevD.107.074506}{{\em Phys. Rev. D}
  {\bfseries 107} no.~7, (2023) 074506},
  \href{http://arxiv.org/abs/2206.15084}{{\ttfamily arXiv:2206.15084
  [hep-lat]}}.

\bibitem{FermilabLattice:2022izv}
{\bfseries Fermilab Lattice, MILC, HPQCD} Collaboration, C.~T.~H. Davies {\em
  et~al.}, ``{Windows on the hadronic vacuum polarization contribution to the
  muon anomalous magnetic moment},''
  \href{http://dx.doi.org/10.1103/PhysRevD.106.074509}{{\em Phys. Rev. D}
  {\bfseries 106} no.~7, (2022) 074509},
  \href{http://arxiv.org/abs/2207.04765}{{\ttfamily arXiv:2207.04765
  [hep-lat]}}.

\bibitem{Bazavov:2023has}
A.~Bazavov {\em et~al.}, ``{Light-quark connected intermediate-window
  contributions to the muon $g-2$ hadronic vacuum polarization from lattice
  QCD},'' \href{http://arxiv.org/abs/2301.08274}{{\ttfamily arXiv:2301.08274
  [hep-lat]}}.

\bibitem{Blum:2023qou}
T.~Blum {\em et~al.}, ``{An update of Euclidean windows of the hadronic vacuum
  polarization},'' \href{http://arxiv.org/abs/2301.08696}{{\ttfamily
  arXiv:2301.08696 [hep-lat]}}.

\bibitem{CMD-3:2023alj}
{\bfseries CMD-3} Collaboration, F.~V. Ignatov {\em et~al.}, ``{Measurement of
  the $e^+e^-\to\pi^+\pi^-$ cross section from threshold to 1.2 GeV with the
  CMD-3 detector},'' \href{http://arxiv.org/abs/2302.08834}{{\ttfamily
  arXiv:2302.08834 [hep-ex]}}.

\bibitem{Stockinger:2006zn}
D.~Stockinger, ``{The Muon Magnetic Moment and Supersymmetry},''
  \href{http://dx.doi.org/10.1088/0954-3899/34/2/R01}{{\em J. Phys. G}
  {\bfseries 34} (2007) R45--R92},
  \href{http://arxiv.org/abs/hep-ph/0609168}{{\ttfamily arXiv:hep-ph/0609168}}.

\bibitem{Blanke:2007db}
M.~Blanke, A.~J. Buras, B.~Duling, A.~Poschenrieder, and C.~Tarantino,
  ``{Charged Lepton Flavour Violation and (g-2)(mu) in the Littlest Higgs Model
  with T-Parity: A Clear Distinction from Supersymmetry},''
  \href{http://dx.doi.org/10.1088/1126-6708/2007/05/013}{{\em JHEP} {\bfseries
  05} (2007) 013}, \href{http://arxiv.org/abs/hep-ph/0702136}{{\ttfamily
  arXiv:hep-ph/0702136}}.

\bibitem{Pospelov:2008zw}
M.~Pospelov, ``{Secluded U(1) below the weak scale},''
  \href{http://dx.doi.org/10.1103/PhysRevD.80.095002}{{\em Phys. Rev. D}
  {\bfseries 80} (2009) 095002},
  \href{http://arxiv.org/abs/0811.1030}{{\ttfamily arXiv:0811.1030 [hep-ph]}}.

\bibitem{Feruglio:2008ht}
F.~Feruglio, C.~Hagedorn, Y.~Lin, and L.~Merlo, ``{Lepton Flavour Violation in
  Models with A(4) Flavour Symmetry},''
  \href{http://dx.doi.org/10.1016/j.nuclphysb.2008.10.002}{{\em Nucl. Phys. B}
  {\bfseries 809} (2009) 218--243},
  \href{http://arxiv.org/abs/0807.3160}{{\ttfamily arXiv:0807.3160 [hep-ph]}}.

\bibitem{Dermisek:2013gta}
R.~Dermisek and A.~Raval, ``{Explanation of the Muon g-2 Anomaly with
  Vectorlike Leptons and its Implications for Higgs Decays},''
  \href{http://dx.doi.org/10.1103/PhysRevD.88.013017}{{\em Phys. Rev. D}
  {\bfseries 88} (2013) 013017},
  \href{http://arxiv.org/abs/1305.3522}{{\ttfamily arXiv:1305.3522 [hep-ph]}}.

\bibitem{Agrawal:2014ufa}
P.~Agrawal, Z.~Chacko, and C.~B. Verhaaren, ``{Leptophilic Dark Matter and the
  Anomalous Magnetic Moment of the Muon},''
  \href{http://dx.doi.org/10.1007/JHEP08(2014)147}{{\em JHEP} {\bfseries 08}
  (2014) 147}, \href{http://arxiv.org/abs/1402.7369}{{\ttfamily arXiv:1402.7369
  [hep-ph]}}.

\bibitem{Calibbi:2014yha}
L.~Calibbi, P.~Paradisi, and R.~Ziegler, ``{Lepton Flavor Violation in Flavored
  Gauge Mediation},''
  \href{http://dx.doi.org/10.1140/epjc/s10052-014-3211-x}{{\em Eur. Phys. J. C}
  {\bfseries 74} no.~12, (2014) 3211},
  \href{http://arxiv.org/abs/1408.0754}{{\ttfamily arXiv:1408.0754 [hep-ph]}}.

\bibitem{Calibbi:2018rzv}
L.~Calibbi, R.~Ziegler, and J.~Zupan, ``{Minimal models for dark matter and the
  muon g-2 anomaly},'' \href{http://dx.doi.org/10.1007/JHEP07(2018)046}{{\em
  JHEP} {\bfseries 07} (2018) 046},
  \href{http://arxiv.org/abs/1804.00009}{{\ttfamily arXiv:1804.00009
  [hep-ph]}}.

\bibitem{Saad:2020ihm}
S.~Saad, ``{Combined explanations of $(g-2)_{\mu}$, $R_{D^{(*)}}$,
  $R_{K^{(*)}}$ anomalies in a two-loop radiative neutrino mass model},''
  \href{http://dx.doi.org/10.1103/PhysRevD.102.015019}{{\em Phys. Rev. D}
  {\bfseries 102} no.~1, (2020) 015019},
  \href{http://arxiv.org/abs/2005.04352}{{\ttfamily arXiv:2005.04352
  [hep-ph]}}.

\bibitem{Calibbi:2020emz}
L.~Calibbi, M.~L. L\'opez-Ib\'a\~nez, A.~Melis, and O.~Vives, ``{Muon and
  electron $g-2$ and lepton masses in flavor models},''
  \href{http://dx.doi.org/10.1007/JHEP06(2020)087}{{\em JHEP} {\bfseries 06}
  (2020) 087}, \href{http://arxiv.org/abs/2003.06633}{{\ttfamily
  arXiv:2003.06633 [hep-ph]}}.

\bibitem{Calibbi:2021qto}
L.~Calibbi, M.~L. L\'opez-Ib\'a\~nez, A.~Melis, and O.~Vives, ``{Implications
  of the Muon g-2 result on the flavour structure of the lepton mass matrix},''
  \href{http://dx.doi.org/10.1140/epjc/s10052-021-09741-1}{{\em Eur. Phys. J.
  C} {\bfseries 81} no.~10, (2021) 929},
  \href{http://arxiv.org/abs/2104.03296}{{\ttfamily arXiv:2104.03296
  [hep-ph]}}.

\bibitem{FileviezPerez:2021lkq}
P.~Fileviez~Perez, C.~Murgui, and A.~D. Plascencia, ``{Leptoquarks and matter
  unification: Flavor anomalies and the muon g-2},''
  \href{http://dx.doi.org/10.1103/PhysRevD.104.035041}{{\em Phys. Rev. D}
  {\bfseries 104} no.~3, (2021) 035041},
  \href{http://arxiv.org/abs/2104.11229}{{\ttfamily arXiv:2104.11229
  [hep-ph]}}.

\bibitem{Lopez-Ibanez:2021yzu}
M.~L. L\'opez-Ib\'a\~nez, A.~Melis, M.~J. P\'erez, M.~H. Rahat, and O.~Vives,
  ``{Constraining low-scale flavor models with (g-2)$_\mu$ and lepton flavor
  violation},'' \href{http://dx.doi.org/10.1103/PhysRevD.105.035021}{{\em Phys.
  Rev. D} {\bfseries 105} no.~3, (2022) 035021},
  \href{http://arxiv.org/abs/2112.11455}{{\ttfamily arXiv:2112.11455
  [hep-ph]}}.

\bibitem{Diaz:2017lit}
B.~Diaz, M.~Schmaltz, and Y.-M. Zhong, ``{The leptoquark
  Hunter\textquoteright{}s guide: Pair production},''
  \href{http://dx.doi.org/10.1007/JHEP10(2017)097}{{\em JHEP} {\bfseries 10}
  (2017) 097}, \href{http://arxiv.org/abs/1706.05033}{{\ttfamily
  arXiv:1706.05033 [hep-ph]}}.

\bibitem{Schmaltz:2018nls}
M.~Schmaltz and Y.-M. Zhong, ``{The leptoquark Hunter\textquoteright{}s guide:
  large coupling},'' \href{http://dx.doi.org/10.1007/JHEP01(2019)132}{{\em
  JHEP} {\bfseries 01} (2019) 132},
  \href{http://arxiv.org/abs/1810.10017}{{\ttfamily arXiv:1810.10017
  [hep-ph]}}.

\bibitem{Allanach:2017bta}
B.~C. Allanach, B.~Gripaios, and T.~You, ``{The case for future hadron
  colliders from $B \to K^{(*)} \mu^+ \mu^-$ decays},''
  \href{http://dx.doi.org/10.1007/JHEP03(2018)021}{{\em JHEP} {\bfseries 03}
  (2018) 021}, \href{http://arxiv.org/abs/1710.06363}{{\ttfamily
  arXiv:1710.06363 [hep-ph]}}.

\bibitem{Allanach:2019zfr}
B.~C. Allanach, T.~Corbett, and M.~Madigan, ``{Sensitivity of Future Hadron
  Colliders to Leptoquark Pair Production in the Di-Muon Di-Jets Channel},''
  \href{http://dx.doi.org/10.1140/epjc/s10052-020-7722-3}{{\em Eur. Phys. J. C}
  {\bfseries 80} no.~2, (2020) 170},
  \href{http://arxiv.org/abs/1911.04455}{{\ttfamily arXiv:1911.04455
  [hep-ph]}}.

\bibitem{Bandyopadhyay:2020wfv}
P.~Bandyopadhyay, S.~Dutta, M.~Jakkapu, and A.~Karan, ``{Distinguishing
  Leptoquarks at the LHC/FCC},''
  \href{http://dx.doi.org/10.1016/j.nuclphysb.2021.115524}{{\em Nucl. Phys. B}
  {\bfseries 971} (2021) 115524},
  \href{http://arxiv.org/abs/2007.12997}{{\ttfamily arXiv:2007.12997
  [hep-ph]}}.

\bibitem{Hiller:2021pul}
G.~Hiller, D.~Loose, and I.~Ni\v{s}and\v{z}i\'c, ``{Flavorful leptoquarks at
  the LHC and beyond: spin 1},''
  \href{http://dx.doi.org/10.1007/JHEP06(2021)080}{{\em JHEP} {\bfseries 06}
  (2021) 080}, \href{http://arxiv.org/abs/2103.12724}{{\ttfamily
  arXiv:2103.12724 [hep-ph]}}.

\bibitem{Huang:2021biu}
G.-y. Huang, S.~Jana, F.~S. Queiroz, and W.~Rodejohann, ``{Probing the RK(*)
  anomaly at a muon collider},''
  \href{http://dx.doi.org/10.1103/PhysRevD.105.015013}{{\em Phys. Rev. D}
  {\bfseries 105} no.~1, (2022) 015013},
  \href{http://arxiv.org/abs/2103.01617}{{\ttfamily arXiv:2103.01617
  [hep-ph]}}.

\bibitem{Asadi:2021gah}
P.~Asadi, R.~Capdevilla, C.~Cesarotti, and S.~Homiller, ``{Searching for
  leptoquarks at future muon colliders},''
  \href{http://dx.doi.org/10.1007/JHEP10(2021)182}{{\em JHEP} {\bfseries 10}
  (2021) 182}, \href{http://arxiv.org/abs/2104.05720}{{\ttfamily
  arXiv:2104.05720 [hep-ph]}}.

\bibitem{Bandyopadhyay:2021pld}
P.~Bandyopadhyay, A.~Karan, R.~Mandal, and S.~Parashar, ``{Distinguishing
  signatures of scalar leptoquarks at hadron and muon colliders},''
  \href{http://dx.doi.org/10.1140/epjc/s10052-022-10809-9}{{\em Eur. Phys. J.
  C} {\bfseries 82} no.~10, (2022) 916},
  \href{http://arxiv.org/abs/2108.06506}{{\ttfamily arXiv:2108.06506
  [hep-ph]}}.

\bibitem{Qian:2021ihf}
S.~Qian, C.~Li, Q.~Li, F.~Meng, J.~Xiao, T.~Yang, M.~Lu, and Z.~You,
  ``{Searching for heavy leptoquarks at a muon collider},''
  \href{http://dx.doi.org/10.1007/JHEP12(2021)047}{{\em JHEP} {\bfseries 12}
  (2021) 047}, \href{http://arxiv.org/abs/2109.01265}{{\ttfamily
  arXiv:2109.01265 [hep-ph]}}.

\bibitem{Parashar:2022wrd}
S.~Parashar, A.~Karan, Avnish, P.~Bandyopadhyay, and K.~Ghosh, ``{Phenomenology
  of scalar leptoquarks at the LHC in explaining the radiative neutrino masses,
  muon g-2, and lepton flavor violating observables},''
  \href{http://dx.doi.org/10.1103/PhysRevD.106.095040}{{\em Phys. Rev. D}
  {\bfseries 106} no.~9, (2022) 095040},
  \href{http://arxiv.org/abs/2209.05890}{{\ttfamily arXiv:2209.05890
  [hep-ph]}}.

\bibitem{MuonCollider:2022xlm}
{\bfseries Muon Collider} Collaboration, J.~de~Blas {\em et~al.}, ``{The
  physics case of a 3 TeV muon collider stage},''
  \href{http://arxiv.org/abs/2203.07261}{{\ttfamily arXiv:2203.07261
  [hep-ph]}}.

\bibitem{Adamov:2018vin}
{\bfseries COMET} Collaboration, R.~Abramishvili {\em et~al.}, ``{COMET Phase-I
  Technical Design Report},'' \href{http://dx.doi.org/10.1093/ptep/ptz125}{{\em
  PTEP} {\bfseries 2020} no.~3, (2020) 033C01},
  \href{http://arxiv.org/abs/1812.09018}{{\ttfamily arXiv:1812.09018
  [physics.ins-det]}}.

\bibitem{Angelique:2018svf}
{\bfseries COMET} Collaboration, J.~Ang\'elique {\em et~al.}, ``{COMET - A
  submission to the 2020 update of the European Strategy for Particle Physics
  on behalf of the COMET collaboration},''
  \href{http://arxiv.org/abs/1812.07824}{{\ttfamily arXiv:1812.07824
  [hep-ex]}}.

\bibitem{Mu2e-II:2022blh}
{\bfseries Mu2e-II} Collaboration, K.~Byrum {\em et~al.}, ``{Mu2e-II: Muon to
  electron conversion with PIP-II},'' in {\em {Snowmass 2021}}.
\newblock 3, 2022.
\newblock \href{http://arxiv.org/abs/2203.07569}{{\ttfamily arXiv:2203.07569
  [hep-ex]}}.

\bibitem{CGroup:2022tli}
{\bfseries C. Group} Collaboration, M.~Aoki {\em et~al.}, ``{A New Charged
  Lepton Flavor Violation Program at Fermilab},'' in {\em {Snowmass 2021}}.
\newblock 3, 2022.
\newblock \href{http://arxiv.org/abs/2203.08278}{{\ttfamily arXiv:2203.08278
  [hep-ex]}}.

\bibitem{Ellis:2016yje}
S.~A.~R. Ellis and A.~Pierce, ``{Impact of Future Lepton Flavor Violation
  Measurements in the Minimal Supersymmetric Standard Model},''
  \href{http://dx.doi.org/10.1103/PhysRevD.94.015014}{{\em Phys. Rev. D}
  {\bfseries 94} no.~1, (2016) 015014},
  \href{http://arxiv.org/abs/1604.01419}{{\ttfamily arXiv:1604.01419
  [hep-ph]}}.

\bibitem{Homiller:2022iax}
S.~Homiller, Q.~Lu, and M.~Reece, ``{Complementary signals of lepton flavor
  violation at a high-energy muon collider},''
  \href{http://dx.doi.org/10.1007/JHEP07(2022)036}{{\em JHEP} {\bfseries 07}
  (2022) 036}, \href{http://arxiv.org/abs/2203.08825}{{\ttfamily
  arXiv:2203.08825 [hep-ph]}}.

\bibitem{Baldini:2018uhj}
A.~Baldini {\em et~al.}, ``{A submission to the 2020 update of the European
  Strategy for Particle Physics on behalf of the COMET, MEG, Mu2e and Mu3e
  collaborations},'' \href{http://arxiv.org/abs/1812.06540}{{\ttfamily
  arXiv:1812.06540 [hep-ex]}}.

\bibitem{ACME:2018yjb}
{\bfseries ACME} Collaboration, V.~Andreev {\em et~al.}, ``{Improved limit on
  the electric dipole moment of the electron},''
  \href{http://dx.doi.org/10.1038/s41586-018-0599-8}{{\em Nature} {\bfseries
  562} no.~7727, (2018) 355--360}.

\bibitem{Alarcon:2022ero}
R.~Alarcon {\em et~al.}, ``{Electric dipole moments and the search for new
  physics},'' in {\em {Snowmass 2021}}.
\newblock 3, 2022.
\newblock \href{http://arxiv.org/abs/2203.08103}{{\ttfamily arXiv:2203.08103
  [hep-ph]}}.

\bibitem{UTfit:2007eik}
{\bfseries UTfit} Collaboration, M.~Bona {\em et~al.}, ``{Model-independent
  constraints on $\Delta F=2$ operators and the scale of new physics},''
  \href{http://dx.doi.org/10.1088/1126-6708/2008/03/049}{{\em JHEP} {\bfseries
  03} (2008) 049}, \href{http://arxiv.org/abs/0707.0636}{{\ttfamily
  arXiv:0707.0636 [hep-ph]}}.

\bibitem{Ferrari:2023slj}
{\bfseries UTfit} Collaboration, F.~Ferrari, ``{Updates in the Unitarity
  Triangle fits with UTfit},''
  \href{http://dx.doi.org/10.22323/1.411.0078}{{\em PoS} {\bfseries CKM2021}
  (2023) 078}.

\bibitem{Goudzovski:2022vbt}
E.~Goudzovski {\em et~al.}, ``{New physics searches at kaon and hyperon
  factories},'' \href{http://dx.doi.org/10.1088/1361-6633/ac9cee}{{\em Rept.
  Prog. Phys.} {\bfseries 86} no.~1, (2023) 016201},
  \href{http://arxiv.org/abs/2201.07805}{{\ttfamily arXiv:2201.07805
  [hep-ph]}}.

\bibitem{Aoki:2021cqa}
K.~Aoki {\em et~al.}, ``{Extension of the J-PARC Hadron Experimental Facility:
  Third White Paper},'' \href{http://arxiv.org/abs/2110.04462}{{\ttfamily
  arXiv:2110.04462 [nucl-ex]}}.

\bibitem{Hisano:1995cp}
J.~Hisano, T.~Moroi, K.~Tobe, and M.~Yamaguchi, ``{Lepton flavor violation via
  right-handed neutrino Yukawa couplings in supersymmetric standard model},''
  \href{http://dx.doi.org/10.1103/PhysRevD.53.2442}{{\em Phys. Rev. D}
  {\bfseries 53} (1996) 2442--2459},
  \href{http://arxiv.org/abs/hep-ph/9510309}{{\ttfamily arXiv:hep-ph/9510309}}.

\bibitem{Ellis:2008zy}
J.~R. Ellis, J.~S. Lee, and A.~Pilaftsis, ``{Electric Dipole Moments in the
  MSSM Reloaded},'' \href{http://dx.doi.org/10.1088/1126-6708/2008/10/049}{{\em
  JHEP} {\bfseries 10} (2008) 049},
  \href{http://arxiv.org/abs/0808.1819}{{\ttfamily arXiv:0808.1819 [hep-ph]}}.

\bibitem{Crivellin:2018qmi}
A.~Crivellin, M.~Hoferichter, and P.~Schmidt-Wellenburg, ``{Combined
  explanations of $(g-2)_{\mu,e}$ and implications for a large muon EDM},''
  \href{http://dx.doi.org/10.1103/PhysRevD.98.113002}{{\em Phys. Rev. D}
  {\bfseries 98} no.~11, (2018) 113002},
  \href{http://arxiv.org/abs/1807.11484}{{\ttfamily arXiv:1807.11484
  [hep-ph]}}.

\bibitem{Parikh:2018huy}
C.~Cesarotti, Q.~Lu, Y.~Nakai, A.~Parikh, and M.~Reece, ``{Interpreting the
  Electron EDM Constraint},''
  \href{http://dx.doi.org/10.1007/JHEP05(2019)059}{{\em JHEP} {\bfseries 05}
  (2019) 059}, \href{http://arxiv.org/abs/1810.07736}{{\ttfamily
  arXiv:1810.07736 [hep-ph]}}.

\bibitem{Nakai:2016atk}
Y.~Nakai and M.~Reece, ``{Electric Dipole Moments in Natural Supersymmetry},''
  \href{http://dx.doi.org/10.1007/JHEP08(2017)031}{{\em JHEP} {\bfseries 08}
  (2017) 031}, \href{http://arxiv.org/abs/1612.08090}{{\ttfamily
  arXiv:1612.08090 [hep-ph]}}.

\bibitem{Okada:1999zk}
Y.~Okada, K.-i. Okumura, and Y.~Shimizu, ``{Mu --\ensuremath{>} e gamma and mu
  --\ensuremath{>} 3 e processes with polarized muons and supersymmetric grand
  unified theories},'' \href{http://dx.doi.org/10.1103/PhysRevD.61.094001}{{\em
  Phys. Rev. D} {\bfseries 61} (2000) 094001},
  \href{http://arxiv.org/abs/hep-ph/9906446}{{\ttfamily arXiv:hep-ph/9906446}}.

\bibitem{Kuno:1999jp}
Y.~Kuno and Y.~Okada, ``{Muon decay and physics beyond the standard model},''
  \href{http://dx.doi.org/10.1103/RevModPhys.73.151}{{\em Rev. Mod. Phys.}
  {\bfseries 73} (2001) 151--202},
  \href{http://arxiv.org/abs/hep-ph/9909265}{{\ttfamily arXiv:hep-ph/9909265}}.

\bibitem{Kitano:2002mt}
R.~Kitano, M.~Koike, and Y.~Okada, ``{Detailed calculation of lepton flavor
  violating muon electron conversion rate for various nuclei},''
  \href{http://dx.doi.org/10.1103/PhysRevD.76.059902}{{\em Phys. Rev. D}
  {\bfseries 66} (2002) 096002},
  \href{http://arxiv.org/abs/hep-ph/0203110}{{\ttfamily arXiv:hep-ph/0203110}}.
  [Erratum: Phys.Rev.D 76, 059902 (2007)].

\bibitem{Fajfer:2012jt}
S.~Fajfer, J.~F. Kamenik, I.~Nisandzic, and J.~Zupan, ``{Implications of Lepton
  Flavor Universality Violations in B Decays},''
  \href{http://dx.doi.org/10.1103/PhysRevLett.109.161801}{{\em Phys. Rev.
  Lett.} {\bfseries 109} (2012) 161801},
  \href{http://arxiv.org/abs/1206.1872}{{\ttfamily arXiv:1206.1872 [hep-ph]}}.

\bibitem{Becirevic:2016zri}
D.~Be\v{c}irevi\'c, O.~Sumensari, and R.~Zukanovich~Funchal, ``{Lepton flavor
  violation in exclusive $b\rightarrow s$ decays},''
  \href{http://dx.doi.org/10.1140/epjc/s10052-016-3985-0}{{\em Eur. Phys. J. C}
  {\bfseries 76} no.~3, (2016) 134},
  \href{http://arxiv.org/abs/1602.00881}{{\ttfamily arXiv:1602.00881
  [hep-ph]}}.

\bibitem{Sakaki:2013bfa}
Y.~Sakaki, M.~Tanaka, A.~Tayduganov, and R.~Watanabe, ``{Testing leptoquark
  models in $\bar B \to D^{(*)} \tau \bar\nu$},''
  \href{http://dx.doi.org/10.1103/PhysRevD.88.094012}{{\em Phys. Rev. D}
  {\bfseries 88} no.~9, (2013) 094012},
  \href{http://arxiv.org/abs/1309.0301}{{\ttfamily arXiv:1309.0301 [hep-ph]}}.

\bibitem{Freytsis:2015qca}
M.~Freytsis, Z.~Ligeti, and J.~T. Ruderman, ``{Flavor models for $\bar{B} \to
  D^{(*)} \tau \bar{\nu}$},''
  \href{http://dx.doi.org/10.1103/PhysRevD.92.054018}{{\em Phys. Rev. D}
  {\bfseries 92} no.~5, (2015) 054018},
  \href{http://arxiv.org/abs/1506.08896}{{\ttfamily arXiv:1506.08896
  [hep-ph]}}.

\bibitem{Cai:2017wry}
Y.~Cai, J.~Gargalionis, M.~A. Schmidt, and R.~R. Volkas, ``{Reconsidering the
  One Leptoquark solution: flavor anomalies and neutrino mass},''
  \href{http://dx.doi.org/10.1007/JHEP10(2017)047}{{\em JHEP} {\bfseries 10}
  (2017) 047}, \href{http://arxiv.org/abs/1704.05849}{{\ttfamily
  arXiv:1704.05849 [hep-ph]}}.

\bibitem{BaBar:2007hvx}
{\bfseries BaBar} Collaboration, B.~Aubert {\em et~al.}, ``{Observation of the
  semileptonic decays $B \to D^{*} \tau^{-} \bar{\nu}$( $\tau^{)}$ and evidence
  for $B \to D \tau^{-} \bar{\nu}$( $\tau^{)}$},''
  \href{http://dx.doi.org/10.1103/PhysRevLett.100.021801}{{\em Phys. Rev.
  Lett.} {\bfseries 100} (2008) 021801},
  \href{http://arxiv.org/abs/0709.1698}{{\ttfamily arXiv:0709.1698 [hep-ex]}}.

\bibitem{Belle:2010tvu}
{\bfseries Belle} Collaboration, A.~Bozek {\em et~al.}, ``{Observation of $B^+
  \to \bar{D}^*0 \tau^+ \nu_\tau$ and Evidence for $B^+ \to \bar{D}^0 \tau^+
  \nu_\tau$ at Belle},''
  \href{http://dx.doi.org/10.1103/PhysRevD.82.072005}{{\em Phys. Rev. D}
  {\bfseries 82} (2010) 072005},
  \href{http://arxiv.org/abs/1005.2302}{{\ttfamily arXiv:1005.2302 [hep-ex]}}.

\bibitem{BaBar:2012obs}
{\bfseries BaBar} Collaboration, J.~P. Lees {\em et~al.}, ``{Evidence for an
  excess of $\bar{B} \to D^{(*)} \tau^-\bar{\nu}_\tau$ decays},''
  \href{http://dx.doi.org/10.1103/PhysRevLett.109.101802}{{\em Phys. Rev.
  Lett.} {\bfseries 109} (2012) 101802},
  \href{http://arxiv.org/abs/1205.5442}{{\ttfamily arXiv:1205.5442 [hep-ex]}}.

\bibitem{BaBar:2013mob}
{\bfseries BaBar} Collaboration, J.~P. Lees {\em et~al.}, ``{Measurement of an
  Excess of $\bar{B} \to D^{(*)}\tau^- \bar{\nu}_\tau$ Decays and Implications
  for Charged Higgs Bosons},''
  \href{http://dx.doi.org/10.1103/PhysRevD.88.072012}{{\em Phys. Rev. D}
  {\bfseries 88} no.~7, (2013) 072012},
  \href{http://arxiv.org/abs/1303.0571}{{\ttfamily arXiv:1303.0571 [hep-ex]}}.

\bibitem{LHCb:2015gmp}
{\bfseries LHCb} Collaboration, R.~Aaij {\em et~al.}, ``{Measurement of the
  ratio of branching fractions $\mathcal{B}(\bar{B}^0 \to
  D^{*+}\tau^{-}\bar{\nu}_{\tau})/\mathcal{B}(\bar{B}^0 \to
  D^{*+}\mu^{-}\bar{\nu}_{\mu})$},''
  \href{http://dx.doi.org/10.1103/PhysRevLett.115.111803}{{\em Phys. Rev.
  Lett.} {\bfseries 115} no.~11, (2015) 111803},
  \href{http://arxiv.org/abs/1506.08614}{{\ttfamily arXiv:1506.08614
  [hep-ex]}}. [Erratum: Phys.Rev.Lett. 115, 159901 (2015)].

\bibitem{Belle:2015qfa}
{\bfseries Belle} Collaboration, M.~Huschle {\em et~al.}, ``{Measurement of the
  branching ratio of $\bar{B} \to D^{(\ast)} \tau^- \bar{\nu}_\tau$ relative to
  $\bar{B} \to D^{(\ast)} \ell^- \bar{\nu}_\ell$ decays with hadronic tagging
  at Belle},'' \href{http://dx.doi.org/10.1103/PhysRevD.92.072014}{{\em Phys.
  Rev. D} {\bfseries 92} no.~7, (2015) 072014},
  \href{http://arxiv.org/abs/1507.03233}{{\ttfamily arXiv:1507.03233
  [hep-ex]}}.

\bibitem{Abdesselam:2016xqt}
A.~Abdesselam {\em et~al.}, ``{Measurement of the $\tau$ lepton polarization in
  the decay ${\bar B} \rightarrow D^* \tau^- {\bar \nu_{\tau}}$},''
  \href{http://arxiv.org/abs/1608.06391}{{\ttfamily arXiv:1608.06391
  [hep-ex]}}.

\bibitem{Tanaka:2012nw}
M.~Tanaka and R.~Watanabe, ``{New physics in the weak interaction of $\bar B\to
  D^{(*)}\tau\bar\nu$},''
  \href{http://dx.doi.org/10.1103/PhysRevD.87.034028}{{\em Phys. Rev. D}
  {\bfseries 87} no.~3, (2013) 034028},
  \href{http://arxiv.org/abs/1212.1878}{{\ttfamily arXiv:1212.1878 [hep-ph]}}.

\bibitem{Asadi:2018sym}
P.~Asadi, M.~R. Buckley, and D.~Shih, ``{Asymmetry Observables and the Origin
  of $R_{D^{(*)}}$ Anomalies},''
  \href{http://dx.doi.org/10.1103/PhysRevD.99.035015}{{\em Phys. Rev. D}
  {\bfseries 99} no.~3, (2019) 035015},
  \href{http://arxiv.org/abs/1810.06597}{{\ttfamily arXiv:1810.06597
  [hep-ph]}}.

\bibitem{Bardhan:2016uhr}
D.~Bardhan, P.~Byakti, and D.~Ghosh, ``{A closer look at the R$_{D}$ and
  R$_{D^*}$ anomalies},'' \href{http://dx.doi.org/10.1007/JHEP01(2017)125}{{\em
  JHEP} {\bfseries 01} (2017) 125},
  \href{http://arxiv.org/abs/1610.03038}{{\ttfamily arXiv:1610.03038
  [hep-ph]}}.

\bibitem{Buras:2004uu}
A.~J. Buras, F.~Schwab, and S.~Uhlig, ``{Waiting for precise measurements of
  $K^{+} \to \pi^{+} \nu \bar{\nu}$ and $K_{L} \to \pi^0 \nu \bar{\nu}$},''
  \href{http://dx.doi.org/10.1103/RevModPhys.80.965}{{\em Rev. Mod. Phys.}
  {\bfseries 80} (2008) 965--1007},
  \href{http://arxiv.org/abs/hep-ph/0405132}{{\ttfamily arXiv:hep-ph/0405132}}.

\bibitem{Grossman:1997sk}
Y.~Grossman and Y.~Nir, ``{$K_L \to \pi^0 \nu\bar{\nu}$ beyond the standard
  model},'' \href{http://dx.doi.org/10.1016/S0370-2693(97)00210-4}{{\em Phys.
  Lett. B} {\bfseries 398} (1997) 163--168},
  \href{http://arxiv.org/abs/hep-ph/9701313}{{\ttfamily arXiv:hep-ph/9701313}}.

\bibitem{KOTO:2020prk}
{\bfseries KOTO} Collaboration, J.~K. Ahn {\em et~al.}, ``{Study of the $K_L
  \to \pi^0 \nu \bar \nu$ Decay at the J-PARC KOTO Experiment},''
  \href{http://dx.doi.org/10.1103/PhysRevLett.126.121801}{{\em Phys. Rev.
  Lett.} {\bfseries 126} no.~12, (2021) 121801},
  \href{http://arxiv.org/abs/2012.07571}{{\ttfamily arXiv:2012.07571
  [hep-ex]}}.

\bibitem{ATLAS:2014vur}
{\bfseries ATLAS} Collaboration, G.~Aad {\em et~al.}, ``{Search for the lepton
  flavor violating decay Z\textrightarrow{}e\ensuremath{\mu} in pp collisions
  at $\sqrt{s}$ TeV with the ATLAS detector},''
  \href{http://dx.doi.org/10.1103/PhysRevD.90.072010}{{\em Phys. Rev. D}
  {\bfseries 90} no.~7, (2014) 072010},
  \href{http://arxiv.org/abs/1408.5774}{{\ttfamily arXiv:1408.5774 [hep-ex]}}.

\bibitem{Becirevic:2017jtw}
D.~Be\v{c}irevi\'c and O.~Sumensari, ``{A leptoquark model to accommodate
  $R_K^\mathrm{exp} < R_K^\mathrm{SM}$ and $R_{K^\ast}^\mathrm{exp} <
  R_{K^\ast}^\mathrm{SM}$},''
  \href{http://dx.doi.org/10.1007/JHEP08(2017)104}{{\em JHEP} {\bfseries 08}
  (2017) 104}, \href{http://arxiv.org/abs/1704.05835}{{\ttfamily
  arXiv:1704.05835 [hep-ph]}}.

\bibitem{Arnan:2019olv}
P.~Arnan, D.~Becirevic, F.~Mescia, and O.~Sumensari, ``{Probing low energy
  scalar leptoquarks by the leptonic $W$ and $Z$ couplings},''
  \href{http://dx.doi.org/10.1007/JHEP02(2019)109}{{\em JHEP} {\bfseries 02}
  (2019) 109}, \href{http://arxiv.org/abs/1901.06315}{{\ttfamily
  arXiv:1901.06315 [hep-ph]}}.

\bibitem{Efrati:2015eaa}
A.~Efrati, A.~Falkowski, and Y.~Soreq, ``{Electroweak constraints on flavorful
  effective theories},'' \href{http://dx.doi.org/10.1007/JHEP07(2015)018}{{\em
  JHEP} {\bfseries 07} (2015) 018},
  \href{http://arxiv.org/abs/1503.07872}{{\ttfamily arXiv:1503.07872
  [hep-ph]}}.

\bibitem{ALEPH:2005ab}
{\bfseries ALEPH, DELPHI, L3, OPAL, SLD, LEP Electroweak Working Group, SLD
  Electroweak Group, SLD Heavy Flavour Group} Collaboration, S.~Schael {\em
  et~al.}, ``{Precision electroweak measurements on the $Z$ resonance},''
  \href{http://dx.doi.org/10.1016/j.physrep.2005.12.006}{{\em Phys. Rept.}
  {\bfseries 427} (2006) 257--454},
  \href{http://arxiv.org/abs/hep-ex/0509008}{{\ttfamily arXiv:hep-ex/0509008}}.

\bibitem{Juttner:2007sn}
A.~Juttner, ``{Progress in kaon physics on the lattice},''
  \href{http://dx.doi.org/10.22323/1.042.0014}{{\em PoS} {\bfseries
  LATTICE2007} (2007) 014}, \href{http://arxiv.org/abs/0711.1239}{{\ttfamily
  arXiv:0711.1239 [hep-lat]}}.

\bibitem{Dowdall:2019bea}
R.~J. Dowdall, C.~T.~H. Davies, R.~R. Horgan, G.~P. Lepage, C.~J. Monahan,
  J.~Shigemitsu, and M.~Wingate, ``{Neutral B-meson mixing from full lattice
  QCD at the physical point},''
  \href{http://dx.doi.org/10.1103/PhysRevD.100.094508}{{\em Phys. Rev. D}
  {\bfseries 100} no.~9, (2019) 094508},
  \href{http://arxiv.org/abs/1907.01025}{{\ttfamily arXiv:1907.01025
  [hep-lat]}}.

\bibitem{Grinstein:2015nya}
B.~Grinstein, ``{TASI-2013 Lectures on Flavor Physics},'' in {\em {Theoretical
  Advanced Study Institute in Elementary Particle Physics}: {Particle Physics:
  The Higgs Boson and Beyond}}.
\newblock 1, 2015.
\newblock \href{http://arxiv.org/abs/1501.05283}{{\ttfamily arXiv:1501.05283
  [hep-ph]}}.

\bibitem{Lenz:2010gu}
A.~Lenz, U.~Nierste, J.~Charles, S.~Descotes-Genon, A.~Jantsch, C.~Kaufhold,
  H.~Lacker, S.~Monteil, V.~Niess, and S.~T'Jampens, ``{Anatomy of New Physics
  in $B - \bar{B}$ mixing},''
  \href{http://dx.doi.org/10.1103/PhysRevD.83.036004}{{\em Phys. Rev. D}
  {\bfseries 83} (2011) 036004},
  \href{http://arxiv.org/abs/1008.1593}{{\ttfamily arXiv:1008.1593 [hep-ph]}}.

\bibitem{Buchalla:1995vs}
G.~Buchalla, A.~J. Buras, and M.~E. Lautenbacher, ``{Weak decays beyond leading
  logarithms},'' \href{http://dx.doi.org/10.1103/RevModPhys.68.1125}{{\em Rev.
  Mod. Phys.} {\bfseries 68} (1996) 1125--1144},
  \href{http://arxiv.org/abs/hep-ph/9512380}{{\ttfamily arXiv:hep-ph/9512380}}.

\bibitem{Buras:2005xt}
A.~J. Buras, ``{Flavor physics and CP violation},'' in {\em {2004 European
  School of High-Energy Physics}}, pp.~95--168.
\newblock 5, 2005.
\newblock \href{http://arxiv.org/abs/hep-ph/0505175}{{\ttfamily
  arXiv:hep-ph/0505175}}.

\bibitem{Golowich:2007ka}
E.~Golowich, J.~Hewett, S.~Pakvasa, and A.~A. Petrov, ``{Implications of $D^0$
  - $\bar{D}^0$ Mixing for New Physics},''
  \href{http://dx.doi.org/10.1103/PhysRevD.76.095009}{{\em Phys. Rev. D}
  {\bfseries 76} (2007) 095009},
  \href{http://arxiv.org/abs/0705.3650}{{\ttfamily arXiv:0705.3650 [hep-ph]}}.

\bibitem{Bazavov:2017weg}
A.~Bazavov {\em et~al.}, ``{Short-distance matrix elements for $D^0$-meson
  mixing for $N_f=2+1$ lattice QCD},''
  \href{http://dx.doi.org/10.1103/PhysRevD.97.034513}{{\em Phys. Rev. D}
  {\bfseries 97} no.~3, (2018) 034513},
  \href{http://arxiv.org/abs/1706.04622}{{\ttfamily arXiv:1706.04622
  [hep-lat]}}.

\end{thebibliography}\endgroup
}

\end{document}